\documentclass[aps,prb,twocolumn,groupedaddress]{revtex4-2}
\usepackage{amssymb}
\usepackage{amsmath}
\usepackage{graphicx}
\usepackage{multirow}
\usepackage{color}
\usepackage{ulem}
\usepackage{cancel}

\def\rev{}
\renewcommand{\sout}[1]{\unskip}

\begin{document}


\title{Excitation and tunneling spectra of a fractional quantum Hall system in the thin cylinder limit}


\author{Jyesta M. Adhidewata}
\email[Email:~]{jyesta\_adhidewata@berkeley.edu}
\affiliation{Department of Physics, University of California, Berkeley}

\author{Joel E. Moore}
\affiliation{Department of Physics, University of California, Berkeley}
\affiliation{Materials Sciences Division, Lawrence Berkeley National Laboratories}



\date{\today}

\begin{abstract}
The excitations of fractional quantum Hall effect (FQHE) states have been largely inaccessible to experimental probes until recently. New electron scanning tunneling microscopy (STM) results from Hu et.al. (2023) show promise in detecting and identifying these excited states via the local density of states (LDOS) spectrum. On a torus, there exists a mapping \rev{from the lowest Landau level states} to a 1D lattice \rev{\sout{Hamiltonian with} with a Hamiltonian that features} \rev{ \sout{center-of-mass or}} dipole moment conservation. In this work, we apply perturbation theory starting from the thin cylinder limit ($L_x \rightarrow \infty, L_y <l_B$ for torus dimensions $L_x$ and $L_y$ \rev{and magnetic length $l_B$})  to obtain an analytical approach to the low-lying neutral and charged excitations of the $\nu =1/3$ FQHE state. Notably, in the thin cylinder we can systematically enumerate all the low-lying excitations by the patterns of 'dipoles' formed by the electron occupation pattern on the 1D lattice. We find that the thin-cylinder limit predicts a significant dispersion of the low-lying neutral excitations but sharpness of the  LDOS spectra, which measure charged excitations. We also discuss connections between our work and several different approaches to the FQHE STM spectra, including those using the composite fermion theory. Numerical exact diagonalization beyond the thin-cylinder limit suggests that the energies of charged excitations remain largely confined to a narrow range of energies, which in experiments might appear as a single peak.
\end{abstract}


\maketitle

\section{\label{sec:intro}Introduction}

The fractional quantum hall effect (FQHE) remains one of the most intriguing phenomena in modern condensed matter physics. In a magnetic field, the Hall conductance of sufficiently clean two-dimensional electron gases shows plateaus at rational multiples of $e^2/h$ \cite{Tsui1982, Girvin1999}. The Landau levels arising from the magnetic field not only cause the quantization of \rev{ \sout{integer quantum Hall effect (IQHE)} the Hall conductivity} \cite{Klitzing1980, Thouless1982}, but also quench the kinetic energy of the electrons, allowing electron-electron interactions to dominate. These interactions leads to formation of incompressible liquid ground states hosting quasiparticles with fractional charge and fractional statistics, i.e. anyons \cite{Arovas1984}.

Unfortunately, \rev{\sout{these many-body interaction terms} the strongly interacting nature of the electrons} means that a straightforward solution to the Hamiltonian is practically impossible. One approach used to understand the FQHE is the construction of trial wave functions, such as the celebrated Laughlin wave function \cite{Laughlin1983} for the $\nu = 1/3$ \rev{\sout{filled} filling} state, which are then extended using the hierarchy construction \cite{Haldane1983, Halperin1984} and the composite fermion approach \cite{Jain1989}. Extending these approaches to excited states can be complex, and there is not an obvious small parameter controlling the accuracy of such calculations for realistic interactions. Excited states and electron tunneling spectra are now of great interest because of the recent demonstration that FQHE states in atomically thin van der Waals materials can be directly accessed by scanning tunneling microscopy~\cite{Hu2023}.

An alternate approach to finding a tractable limit from which to begin analyzing the FQHE is the thin-torus limit~\cite{Bergholtz2005, Bergholtz2008, Tao1983}, where one considers a torus with one of the dimensions taken to be very small compared to the magnetic length. 
In the Landau gauge, the electron-electron interaction Hamiltonian can be written in second quantized language as a one-dimensional fermion chain with center-of-mass-conserving two-body electron interaction terms\cite{Seidel2005, Bergholtz2005,Bergholtz2008, Nakamura2010}. 
There exist well-known parent Hamiltonians with pseudopotentials for which FQHE trial wavefunctions become exact, and in many cases some level of continuity has been established between parent and physical Hamiltonians \cite{Haldane1985a, Chen2015}. By tuning the interaction terms in accordance with the Haldane pseudopotential \cite{Trugman1985}, we would then obtain a parent Hamiltonian with the FQHE trial states as its ground state \cite{Haldane2011, Chen2015, Chen2019}. The thin torus limit can be viewed as another way of tuning the 2D Hamiltonian to obtain a simpler ground state - in this limit the interaction coefficients fall off quickly with distance so we can obtain a truncated second quantized Hamiltonian with weak off-diagonal terms.

One-dimensional systems do not have the full complexity of particle statistics present in two dimensions, and the Tao-Thouless limit is viewed now as describing a charge density wave rather than the incompressible quantum liquid. However, there is ample evidence that the 1D charge density wave (CDW) state in the TT limit connects adiabatically to the Laughlin state as the small dimension of the torus grows and the amplitude of the CDW becomes exponentially small \rev{\cite{Rezayi1994, Seidel2005, Bergholtz2008}}.

For a small enough circumference of the torus, the off-diagonal interaction terms in this Hamiltonian decay exponentially with hopping range, which allows us to discard terms involving longer jumps\rev{\sout{.}, significantly simplifying the problem.\sout{For a certain potential this truncated Hamiltonian has an exact analytic zero mode solution} In fact, with fine tuned pseudopotentials, the truncated Hamiltonian can give an exact analytic zero mode solution for certain FQHE states beyond the ultrathin limit \cite{Bergholtz2005, Nakamura2012, Papic2014}. For example, in Ref.~\onlinecite{Papic2014}, using the bosonic Read-Rezayi series Hamiltonian, an analytic solution was found for the bosonic Laughlin $\nu=1/2$ state and Moore-Read $\nu = 5/2$ state}. Previous works have \rev{also} successfully used this truncated Hamiltonian to calculate the ground states of the $\nu = 1/2$ \rev{\sout{\cite{Bergholtz2005}}} and $\nu=1/3$ \rev{\sout{\cite{Nakamura2012}}} states in the TT limit and connect them with important bulk phases - the Luttinger liquid for \rev{\sout{Ref. \onlinecite{Bergholtz2005}} the $\nu=1/2$ phase \cite{Bergholtz2005}} and the Haldane/large-D phase of 1D spin chains for \rev{\sout{Refs. \onlinecite{Nakamura2012} and \onlinecite{Nakamura2010}} the $\nu = 1/3$ phase \cite{Nakamura2012, Nakamura2010}}. This 1D lattice Hamiltonian also has a peculiar center of mass/dipole moment conservation arising from conservation of momentum under a magnetic field. This dipole moment conservation would constrain the movement of electrons and quasiparticles in the system, leading to features such as Hilbert space fragmentation \cite{Sala2020} or a subdiffusive hydrodynamics \cite{Zerba2024}.

Previous works have mostly used the thin torus model to work with either the zero modes or the ground state of a FQHE system (\cite{Bergholtz2005, Jansen2008, Nakamura2010, Jansen2012, Rotondo2016}) or to approximate the band gap of the system (\cite{Weerasinghe2014, Mazaheri2015, Weerasinghe2016, Nachtergaele2021}). In this work, we would like to study the excitations of the $\nu = 1/3$ Laughlin state in the thin torus limit. We specialize further to a thin cylinder limit, where we take the other dimension of the torus to be infinitely large, to obtain the thermodynamic limit of the system. In particular, recent STM experiments by Hu et.al. \cite{Hu2023} have shown promise in probing the charged excitations of a FQHE system. To approach these excitations analytically, we treat the (off-diagonal) hopping terms in the lattice Hamiltonian as a small perturbation to the on-diagonal static terms of the Hamiltonian, similar to the approach in \cite{Nakamura2012} and \cite{Weerasinghe2014}. Using this approach, we find the dispersions of the neutral excitations of the system, which are analogous to the magneto-roton mode \cite{Girvin1986, Repellin2014, Repellin2015}, while for the charged excitations we show that the dipole conservation prevents a broadening of their dispersions. This leads to a STM spectrum with a sharp main peak similar to the recent experimental result. We also performed exact diagonalization on the lattice Hamiltonian as a check and comparison. As  state-of-the-art numerics on fractional quantum Hall states often use extended cylinders or tori and the density-matrix renormalization group method \cite{Shibata2001,Zaletel2013} (and the same is true for a recent study with a quantum computer \cite{Shen2025}), our results also provide an assessment of the utility of future studies using those techniques, as we discuss in closing.

The paper is organized as follows: in Section \ref{sec:model} we introduce the basics of the thin cylinder mapping and the effective 1D lattice Hamiltonian (Subsection \ref{ssec:TTLim}), including the behavior of the interaction coefficients for our choice of potential (Subsection \ref{ssec:CoulombPot}). In Section \ref{sec:res} we present our general results: in Subsection \ref{ssec:uncharged} we used perturbation theory and exact diagonalization \rev{\sout{on the $\nu = 1/3$ ground state} to classify the neutral excitations of the $\nu = 1/3$ ground state and their respective energies}, while in Subsection \ref{ssec:STM} we discuss the charged excitations and their predicted STM spectra. Finally in Subsection \ref{ssec:2Dlimit} we discuss the relevance of our results to larger cylinder or torus size and ultimately the isotropic 2D limit.

\section{\label{sec:model}Model}

In this section, we provide a brief overview of the Tao-Thouless limit and the approach that we use to calculate the interaction coefficients in the cylinder for the long-ranged Coulomb interaction.

\subsection{\label{ssec:TTLim}Tao-Thouless limit} 

\begin{figure*}
\includegraphics[width=0.99\textwidth]{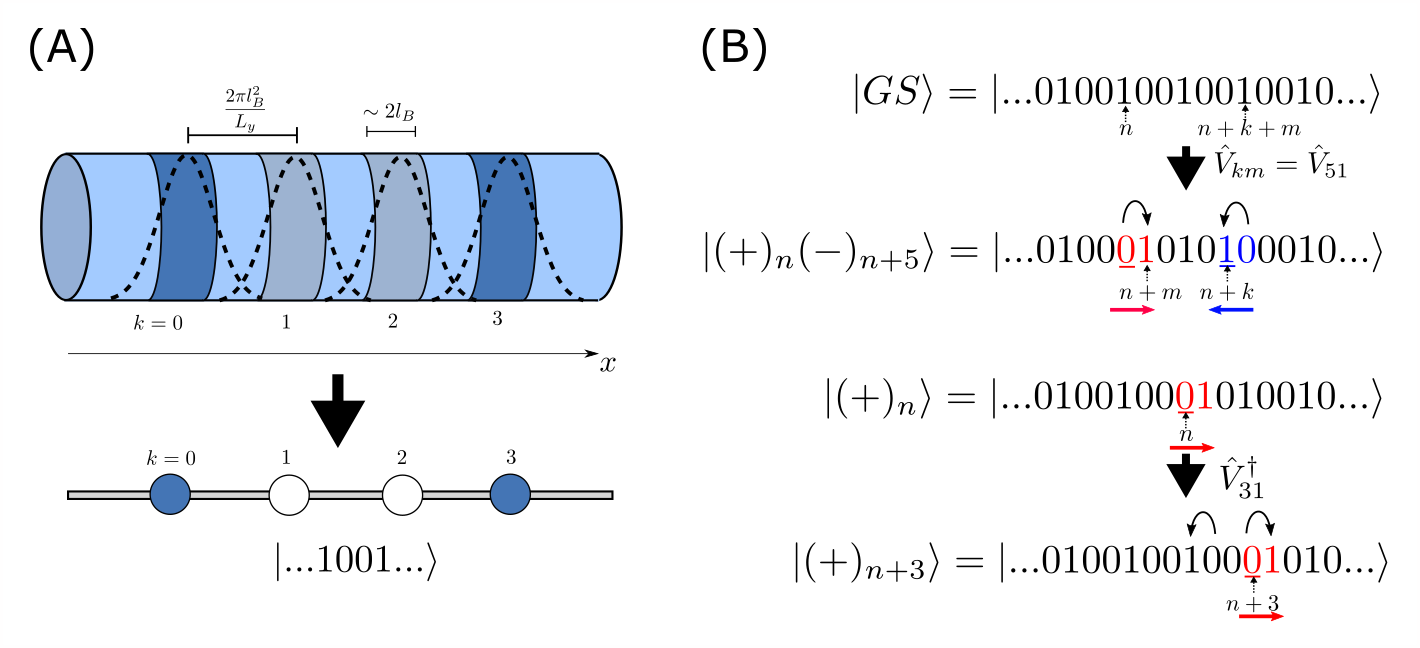}%
\caption{ (A) A schematic of the cylinder/torus to lattice mapping. On the cylinder the single electron wave functions form a Gaussian of width $\sim 2l_B$ in the $x$ direction. With our choice of gauge, the Gaussians are centered at $2\pi l_B^2 k/L_y, k \in 0,1,...,N_{\phi}-1$, forming a 1D lattice. The darker stripes (and the corresponding filled circles in the lattice picture below it) signify the occupied states. In a many body system we can then write down the state in terms of the occupation of these Gaussians. When $L_y \rightarrow 0$, the separation between sites increases and the overlap between wave functions decreases exponentially, allowing us to ignore the hopping terms. (B) The action of the hopping term of the Hamiltonian on the many body state. (Top) Applying the hopping term to the ground state produces a state with a dipole pair (see explanation in Section \ref{ssec:uncharged}). (Bottom) The $P=1$ sector, which are unreachable from the ground state. Applying $\hat{V}_{31}$ to this state moves the dipole to an adjacent position, creating another state with the same energy.\label{fig:Cyl}}
\end{figure*}

First we review how to get to the 1D lattice model on a torus of size $L_x \times L_y$. Because of the magnetic field, the translation operators in the $x$ and $y$ directions do not commute: $t_x(a) t_y(b) = t_y(b) t_x(a) e^{iab/l_B^2}$ \cite{Girvin2019}, where $l_B = \sqrt{\hbar c/eB}$ is the magnetic length. Consistency with the periodic boundary condition then requires $L_x L_y/2\pi l_B^2 \in \mathbb{Z}$. This sets the number of magnetic flux going through the torus to be an integer $N_{\phi} = L_x L_y/(2\pi l_B^2)$. \rev{We will assume that the magnetic field is large enough that we can ignore the higher Landau levels.}. Working in the Landau gauge with $\mathbf{A} = Bx ~\mathbf{\hat{y}}$, \rev{\sout{in} on} the torus the lowest Landau level single electron wave functions \rev{with $y$-momentum index $k \in \mathbb{Z}$}  are given by \cite{Repellin2014}
\begin{equation}
    \phi_k(\mathbf{r}) = \frac{1}{\sqrt{\sqrt{\pi} L_y l_B}} \sum_{l \in \mathbb{Z}} e^{i\frac{2\pi l_B^2}{L_y}(k+lN_{\phi})y} e^{-\frac{1}{2l_B^2}(x-\frac{2\pi l_B}{L_y}(k+lN_{\phi}))^2}
    \label{eq:LLelectron}
\end{equation}
We see that in the Landau gauge the wave functions form Gaussians of width $\sim l_B$ \rev{along the $x-$ direction}, centered at \rev{$x = \frac{2\pi l_B^2}{L_y} \times (k~\mathrm{mod} ~N_{\phi})$}, forming a 1D lattice of sites indexed by the $y$-momentum \rev{indices} $k$ (see Figure~\ref{fig:Cyl}(A)). Assuming a completely spin polarized system, each $k$, and thus each site, can only be occupied by one electron. When $L_y$ is smaller than $2\pi l_B$, the width of the Gaussians become much smaller than the distance between them. This vanishing overlap between wave functions will exponentially suppress the higher-order hopping terms, such that the density-density interaction terms will be dominant. In the $L_y \rightarrow 0$ limit we would have a \rev{\sout{static} classical} Hamiltonian \rev{(in the sense that the occupational basis states are its eigenstates)} with only density-density interactions.  

Applying this lattice construction to many-body states, we then form the Fock space spanned by $\vert n_1,n_2, ...,n_k,n_{k+1},... \rangle$for multiple electrons, with $n_k = 0$ or $1$ denoting the occupation of the $k$-th lattice site. Introducing the electron creation and annihilation operators $\hat{c}_k^{\dagger},~\hat{c}_k$, which creates/annihilates an electron with momentum number $k$ (or equivalently, in the $k$-th lattice site), we can then write the second quantized Hamiltonian \cite{Bergholtz2008}:
\begin{equation}
    H = \frac{1}{2} \sum_{\rev{n},k,m = 0}^{N_{\phi}-1} V_{km} \hat{c}_{n+k}^{\dagger} \hat{c}_{n+m}^{\dagger} \hat{c}_{n+k+m} \hat{c}_n
    \label{eq:HamSym}
\end{equation}

\begin{widetext}
where the interaction coefficients $V_{km}$ \rev{(with $k, m \in \mathbb{Z}$)} can be derived from our choice of interaction potential $V(r)$:
\begin{eqnarray}
    V_{km} = &&\frac{1}{2}\int d^2\mathbf{r}_1 d^2\mathbf{r}_2~  V(\vert \mathbf{r}_1 - \mathbf{r}_2\vert ) \left[\phi^*_{k}(\mathbf{r}_1) \phi^*_{m} (\mathbf{r}_2) \phi_{k+m}(\mathbf{r}_1) \phi_0(\mathbf{r}_2) \nonumber - \phi^*_{k}(\mathbf{r}_1) \phi^*_{m}(\mathbf{r}_2) \phi_{k+m}(\mathbf{r}_2) \phi_0(\mathbf{r}_1)\right] + \mathrm{h.c.}
\end{eqnarray}
\end{widetext}
\rev{While in the Hamiltonian (Eq.~\eqref{eq:HamSym}) we only use coefficients with positive $k$ and $m$, in general we allow $k$ (and $m$) to take negative values, with $V_{-k,m} = V_{km}^* = V_{km}$ as we shall show later. \sout{The coefficients $V_{km}$ obey $V_{km} = -V_{mk}$ from fermion antisymmetrization so we can rewrite the Hamiltonian to only contain terms with $m>k$} Fermionic antisymmetry forces that only the
anti-symmetric part of the interaction matrix elements to be relevant. Using $V_{km} = -V_{mk}$ we can then rewrite the Hamiltonian to only contain terms with $m>k$ ~\cite{Bergholtz2008}}
\begin{eqnarray}
    H  = \sum_{n} \sum_{k=0}^{N_{\phi}/2} \sum_{k<m\leq N_{\phi}/2} \frac{V_{km}}{1+\delta_{k,\frac{N_{\phi}}{2}}} \hat{c}_{n+k}^{\dagger} \hat{c}_{n+m}^{\dagger} \hat{c}_{n+k+m} \hat{c}_n \nonumber \\ \label{eq:HamAsym} 
\end{eqnarray}
In this form it is clear that for $m\neq 0$ $V_{km}$ is the hopping energy for the process where two electrons separated by $k+m$ sites hop by the same distance ($m$ sites) in opposite directions, such that their total center of mass/total dipole moment is conserved. The $m=0$ term, meanwhile, \rev{\sout{is just the Coulomb interaction term between two electrons separated by $k$.} is just the density-density interaction term between electrons in site $n$ and site $n+k$. } 

This Hamiltonian has several important symmetries. Obviously it conserves particle number $\hat{N} = \sum_{k} \hat{c}_k^{\dagger} \hat{c}_k$. The translation symmetries of the Hamiltonian are a bit more subtle (For a complete explanation, see Haldane (1985) \cite{Haldane1985}). Clearly, the Hamiltonian as written in Eq.~\eqref{eq:HamAsym} not invariant under continuous translations, even though our system should be fully translationally invariant. This symmetry breaking is related to the magnetic field and our choice of gauge. Periodic boundary conditions require only \textit{physical quantities} to be invariant after translation around the torus, which allows for a difference in phase after translation - generally we have $ t_i (L_i) \vert \psi \rangle = e^{i\theta_i} \vert \psi \rangle$. Then, for our system to be consistent, all states and operators have to be formed from eigenstates of $t_i(L_i)$ \textit{with the same} $e^{i\theta_i}$, which means we have to choose a set $\theta_i$ and enforce periodic boundary conditions using that $\theta_i$ for all states. When we set $\mathbf{A} = Bx~ \mathbf{\hat{y}}$ (instead of $\mathbf{A} = B(x-a) \mathbf{\hat{y}}$ for some constant $a$), we are choosing $\theta_i = 0$, which gives Eq.\eqref{eq:LLelectron} as our basis wave functions. The states with different $\theta_y$ can be obtained by translating  $\phi_{\mathbf{k}}(\mathbf{r})$ by some $x \in (0, 2\pi l_B^2/L_y)$, which is equivalent to shifting the gauge by the same amount. Thus our choice of gauge breaks the translation symmetry in the $x$ direction, allowing us to write the Hamiltonian in lattice form. 

With $\theta_i$ fixed for the single-electron problem, we can move to the many-particle problem, and define center-of-mass translation operators $T_i(a) = \prod_n^N t^n_i(a)$, where the translation operator $t^n_i$ translates the $n$-th electron in the system by the same amount $a$. For this translation operator to respect the fixed $\theta_i$ that we chose before, we require $\mathbf{a} \in \frac{2\pi l_B^2}{L_y} m ~\mathbf{\hat{x}} + \frac{2\pi l_B^2}{L_x} n~\mathbf{\hat{y}}$ with $m,n \in \mathbb{Z}$. With this restriction, it is clear that both $T_i$ commute with our Hamiltonian. However, $T_x \equiv T_x(\frac{2\pi l_B^2}{L_y})$ and $T_y \equiv T_y(\frac{2\pi l_B^2}{L_x})$ do not commute with each other - for a $\nu = 1/m$ state we have in fact \cite{Bergholtz2008, Repellin2014}
\begin{eqnarray}
    T_x T_y =  e^{2\pi i/m}T_y T_x
\end{eqnarray}
One completely commuting set of operators would then be $\hat{H}, T_x^m$ and $T_y$. The quantum numbers for this set can then be derived from the periodic boundary conditions: $T_i^{N_{\phi}} =  1$, which gives eigenvalues of $e^{i 2\pi mq/N_{\phi}} = e^{i2\pi q/N}, q = 0,1,..,N-1$ for $\rev{T_x^m}$ and $e^{i 2\pi P/N_{\phi}}, P = 0,1,.., N_{\phi}-1$ for $T_y$. From the commutation relation, applying $T_x$ to a state changes $P$ by $N$, while from the lattice Hamiltonian we see that the same thing happened when we shift $\hat{c}_k \rightarrow \hat{c}_{k+1}$, as expected. It is clear then that $q$ can be taken as the x-momentum of the many body state, while $P$ is \rev{nothing} other than the sum of $y$-momentum of all electrons, $\hat{P} = \sum k\hat{c}^{\dagger}_k \hat{c}_k$ \rev{(modulo $N_{\phi}$ on the torus)}, and  this is the source of the center-of-mass/dipole conservation of our Hamiltonian.

Fourier transforming the potential into $\tilde{V}(\mathbf{q}) = \int d^2 \mathbf{r} e^{i\mathbf{q}\cdot \mathbf{r})} V(\mathbf{r})$, we obtain the following expression
\begin{widetext}
\begin{eqnarray}
    V_{km} = \frac{1}{L_x L_y} && \left\{ \sum_{\rev{l \in \mathbb{Z},~n \geq \delta_{m,0}}} e^{-\frac{2\pi^2 l_B^2}{L_x^2}n^2} e^{-\frac{2\pi^2 l_B^2}{L_y^2}(m-l N_{\phi})^2} \cos{\frac{2\pi k n}{N_{\phi}}} \tilde{V}(\frac{2\pi l_B}{L_x}n,\frac{2\pi l_B}{L_y}(m-l N_{\phi})) \right. \nonumber \\
    -&&\left. \sum_{\rev{l \in \mathbb{Z},~n \geq \delta_{k,0}}} e^{-\frac{2\pi^2 l_B^2}{L_x^2}n^2} e^{-\frac{2\pi^2 l_B^2}{L_y^2}(k-l N_{\phi})^2} \cos{\frac{2\pi m n}{N_{\phi}}} \tilde{V}(\frac{2\pi l_B}{L_x}n,\frac{2\pi l_B}{L_y}(k-l N_{\phi})) \right\}
    \label{eq:VkmFourier}
\end{eqnarray}
\end{widetext}

Here, adding a uniform positive neutralizing background to the system cancels out the $\tilde{V}(0)$ terms from the sum. In this form, it is clear that $V_{km} = -V_{mk}$, and $V_{-k,m} = V_{km}$. If we write the Hamiltonian as in \eqref{eq:HamAsym} where we restrict $k>m$, $V_{km}$ will fall off as $\exp{(\frac{-2\pi^2 l_B^2}{L_y^2} m^2)}$ such that the \rev{\sout{static} non-hopping density-}density interaction term will be dominant. Thus, in the $L_y \rightarrow 0$, we effectively have a \rev{\sout{static} classical} Hamiltonian - reducing the problem to a \rev{\sout{classical}} electrostatic problem. It has been shown for the Laughlin $\nu = 1/m$ state that the energy is minimized when the distances between the neighboring electrons are equal to each other, so that the ground state is formed by the charge density wave (CDW) state where we have one electron every $m$ lattice site\cite{Tao1983, Seidel2005, Bergholtz2005, Bergholtz2008, Nakamura2012}. While this is effectively a 'classical' state with no dynamics of its own, it has been shown \cite{Bergholtz2008} that for a short-ranged pseudopotential this simple CDW ground state is adiabatically connected to the Laughlin wave function describing the bulk Laughlin state. To start seeing what the dynamics of the thin system has to say about the planar system, we need to reintroduce $V_{km}$ for $m \neq 0$ by increasing $L_y$. In order to keep the system computationally tractable, we set $L_y$ to a non-zero but small value ($L_y < 2\pi l_B$). With a small $L_y$, since $\hat{V}_{km}$ are proportional to $\exp{\left(-\frac{2\pi^2 m^2}{L_y^2}\right)}$ the $V_{km}, m> 0$ terms are still much smaller than $V_{k0}$. We can then treat the hopping terms as a perturbation to the TT ground state. 

Formally we set $\lambda = e^{-2\pi^2/L_y^2}$ as a small parameter and expand the energies and wave function to the lowest order of $\lambda$. While unusual, this kind of perturbation expansion has been successfully used before to study the excitation gaps of the Haldane-Rezayi and Gaffnian states \cite{Weerasinghe2014}. As we shall show later, second-order expansion in $V_{k1}$ ($\sim \mathcal{O}( e^{-4\pi^2/L_y^2})$) is enough to obtain a non-zero correction in the energy, so we can discard $V_{k2} \sim \mathcal{O}(e^{-8\pi^2/L_y^2})$ and higher $m$ terms at least initially. 

\subsection{\label{ssec:CoulombPot}\rev{Interaction coefficients}}

\rev{In this work, we choose to use the physical Coulomb potential on a cylinder, that is, 
\begin{eqnarray}
    V(\mathbf{r}) = \frac{e^2}{\epsilon\sqrt{x^2 + \frac{L_y^2}{2\pi^2}(1 - \cos{\frac{2\pi y}{L_y}})}}
\end{eqnarray}
We set the permittivity $\epsilon$ to $1$ for the rest of the paper. Note that when $L_y \rightarrow \infty$ this reduces to the usual Coulomb potential on the plane $\sim (x^2+y^2)^{-1/2}$. Using a physical cylinder geometry for the Coulomb potential instead of a 2D Coulomb potential with periodic boundary condition on the $y$-axis makes it easier to deal with the divergences inherent in using a $\sim 1/x$ potentials. Further discussion on the renormalization procedures we used for the potential is relegated to Appendix A.} 

\begin{figure}
\includegraphics[width=0.52\textwidth]{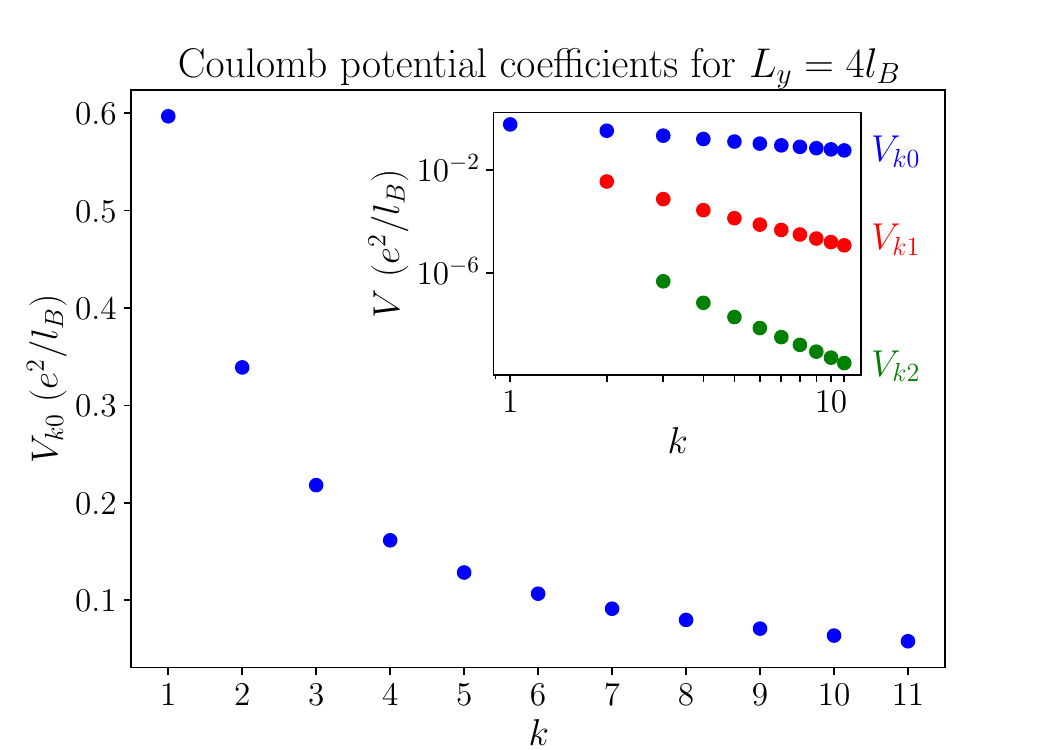}%
\caption{The interaction coefficients for our choice of potential with $L_y = 4l_B$, in units of $e^2/l_B$ (Gaussian units, setting $\epsilon=1$). The main plot shows the static pairwise interaction $V_{k0}$\rev{. The} inset \rev{shows $V_{k0}$ and the hopping terms $V_{k1}$ and $V_{k2}$ on a log-log plot.} From the plot we see that $V_{km}$ has a power law dependence on $k$. \rev{The coefficients for higher $m$ are order of magnitudes smaller, reflecting the exponential dependence of $V_{km}$ on $m$ at small $L_y$}. For our choice of $L_y$, $V_{k1} \leq V_{21} \ll V_{k0}$ for most values of $k$.  \label{fig:Vkm}}
\end{figure}

\rev{The Coulomb potential has a different behavior compared to the short-ranged potentials usually used in the algebraic Hamiltonian construction used in Refs.~\cite{Weerasinghe2014, Chen2015, Chen2019}. Unlike the Trugman-Kivelson potential \cite{Trugman1985} and the contact potential ($V(x) \sim \delta(x)$), which falls off exponentially with both $k^2$ and $m^2$, here the direct (exchange) part of the interaction have a power law dependence on $k$ ($m$) and an exponential dependence on $m$ ($k$): 
\begin{eqnarray}
    V_{km} \sim e^{-\frac{2\pi^2 l_B^2 m^2}{L_y^2}} k^{-\alpha(m)} - e^{-\frac{2\pi^2 l_B^2 k^2}{L_y^2}} m^{-\alpha(k)}
\end{eqnarray}
For $L_y = 4l_B$ we have $\alpha(0) = 0.991$, $\alpha(1) = 3.33$, and $\alpha(2) = 5.54$ (see Fig. \ref{fig:Vkm}). Since we've rewritten our Hamiltonian to only contain $V_{km}$ with $k>m$, for the small $L_y$ regime in which we are most interested in, the first term will dominate the exchange term. The exponential factor in the first term, will in turn suppress the higher hopping terms with $m>1$. This allows us to discard the $m>1$ hopping terms while retaining the longer ranged $m=1$ terms with higher $k$, allowing us to see the dynamics of the system by just including the $V_{k1}$ terms as our perturbation.}

\rev{For small $L_y$, the  $V_{k0}$ interaction coefficients obey the concavity condition: $V_{k+1,0} - 2V_{k0} + V_{k-1,0} >0$, which leads to the charge density wave ground state in the $L_y \rightarrow 0$ limit \cite{Bergholtz2008}. When $L_y$ is increased, the exchange term (second term in Eq.~\eqref{eq:VkmFourier}) would start to become comparable with the direct term, reducing the value of $V_{k0}$ for small $k$ and breaking the concavity condition. However, for most of this paper we will only use values of $L_y$ small enough to maintain the concavity condition. Physically, the reason the exchange force start to become important at larger $L_y$ is because $L_y$ determines the spacing between the sites; at large $L_y$ the spacing is small enough such that two sites separated by a small $k$ would overlap with each other, significantly increasing the exchange force.
}

\section{\label{sec:res}Excitations on the thin cylinder limit}

\subsection{\label{ssec:uncharged}Perturbation theory for the charge-neutral sector}

Before studying the full Hamiltonian of the system, here we first look at the extreme TT limit where the hopping terms are suppressed so that we can enumerate all the possible excitations of the system. As previous works have shown \cite{Bergholtz2008}, for potentials obeying the concavity condition, $V_{k+1,0} - 2 V_{k0} + V_{k-1,0}>0$, the ground state of the TT limit is the CDW state with one electron every $1/\nu$ sites. Writing the Fock space of the system as $\vert ...n_{i-1} n_i n_{i+1} ... \rangle$ where $n_i = 0,1$ denotes the occupation of the $i$-th state, the ground state for the $\nu = 1/3$ can be written as
\begin{eqnarray}
    \vert...010010010010010... \rangle
\end{eqnarray}

As shown in \cite{Bergholtz2008} the low-lying quasihole and quasiparticle excitations of the TT system are domain walls between regions with electrons shifted by a single site:
\begin{eqnarray}
    \vert (-\frac{e}{3})_i \rangle = \vert ...010010\underset{i}{\vert}100100100... \rangle \\
    \vert (+\frac{e}{3})_i \rangle = \vert ...0100100\underset{i}{\vert}01001001...\rangle
\end{eqnarray}
These quasiparticles carry a fractional charge of $\pm e/3$. A quasiparticle-quasihole pair separated by $s/\nu$ lattice sites (i.e. $\vert 010\vert\underbrace{100100100}_{3s}\vert0100$) has an excitation energy of
\begin{eqnarray}
    E_{ph}(s) = \sum_{k=1}^{s-1} k \Delta^2 V_{k/\nu,0} + s \sum_{s \leq k} \Delta^2 V_{k/\nu,0}
    \label{eq:Edomain}
\end{eqnarray}
where $\Delta^2 V_{k0} \equiv V_{k+1,0} - 2 V_{k0} + V_{k-1,0}$. In particular, a minimally separated dipole pair has an energy of $\sum_{k} \Delta^2 V_{3k,0} \equiv E_{ph}$. Note that all of these low-lying excitations have different total dipole moment $\hat{P} = \sum_k k\hat{c}^{\dagger}_k \hat{c}_k$. Defining the ground state to have $P=0$, a pair separated by $s/\nu$ sites has a total  of $P=s$. This means that even when the $\hat{V}_{km}$ hoppings are activated these states are unreachable from each other. In particular, a single quasiparticle domain wall has $P = \infty$ which implies that it is impossible for a quasiparticle to move freely through just the $\hat{V}_{km}$ electron-electron interactions: only quasiparticle-quasihole pairs can move around.

In order to do perturbation theory with $\hat{V}_{k1}$, we need to enumerate all the states that are reachable from these lowest-lying states through this hopping. The $V_{km}$ hoppings create two quasihole-quasiparticle pairs in opposite direction, and each of these minimally separated pairs \rev{\sout{can be considered as} forms} a dipole with dipole moment $\pm m$. For example, $\hat{V}_{51} \vert1001001001001 \rangle \sim$\rev{\sout{$\vert 100010101000\rangle =$}} $ \vert 100 \underset{+\frac{e}{3}}\vert \overset{\curvearrowright}{\textcolor{red}{\underline{0}1}}0 \underset{-\frac{e}{3}}\vert10\underset{-\frac{e}{3}} \vert \overset{\curvearrowleft}{\textcolor{blue}{\underline{1}0}}0\underset{+\frac{e}{3}}{\vert}01 \rangle$ has two dipoles with dipole moment $+1$ and $-1$ at the underlined positions (see also Figure~\ref{fig:Cyl}(B)). It is then more convenient for our perturbation theory approach to express the excited states in terms of these quasiparticle dipoles. Note that in this paper the dipole moment used is the moment of the electron \textit{density}, which is opposite of the charge dipole moment.

\begin{figure}
\includegraphics[width=0.49\textwidth]{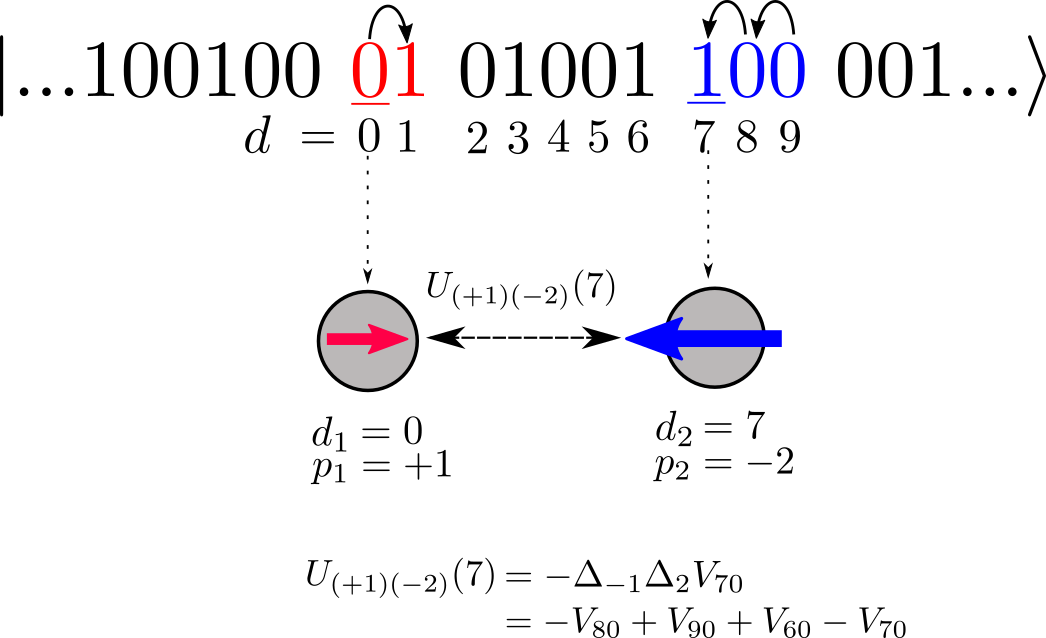}%
\caption{\rev{The dipole mapping used in our work. A dipole of moment $p$ is formed when an electron is moved by $p$ sites from its initial position. We define the position of a dipole $d$ in such a way that $d$ is leftmost orbital that does not follow the ultrathin ground state pattern. In other words, if an electron is moved to the right (positive $p$, red in the figure above) then $d$ is the initial position of the electron while if an electron is moved to the left (negative $p$, blue in the figure above) then $d$ is the final positon of the electron. By defining the position in this way we can write down a simple expression for the effective interaction potential between two dipoles (bottom).}
\label{fig:DipDip}}
\end{figure}

\rev{\sout{Let us denote the state with one dipole with moment $p$ extending from $i$ to $i + \vert p\vert$ as $\vert p_i \rangle$, where $p$ is positive if the electron is moved to the right and negative when its moved to the left.} We will define these dipoles in the following way: given a charge density wave TT ground state, we have a dipole of moment $p$ if an electron is moved by $p$ sites from its initial position. To mark the position $d$ of the dipoles we use the following convention: if $p>0$, i.e. the electron is moved to the right, then we set $d$ to be the initial position of the electron. Otherwise, if the electron is moved to the left, then $d$ is the final position of the electron (see Fig.~\ref{fig:DipDip}). That is, in the Fock space basis, the position of the dipole is defined as the leftmost orbital occupied by the dipole. The state containing $N$ dipoles with moments $p_0, p_1, ..., p_{N-1}$ and positions $d_0, d_1, ..., d_{N_1}$ can then be written as $\vert (p_0)_{d_0} (p_1)_{d_1} ... (p_{N-1})_{d_{N-1}} \rangle$} For  example, \rev{with the underlined site showing their position}, $\vert (+1)_5\rangle = \vert 00100~\overset{\curvearrowright}{\textcolor{red}{{{\underset{5}{{\underline{0}}}{1}}}}}~{0}1\rangle, \vert (-2)_0 \rangle = \vert{\overset{\curvearrowleft \curvearrowleft}{\textcolor{blue}{{\underset{0}{\underline{1}}{00}}}}}~0001001 \rangle $. 
Since the majority of dipoles in our unperturbed states are of dipole moment $\pm 1$, we abbreviate them with just $(+)_i$ or $(-)_i$. Our unperturbed states will then be denoted by listing all the dipoles, so for example $\vert (+2)_2(+)_5 (-)_7 \rangle = \vert 00 ~\overset{\curvearrowright \curvearrowright}{\textcolor{red}{{\underset{2}{\underline{0}}01}}} ~\overset{\curvearrowright}{\textcolor{red}{{\underset{5}{\underline{0}}1}}}~ \overset{\curvearrowleft}{\textcolor{blue}{{\underset{7}{\underline{1}}0}}} \rangle$. There is a bit of ambiguity in this convention, for example, $\vert ...0110... \rangle$ might be written as $\vert (+)_0 (-)_2 \rangle$ or $\vert (+2)_0 (-2)_1 \rangle$. To deal with this we always choose the representation with the smallest number of dipoles and the least $\sum_i \vert p_i \vert$, so the above example would be $\vert (+)_0 (-)_2 \rangle$. 

Previous works in the TT limit have developed similar mappings to a dipole system \cite{Bergholtz2005} or a similar spin-1 system \cite{Nakamura2012}, although they only considered dipoles of dipole moment 1, which allows some simplification of the Hamiltonian. There are hints of a connection between this dipole mapping and a previous field theory based approach to the FQHE \cite{Shankar1997, Read1998, Pasquier1998} in which the composite fermions responsible for the FQHE gains a dipolar charge. These connections might be explored in a future work, but here we emphasize that we are simply using the dipole mapping as a matter of mathematical convenience. With this dipole mapping, we can describe all excitations in the TT limit simply by enumerating all possible dipole configurations, and, given a dipole configuration, we can calculate the unperturbed energy by treating the dipoles as interacting particles with a pairwise interaction energy . \rev{\sout{$U_{p_i p_j}(d_i - d_j)$}}\rev{Consider two dipoles with position $d_i$ and $d_j$ (as defined in the previous paragraph) and moments $p_i$ and $p_j$. Then the interaction energy is given by
\begin{eqnarray}
    U_{p_i p_j}(d_i - d_j) = \mathrm{sgn}{(p_i p_j)} \Delta_{-\vert p_i \vert} (\Delta_{\vert p_j \vert} V_{d_j - d_i, 0})
\end{eqnarray}
}  
\rev{Here we use the finite difference notation where $\Delta_p V_{d0} \equiv V_{d+p,0} - V_{d0}$, and $\Delta_p(\Delta_{q}V_{d0}) = \Delta_p(V_{d+q,0} - V_{d0}) = V_{d+p+q,0} - V_{d+q,0} - V_{d+p,0} + V_{d0}$. Remember that $V_{d0}$ can take negative values of $d$ with $V_{-d,0} = V_{d0}$. When both dipoles are of moment $\pm 1$ we have an interaction that depends only on the distance $d = \vert d_j - d_i \vert$ between the dipoles
\begin{eqnarray}
    U_{p_i p_j}(d) = \mathrm{sgn}{(p_i p_j)}(- \Delta^2 V_{d0})
\end{eqnarray}
For a concave potential, $\Delta^2 V_{k0} > 0$. We can see then that in our system dipoles that point at the same direction have an attractive effective interaction while dipoles that point at the opposing direction have a repulsive effective interaction.}

The energy of a dipole configuration with respect to the ground state is then just the sum of the energy required to create these dipoles and the total interaction energy between dipoles:  
\begin{eqnarray}
    \Delta E = \sum_{i} E_{dip}(p_i) + \sum_{i} \sum_{j>i} U_{p_i p_j}(d_i - d_j)
\end{eqnarray}

The (unperturbed) energy required to create a dipole of moment $p$ is given by 
\begin{eqnarray}
    E_{dip}(p) = \sum_{k=1}^{\infty} \Delta^2_p V_{3k,0}
\end{eqnarray}
where $\Delta^2_p V_{k0} = V_{k-p,0} - 2V_{k0} + V_{k+p,0}$. Note that $E_{ph}(0) = E_{dip}(1)$. 
\rev{
}
\rev{\sout{where $\Delta_p V_{d0} = V_{d+p,0} - V_{d0}$}}. \rev{\sout{Specifically, for two dipoles of moment 1, if they point to the same direction, we would obtain an attractive effective interaction, $U_{\pm \pm}(d) = -\Delta^2 V_{d0}$, while dipoles that point in the opposite direction have a repulsive effective interaction, $U_{\pm \mp}(d) = \Delta^2 V_{d0}$.}}

Above the ground state, the lowest-lying excitations in terms of energy are the states made of one quasiparticle-quasihole pair discussed earlier (in our dipole notation, $\vert (\pm)_{i} (\pm)_{i+3}... (\pm)_{i+3(P-1)} \rangle $), up to the state with a semi-infinite electron domain, i.e., just one domain wall ($\vert (\pm)_i (\pm)_{i+3} (\pm)_{i+6} .... \rangle$). All of these states, however, belong to different $P$ sectors. Within a $P$ sector we can show that these quasiparticle-quasihole pairs are always the lowest energy states through the following argument: 

For a state with a finite number of dipoles, the energy required to create a new dipole pair ($2E_{ph}(0)$ and the repulsion between the dipole pair) is always larger than the possible energy reduction from dipole-dipole interactions, so the lowest energy state would be composed of exactly $P$ dipoles of the same type. The interaction energy between these dipoles are then strictly attractive, so the lowest energy state is reached when each dipole are as close as possible to each other, i.e., forming a continuous domain of length $3(P-1)$. We note that by translation invariance these lowest energy states are highly degenerate in this TT limit since for a cylinder/torus with $N$ electrons there are $N$ possible spot to place the dipole. However, as we shall see, introducing the hopping elements $V_{k1}$ to this Hamiltonian will allow the dipoles to move around, breaking this degeneracy. Following our argument, the next lowest spectra in each $P$ sector would be formed out of states with non-zero separation between elementary dipoles, which are then separated by some gap from the higher energy states, namely those with dipoles of the opposing kind. 

The $P = 0$ and $P=1$ sectors are somewhat special in that there are no states in these sector which are composed of just one type of dipole other than the lowest energy states. The lowest energy state of the $P=0$ sector is just the ground state while for the $P=1$ sector it is the (degenerate) state with just one dipole, $\vert (+)_i \rangle = \vert ...00100~\overset{\curvearrowright}{\textcolor{red}{\underset{i}{\underline{0}}{1}}}~0100... \rangle$. The degeneracy of the $P=1$ sector is later broken by first order perturbation in the hopping coefficients $V_{k1}$. The lowest lying excitations of these sector are those with one extra dipole pair, e.g. $\vert ...0010~\overset{\curvearrowleft}{\textcolor{blue}{\underline{1}0}}~00100~\overset{\curvearrowright}{\textcolor{red}{\underline{0}1}}~0100...\rangle$ (for $P=0$) or $\vert ...00100~\overset{\curvearrowright}{\textcolor{red}{\underline{0}1}}~0~\overset{\curvearrowright}{\textcolor{red}{\underline{0}1}}~\overset{\curvearrowleft}{\textcolor{blue}{\underline{1}0}}~00100100...\rangle$ (for $P=1$), followed by those with two extra dipole pair, etc. For $P>1$, we also have states composed of $P$ similar dipoles that are not adjacent to each other. For example, for $P=3$, we have $\vert 00100~\overset{\curvearrowright}{\textcolor{red}{\underline{0}1}}~00100~\overset{\curvearrowright}{\textcolor{red}{\underline{0}1}}~00100100~\overset{\curvearrowright}{\textcolor{red}{\underline{0}1}} \rangle$. The number of these states grows very quickly with $P$ - for an infinite cylinder there are indeed an infinite number ($N!/P!(N-P)!$) of non-degenerate states of this kind, so we will not attempt to enumerate them all here. We give a list of the accessible low lying neutral excitations with some examples in Table \ref{tab:NeutExc}.

\begin{table*}[]
    \centering
    \begin{tabular}{l|l|p{0.3\textwidth}|l}
      Sector   & Excitations & Energy & Example \\
      \hline
      \multirow{3}{4em}{$P=0$}   & $\vert GS \rangle$ & $0$ & $\vert..010010010..\rangle$ \\
      & $\vert (+)_i (-)_{i+k} \rangle$ & $2E_{ph} + \Delta^2 V_{k0}$ & $\vert ..0100~\overset{\curvearrowright}{\textcolor{red}{\underset{i}{\underline{0}}1}}~010~\overset{\curvearrowleft}{\textcolor{blue}{\underset{i+k}{\underline{1}0}}}~00.. \rangle$ \\
      & $\vert (+)_i (-)_{i+k} (+)_j (-)_{j+p} \rangle$ & $4E_{ph} + \Delta^2 V_{p0} + \Delta^2 V_{k0} - \Delta_p \Delta_{-k} (\Delta^2 V_{j-i,0})$ & $\vert ..00~\overset{\curvearrowright}{\textcolor{red}{\underset{i}{\underline{0}}1}}~\overset{\curvearrowleft}{\textcolor{blue}{\underset{i+k}{\underline{1}0}}}~0~\overset{\curvearrowleft}{\textcolor{blue}{\underset{j}{\underline{1}}0}}~00~\overset{\curvearrowright}{\textcolor{red}{\underset{j+p}{\underline{0}1}}}~001..\rangle$ \\
      \hline
      \multirow{2}{4em}{$P=\pm 1$} & $\vert (\pm)_i \rangle$ & $E_{ph}$ & $\vert ..00100~\overset{\curvearrowright}{\textcolor{red}{\underset{i}{\underline{0}}1}}~01001... \rangle$ \\
      & $\vert (\pm)_i (+)_j (-)_{j+k} \rangle$ & $3E_{ph}+\Delta^2 V_{k0} \pm \Delta^2 V_{j+k-i,0} \mp \Delta^2 V_{j-i,0}$ & $\vert ..00100~\overset{\curvearrowright}{\textcolor{red}{\underset{i}{\underline{0}}1}}~0~\overset{\curvearrowright}{\textcolor{red}{\underset{j}{\underline{0}}1}}~\overset{\curvearrowleft}{\textcolor{blue}{\underset{j+k}{\underline{1}0}}}~001... \rangle$\\
      \hline
      \multirow{3}{4em}{$\vert P \vert > 1$} & $\vert \prod_{n=0}^{P-1} (\pm)_{i+3n} \rangle$ & $E_{ph}(P-1)$ & $\vert ..0100~\overset{\curvearrowright}{\textcolor{red}{\underset{i}{\underline{0}}1}}~0~\overset{\curvearrowright}{\textcolor{red}{\underline{0}1}}~0~\overset{\curvearrowright}{\textcolor{red}{\underset{i+3n}{\underline{0}1}}}~01001...\rangle$ \\
      & $\vert \underbrace{(\pm)_{a_1} (\pm)_{a_2}... (\pm)_{a_{P}}}_{P} \rangle$  & $PE_{ph} - \sum_{i,j \in{\{a_i\}}} \Delta^2 V_{j-i,0}$ & $\vert ..0100~\overset{\curvearrowright}{\textcolor{red}{\underset{a_1}{\underline{0}1}}}~0~\overset{\curvearrowright}{\textcolor{red}{\underset{a_2}{\underline{0}1}}}~0100100~\overset{\curvearrowright}{\textcolor{red}{\underset{a_3}{\underline{0}1}}}~01... \rangle$   \\
      &$\vert \prod_{n=0}^{P-1} (\pm)_{i+3n} (+)_j (-)_{j+k}$& $E_{ph}(P-1) + 2E_{ph} + \Delta^2 V_{k0} + \sum_{n}^{P-1} (\Delta^2 V_{j+k-i-3n,0} - \Delta^2 V_{j-i-3n,0})$ & $\vert..0100~\overset{\curvearrowright}{\textcolor{red}{\underset{i}{\underline{0}1}}}~0~\overset{\curvearrowright}{\textcolor{red}{\underset{i+3}{\underline{0}1}}}~0100100~\overset{\curvearrowright}{\textcolor{red}{\underset{j}{\underline{0}}1}}~\overset{\curvearrowleft}{\textcolor{blue}{\underset{k}{\underline{1}}0}}~00... \rangle$ \\
      \hline
    \end{tabular}
    \caption{List of neutral excitations in the ultrathin limit and their unperturbed energies measured with respect to the ground state. The excitations in each sector are ordered roughly with increasing energy.}
    \label{tab:NeutExc}
\end{table*}

Now that we have a way to list all of the low-lying excitations and their unperturbed energies, we can apply the hopping terms $\hat{H}_1 = \sum_{n,k} V_{k1}  \hat{c}^{\dagger}_{n+k} \hat{c}^{\dagger}_{n+1} \hat{c}_{n+k+1} \hat{c}_n $ as a perturbation to the TT limit ($L_y \rightarrow 0$) Hamiltonian $\hat{H}_0 = \sum_{n,k} V_{k0} \hat{n}_{n+k} \hat{n}_n$. As mentioned before, $V_{k1}$ is proportional to $\lambda =e^{-2\pi^2 l_B^2/L_y^2}$, which falls off quickly when $L_y \leq 2\pi l_B^2$, so formally we can treat $\lambda$ as a small parameter and expand $\hat{H}_1 \sim \lambda (V_{k1}/\lambda)$ in terms of powers of $\lambda$. With the Hamiltonian written as $\hat{H} = \hat{H_0} + \lambda \hat{H}_1$, the first order correction to the energy of states $\vert \{\psi\} \rangle$is given by \cite{Griffiths2018}
\begin{eqnarray}
    E^{(1)}_{\{\psi\}} = \langle \tilde{\psi} \vert H \vert \tilde{\psi} \rangle
\end{eqnarray}
where $\vert \tilde{\psi} \rangle = \vert \psi \rangle$ for a non-degenerate $\psi$ - otherwise it is a state in the subspace spanned by $\{ \psi\}$ which diagonalizes $\hat{H}_1$

Expanding to the first order, we found no energy correction $\langle \psi\vert \hat{H}_1\vert \psi \rangle$ to the $P=0$ sector. However, in the lowest energy states of the $P>0$ sector, which contain $P$ 1-dipoles, we see that the hopping $\hat{V}_{k1} = \sum_n V_{k1} \hat{c}^{\dagger}_{n+k} \hat{c}^{\dagger}_{n+1} \hat{c}_{n+k+1} \hat{c}_n$ can move a single dipole by $k$ sites. In the P=1 sector specifically, we have $\langle (\pm)_{a} \vert \hat{H} \vert (\pm)_{b} \rangle = V_{a-b,1}$, or, in matrix form,
\begin{equation}
    \begin{bmatrix}
         0 && V_{31} && V_{61} && V_{91} && ... \\
         V_{31} && 0 && V_{31} && V_{61} && ... \\
         V_{61} && V_{31} && 0 && V_{31} && ... 
    \end{bmatrix}
\end{equation}
where the basis for the $i$-th column of the vector is $\vert (\pm)_{3i} \rangle$. For a concrete example, $\hat{V}_{61} \vert ...00~\textcolor{red}{\underline{0}1}~01001001... \rangle = V_{61}\vert ...00\overset{\curvearrowleft}{\underline{1}0}0100~\overset{\curvearrowright}{\textcolor{red}{\underline{0}1}}~01...\rangle$ 
\rev{\sout{Using translation invariance we can easily diagonalize this matrix with the Bloch eigenvector $\vert P=1,q \rangle = \sum_{n}^{\infty} e^{iqn} \vert (\pm)_{3n} \rangle$, giving a first-order energy correction of} On the torus with periodic boundary condition, we can easily diagonalize this matrix with the Bloch eigenvector $\vert P=1,q \rangle = \sum_{n}^{N} e^{i 2\pi qn/N} \vert (\pm)_{3n} \rangle$, giving a first order energy correction of}
\begin{eqnarray}
    E^{(1)}_{P=1}(q) = \sum_{n=1}^{\rev{N/2}} 2V_{3n,1} \cos{\rev{\frac{2\pi qn}{N}}}
    \label{eq:neutraldisp1}
\end{eqnarray}

\rev{On the cylinder, we can take the limit $L_x,~N \rightarrow \infty$ in order to obtain a continuous dispersion
\begin{eqnarray}
    E^{(1)}_{P=1} (p_x) = \sum_{n=1}^{\infty} 2 V_{3n,1} \cos{\left(\frac{6 \pi nl_B^2}{L_y}p_x\right)}
\end{eqnarray}
where $p_x = \frac{2\pi q}{L_x}$ is the $x$-momentum. For ease of comparison with our exact diagonalization calculation on the torus, the $x$-momentum of our calculation will be expressed in $2\pi q/N = 6\pi l_B^2 p_x /L_y$
}

For $\vert P \vert > 1$, the situation is slightly different. Each hopping term can only move one electron at the time, so the off diagonal element of the perturbation Hamiltonian is non-zero only between adjacent states, i.e. $\langle (\pm)_i ... (\pm)_{i+3(P-1)} \vert \hat{H} \vert (\pm)_j ... (\pm)_{j\pm3(P-1)} \rangle = \delta_{i,j \pm 1} V_{3P,1}$, for example, $\hat{V}_{61} \vert ...00~\textcolor{red}{\underline{0}1}~0~\textcolor{red}{\underline{0}1}~01001...\rangle = V_{61} \vert ...00\overset{\curvearrowleft}{\underline{1}0}0~\textcolor{red}{\underline{0}1}~0~\overset{\curvearrowright}{\textcolor{red}{\underline{0}1}}~01...\rangle$. Then the correction for the energy will be given by
\begin{eqnarray}
    E^{(1)}_{P\rev{>1}}  ~= &&2V_{3P,1} \cos{\rev{\frac{2\pi q}{N}}} \nonumber \\
    ~\rev{{=}} && 2V_{3P,1} \cos {\left(\frac{6\pi l_B^2}{L_y}p_x\right)}
    \label{eq:neutraldispP}
\end{eqnarray}

Moving to second-order corrections, we have $E^{(2)}\sim \sum_{m\neq n} \vert\langle m \vert \hat{H}_1 \vert n \rangle\vert^2/\Delta E$ \cite{Griffiths2018}. In the $P=0$ sector, we see that the ground state can reach the states with one dipole pair $\vert (-)_i (+)_j \rangle$ through one hopping. We have, for $k=2,5,8,...$, $\hat{V}_{k1} \vert GS \rangle = \sum_{n} \vert (+)_{3n} (-)_{3n+k}\rangle$, while for $k=4,7,10,..$, $\hat{V}_{k1}^{\dagger} \vert GS \rangle = \sum_{n} \vert (+)_{3n} (-)_{3n-k}\rangle$. The unperturbed excitation energy of these states is equal to the energy required to create two dipoles plus the interaction between the dipoles, $2E_{ph} + \Delta^2 V_{k0}$, so the second-order correction is
\begin{eqnarray}
    E_{GS}^{(2)} &&= \sum_{i,j}  \frac{\vert \langle (-)_i (+)_j \vert \hat{H}_1 \vert GS \rangle \vert^2}{-\Delta E_{ij}}\\
    &&= -N \sum_{k\in 3\mathbb{Z}-1} \frac{V_{k1}^2}{2E_{ph} + \Delta^2 V_{k0}}
\end{eqnarray}
Note that the left-right dipole pair states $\vert (-)_i (+)_j \rangle$ are highly degenerate, with $N$ different states for each hopping range $k = \vert j - i\vert$. This means that the second-order correction would actually diverge in the thermodynamic limit. However, as we will soon see, the second-order corrections to all the excited states will also contain terms that are proportional to $N$ such that the difference between the ground state and the excited states will remain finite.

Now consider the second-order correction to the states with a single dipole pair $\vert (+)_{3n} (-)_{3n+k} \rangle$ with $k = ...,-7,-4,-1,2,5,8,...$. Since these states are degenerate, the correction is given by diagonalizing the matrix $U^k$ with elements given by
\begin{equation}
    U^k_{ij} = \sum_n \frac{\langle (+)_{3i} (-)_{3i + k} \vert \hat{H}_1 \vert \psi_n \rangle \langle \psi_n \vert \hat{H}_1 \vert (+)_{3j} (-)_{3j + k} \rangle}{2 E_{ph} + \Delta^2 V_{k0} - E_n}
\end{equation}
where $\vert \psi_n \rangle$ ranges over the ground state and all the other 1st and 2nd excited states in the sector. 

Let us first consider the diagonal elements of this matrix. There are several ways for the configuration $\vert l_i (+)_j \rangle$ to get back to itself in two hoppings: through the ground state, through another 1st excited state, or through a 2nd excited state, which might be a state with two dipole pairs or a state with a dipole of moment $\pm 2$. Schematically,
\begin{widetext}
\begin{eqnarray}
    U_{ii}^k = &&\frac{\vert \langle (+)_{3i}(-)_{3i + k}  \vert \hat{H}_1\vert GS \rangle \vert^2}{\Delta E_{GS}} + \frac{\vert \langle (+)_{3i} (-)_{3i+k} \vert \hat{H}_1 \vert (+2)_{3i} (-2)_{3i+k-1} \rangle\vert^2}{\Delta E_{\pm2,\pm}} \nonumber \\
     &&+\sum_{n} \frac{\vert \langle (+)_{3i}(-)_{3i + k} \vert \hat{H}_1 \vert (+)_{3i+3n}(-)_{3i+k} \rangle \vert^2}{\Delta E_{\pm,\pm}} + \sum_{p} \frac{\vert \langle(+)_{3i} (-)_{3i+k} \vert \hat{H}_1 \vert (\pm2)_{3i} (\mp)_{3i\pm k -1} \rangle \vert^2}{\Delta E_{+2-,\pm}} \nonumber \\
    &&+ \sum_{n} \sum_{p\in 3\mathbb{Z}-1} \frac{\vert \langle (+)_{3i} (-)_{3i + k}\vert \hat{H}_1 \vert (+)_{3i} (-)_{3i+k} (+)_{3n} (-)_{3n + p}\rangle \vert^2}{\Delta E_{\pm \pm, \pm}}
\end{eqnarray}
The full expression for the diagonal matrix elements is available in Appendix B. Here, we would like to point out that the last term above has two summations, so this term will actually diverge in the thermodynamic limit. However, subtracting the $O(N)$ second order correction to the ground state energy we obtain a summation that is convergent:
\begin{eqnarray}
    U_{ii}^k - E_{GS}^{(2)} \sim \sum_{p \in 3\mathbb{Z}-1} \frac{2V_{p1}^2}{2E_{ph} + \Delta^2 V_{p0}} + \sum_{p \in 3\mathbb{Z}-1} \sum_{\substack{n \in 3 \mathbb{Z} \\ n \neq 0,k+1 \\n+p+1 \neq 0,k}} \frac{V_{p1}^2 (-\Delta_p \Delta_{-k}(\Delta^2 V_{n0}))}{2E_{ph} + \Delta^2 V_{p0} - \Delta_p \Delta_{-k}(\Delta^2 V_{n0})}
\end{eqnarray}
Since $V_{n0} \sim 1/n$, $\Delta_p \Delta_{-k}(\Delta^2 V_{n0}) = \Delta^2 V_{n+p,0} - \Delta^2 V_{n+p-k,0} - \Delta^2 V_{n,0} + \Delta^2 V_{n-k,0} \sim 1/n^4$ and the sum over $n$ is always convergent.

Meanwhile, for the off-diagonal element, there are three ways to jump from $\vert (+)_i (-)_{i + k} \rangle$ to $\vert (+)_j (-)_{j - k}  \rangle$: through the ground state, through another one-pair state with different $k$, and through a second excited state:
\begin{align}
    00100~\textcolor{red}{\underline{0}1}~010~\textcolor{blue}{\underline{1}0}~001 \rightarrow 00100\overset{\curvearrowleft}{{1}0}010\overset{\curvearrowright}{0{1}}001 \rightarrow 00100100~\overset{\curvearrowright}{\textcolor{red}{\underline{0}1}}~010~\overset{\curvearrowleft}{\textcolor{blue}{\underline{1}0}} ~&\sim V_{51}^{\dagger} V_{51} \\
    00100~\textcolor{red}{\underline{0}1}~\textcolor{blue}{\underline{1}0}~001001 \rightarrow 00100~\textcolor{red}{\underline{0}1}~\overset{\curvearrowright}{01}0~\overset{\curvearrowleft}{\textcolor{blue}{\underline{1}0}}~001 \rightarrow 00100\overset{\curvearrowleft}{10}0~\overset{\curvearrowright}{\textcolor{red}{\underline{0}1}}~\textcolor{blue}{\underline{1}0}~001~ &\sim V_{31} V_{31}^{\dagger} \\
    0~\textcolor{blue}{\underline{1}0}~00~\textcolor{red}{\underline{0}1}~01001001 \rightarrow 0~\textcolor{blue}{\underline{1}0}~00~\textcolor{red}{\underline{0}1}~\overset{\curvearrowleft}{\textcolor{blue}{\underline{1}0}}~00~\overset{\curvearrowright}{\textcolor{red}{\underline{0}1}}~01 \rightarrow 0\overset{\curvearrowright}{01}00\overset{\curvearrowleft}{10}~\textcolor{blue}{\underline{1}0}~00~\textcolor{red}{\underline{0}1}~01 ~ &\sim V_{41}^{\dagger} V_{41} 
\end{align}
Note that the last process can only connect states that are separated by more than $k+1$ sites. Schematically then, the off-diagonal element would be
\begin{eqnarray}
    U_{ij}^k = &&\frac{\langle (+)_{3i} (-)_{3i + k} \vert \hat{H}_1 \vert GS \rangle\langle GS \vert \hat{H}_1\vert (+)_{3j} (-)_{3j + k}\rangle}{\Delta E_{GS}} + \frac{\langle (+)_{3i} (-)_{3i + k} \vert \hat{H}_1 \vert (+)_{3i} (-)_{3j+k}\rangle \langle (+)_{3j} (-)_{3j + k} \vert \hat{H}_1 \vert (+)_{3i} (-)_{3i + k} \rangle}{\Delta E_{\pm,\pm}} \nonumber \\
    &&+ \frac{\langle (+)_{3i} (-)_{3i + k}\vert \hat{H}_1 \vert (+)_{3i} (-)_{3i + k}(+)_{3j} (-)_{3j + k} \rangle \langle (+)_{3i} (-)_{3i + k} (+)_{3j} (-)_{3j + k} \vert \hat{H}_1 \vert (+)_{3i} (-)_{3i + k} \rangle}{\Delta E_{\pm \pm, \pm}}
\end{eqnarray}
where the last two terms vanish when $3\vert i-j \vert = \vert k+1 \vert$.   We then diagonalize $U^k - E_{GS}^{(2)} I$. Switching our basis to $\vert 0; \psi_k(q) \rangle = \frac{1}{\sqrt{N}}\sum_{n=1}^N e^{i \rev{2 \pi q n/N}} \vert (+)_{n} (-)_{n + k} \rangle $ where $k\in 3\mathbb{Z} - 1$, we obtain the second order correction,
\begin{eqnarray}
        E^{(2)}_k(q) - E_{GS} &&= (U_{ii}^k - E_{GS}) \nonumber
        \\ &&+ 2 \sum_{n=1}^{N/2} \left(\frac{-V_{k1}^2 \Delta_k \Delta_{-k} (\Delta^2 V_{3n,0})(1-\delta_{3n,\vert k+1 \vert})}{(2 E_{ph} + \Delta^2 V_{k0})(2E_{ph} + \Delta^2V_{k0} - \Delta_{k} \Delta_{-k} (\Delta^2 V_{3n,0}))}  \right) \cos {\rev{\frac{2\pi qn}{N}}}  \nonumber\\
        &&+ 2\sum_{n\neq0 } \left( \frac{V_{3n,1}^2(1-\delta_{3n,\vert k+1 \vert})}{\Delta^2 V_{k0} - \Delta^2 V_{k+3n,0}} \right) \cos{\rev{\frac{2\pi qn}{N}}} + \frac{2V_{k1}^2}{2E_{ph} + \Delta^2 V_{k0}} \cos{\rev{\frac{2\pi q \vert k+1 \vert}{3N}}}
\end{eqnarray} 
\end{widetext}

In the cylinder limit, we take $N \rightarrow \infty$ and replace $2\pi q/N$ with  $6\pi l_B^2 p_x/L_y$ as before. Since $V_{n1}$ and $\Delta^2 V_{n0}$ falls off quicker than $1/n$ the infinite sums are convergent. For the sake of consistency we also include the second order correction to the $P=1$ particle-hole state in Appendix B. Of course, this correction would be dominated by the first order correction. 

\begin{figure*}
\includegraphics[width=0.99\textwidth]{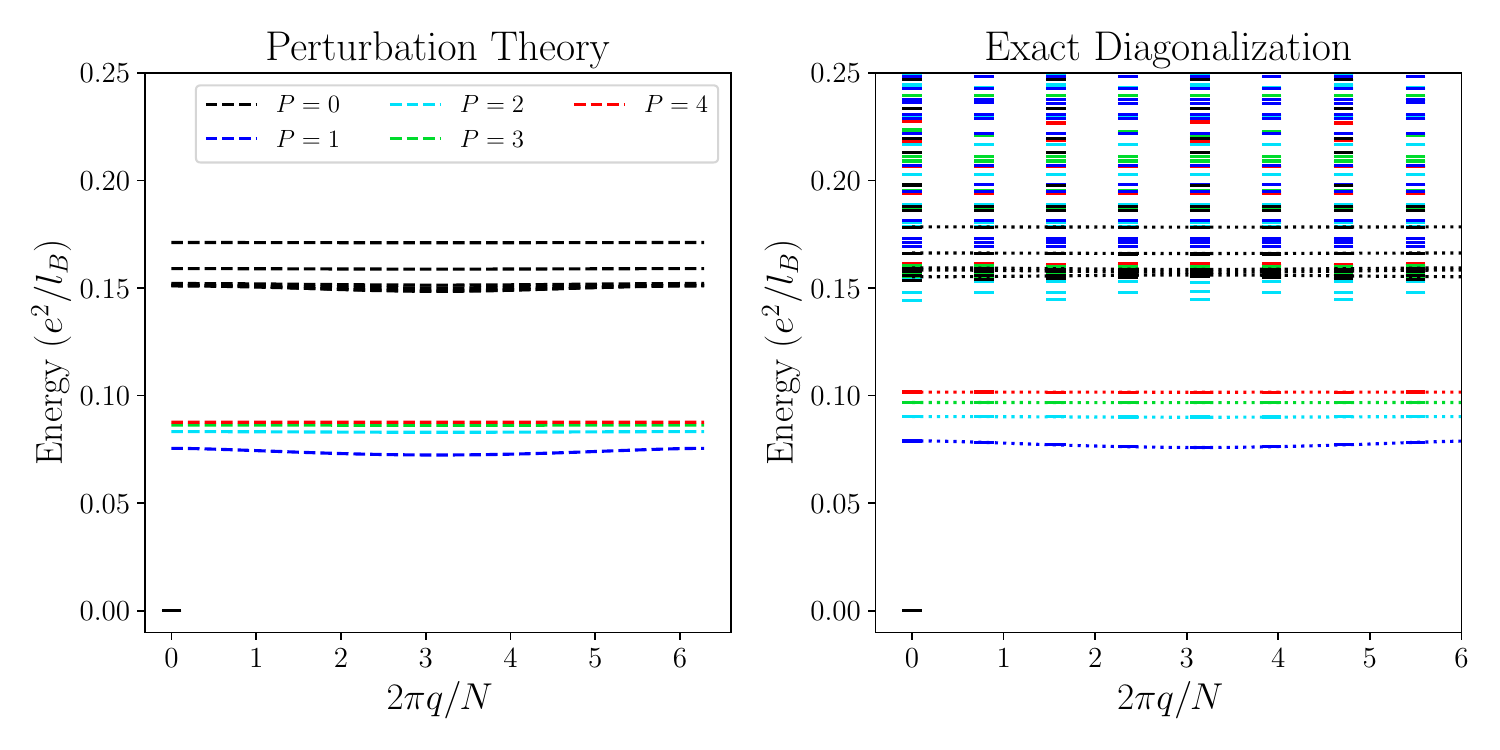}%
\caption{The energy spectrum for the charge-neutral, $P=0$ to $P=4$ sector, $L_y = 4l_B$. Color corresponds to the different dipole moment sectors. The ground state belongs to the black $P=0$ sector. Note that the lowest excited $P=1$ state would become the magnetoroton state as we move away from the thin torus limit. (Left) Perturbation theory calculation for the lowest energy states in $P=0$ to $P=4$ sector and the next-lowest energy states in the $P=0$ sectors - the dipole-dipole $\vert (+)_i (-)_{i+k} \rangle$ states. (Right) Exact diagonalization results with $N_{\phi} = 24$ lattice sites and $N=8$ electrons on the torus. \rev{The energies are shifted by a constant amount such that the ground state is at 0 energy.} The dotted lines corresponds to the perturbation theory result, shifted in position to account for finite size effects.  \label{fig:SpecFull}}
\end{figure*}

Using our expression for the corrected energies, we plot the energies of the low-lying excitations as a function of momentum in Figure \ref{fig:SpecFull}. Here we pick $L_y = 4l_B$, which gives $\lambda = e^{-2\pi^2l_B^2/L_y^2} = 0.291$. As comparison, we also performed an exact diagonalization calculation on a torus with $N_{\phi} = 24$ lattice sites and $N=8$ electrons, using the same potentials. Comparing our calculation with the exact diagonalization results, we see that our calculation captures the important  qualitative details: above the ground state, by about $E_{ph} = E_{dip}(1) = 0.073$, are the low-lying quasiparticle-quasihole states composed of $P$ dipoles $\vert (\pm)_i (\pm)_{i+3} ... (\pm)_{i+3(P-1)} \rangle$. Meanwhile, in the $P=0$ sector the next lowest-lying states, separated by a gap of around $2E_{ph}$, are the dipole-dipole states $\vert (+)_i (-)_{i+k} \rangle$, where states with lower $k$ have higher energy. The additional $P=0$ (black lines) states in the exact diagonalization plot are states with more than one dipole pair ($\vert (+)_i (-)_{i+p} (+)_j (-)_{j+q} \rangle$) or with a larger than $\vert p \vert =1$ dipole moment, while the additional $P>0$ (colored lines) states are either states with an extra dipole pair, or (like the cyan lines in the bottom of the upper excited states) states where the dipoles are not adjacent to each other e.g. $\vert (+)_i (+)_{i+3k}\rangle, k>1$. We note that on our torus with $24$ lattice sites the ground state has a total dipole moment of $\langle \hat{P} \rangle_{\mathrm{torus}} = 4,8$ or $12$, while the values of $P$ given in Fig.~\ref{fig:SpecFull} correspond to the dipole moment in the cylindrical system. In effect, we shifted $\langle \hat{P} \rangle_{\mathrm{torus}}$ by $-4$, $P = \langle \hat{P} \rangle_{\mathrm{torus}} - 4$ so that $P_{GS,\mathrm{torus}} = 0$, and so on for the states with higher momenta.

\begin{figure}
\includegraphics[width=0.49\textwidth]{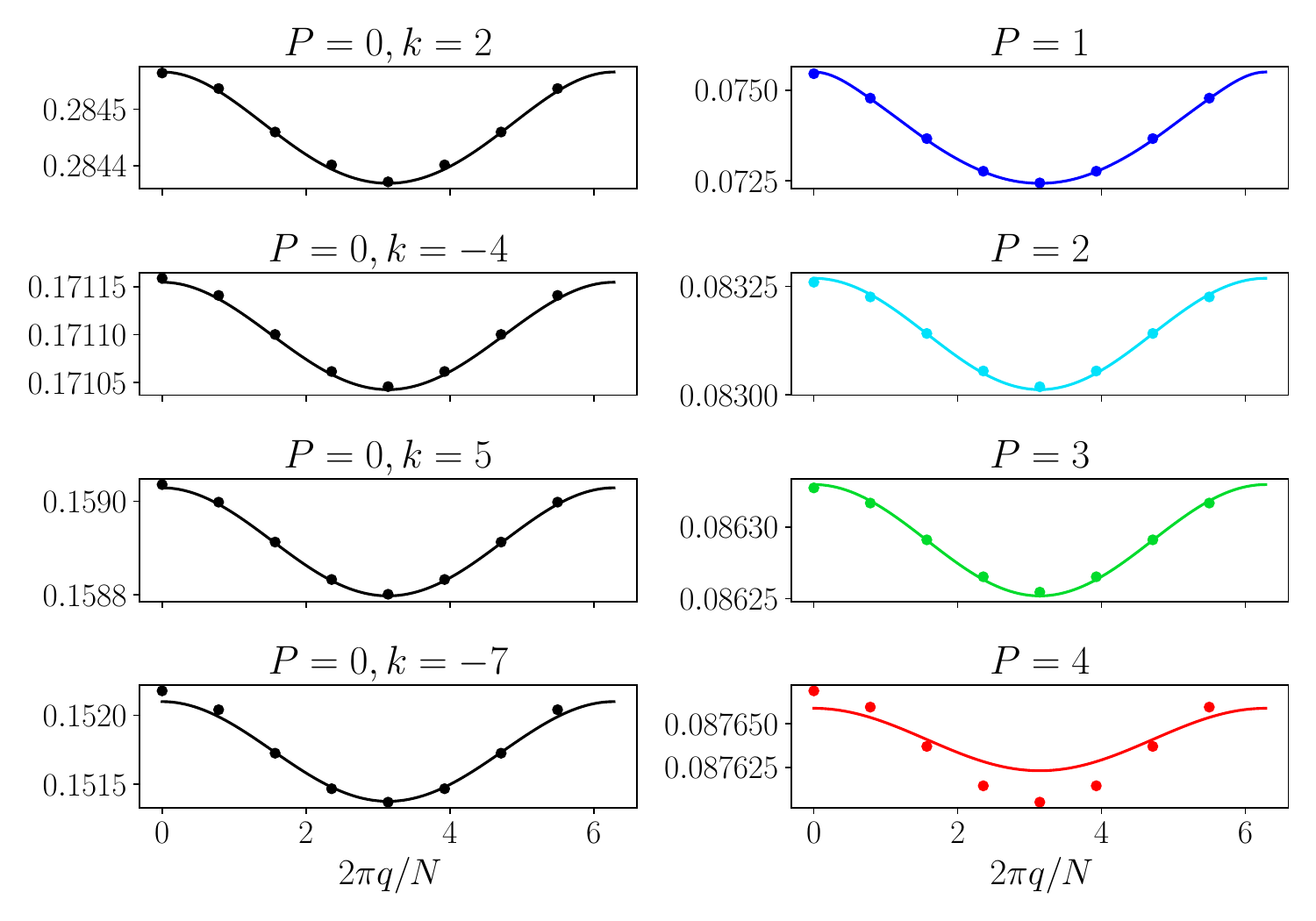}%
\caption{The dispersion calculated by perturbation theory (lines) and by exact diagonalization (dots), for the the $\vert (+)_i (-)_{i+k} \rangle$ states which are in the $P=0$ sector (left) and the $\vert (\pm)_i (\pm)_{i+3} ... (\pm)_{i+3(P-1)}\rangle$ states which are in the $P=1-3$ sector (right). The y axes show the energy in units of $e^2/\epsilon l_B$. We obtain very good agreement except for the $P=4$ case, where the dispersion becomes very small such that higher-order contributions cannot be neglected.  
\label{fig:Disp}}
\end{figure}

Numerically, the positions of the excited states do not cleanly match each other. This is attributable to finite size effects - notably, due to periodic boundary conditions in the ED system for $k>12$ we have $V_{km} = V_{N_{\phi}-k,m}$ and the sum in Eq.\ref{eq:Edomain} is terminated on $k=4$. If we shift the energies of our perturbation theory calculation to take this into account, we can see that the lines match up (dotted lines in the ED plot). Shifting the ED energies in this manner, we zoom in to each of the excitation energies to show more details in Fig. ~\ref{fig:Disp}. Here $k$ refers to the dipole-dipole separation in the dipole pair states $\vert (+)_i (-)_{i+k} \rangle$. We see that the results match closely for $P<4$. This is because in our perturbation theory calculation the dispersion is proportional to $V_{3P,1}$, which decays quickly with $P$, so that for higher $P$ the higher order terms in the perturbation expansion and the $V_{k2}, V_{k3}$ terms can no longer be ignored.

Meanwhile, in the $P=0$ sector the leading order in the dispersion is proportional to $V_{31}^2/(2E_{ph} + \Delta^2 V_{k0})$ so state lower in energy, i.e., those with larger $k$ also has a larger dispersion. In accordance with previous predictions \cite{Bergholtz2008, Jansen2008, Jansen2012, Nachtergaele2021}, we observe a nonvanishing gap between the ground state and the lowest excited states.   

One final remark on these neutral excitations: the eigenstates for the lowest lying excitations are "dipole waves" \rev{ \sout{$\sum e^{ip n} \vert \prod_i^{P-1}(\pm)_{n+3i} \rangle$. We can then identify the lowest energy one with $P=1$ with the well-known magnetoroton neutral excitations in the 2D limit [33]-[35]  which can be obtained from the single-mode approximation, assuming a wave function of form  $\sim \sum_{k} e^{ip_x q l_B^2} \hat{c}^{\dagger}_{k+q} \hat{c}_k \vert GS \rangle$.} $\sum e^{i 2\pi q n/N} \vert (\pm)_{3n} \rangle$ which can be written as $\hat{\rho}_{(q,P)} \vert GS \rangle$ where
\begin{eqnarray}
    \hat{\rho}_{(q,P)} = \sum_n e^{i \frac{2 \pi q}{N_{\phi}}n} \hat{c}_{n+P}^{\dagger}\hat{c}_n 
\end{eqnarray}
is, up to some normalization, the Fourier transform of the electron density operator, projected to the lowest Landau level
\begin{eqnarray}
    \tilde{\rho}(\mathbf{r}) = \sum_{k,k'} \phi^*_{k}(\mathbf{r}) \phi_{k'}(\mathbf{r}) \hat{c}_{k}^{\dagger} \hat{c}_{k'}
\end{eqnarray}
This is the form of the variational wavefunction used in Girvin, MacDonald, and Platzmann's single mode approximation (SMA)~\cite{Girvin1986}. The SMA is known to be a good approximation of the low-lying neutral excitation of the FQHE state, known as the magneto-roton \cite{Girvin1986, Repellin2014, Repellin2015}. We can then identify the low-lying $P>0$ states as the manifestation of this magneto-roton state.}

\rev{The position of the magneto-roton minimum has been predicted to be robust with respect to microscopic details \cite{Golkar2016, Balram2017}. For the $\nu=1/3$ state, various calculations have found a minima at a dimensionless wave vector $pl_B$ of 1.28 (Golkar et. al.'s composite fermion Fermi liquid calculation \cite{Golkar2016}), 1.40 (Balram and Pu's microscopic composite fermion calculation \cite{Balram2017}), and 1.60 (Simon and Halperin's Chern-Simons theory with random phase approximation \cite{Simon1993}). For the torus, the relevant wave-vector is given by $p = \sqrt{p_x^2 + p_y^2}$, where the momenta $p_x$ and $p_y$ are given by the eigenvalues of the translation operators $\hat{T}_x$ and $\hat{T}_y$ as usual. With the SMA magneto-roton state we have $\hat{T}_x(6\pi l_B^2/L_y) \vert P=1,q  \rangle = e^{i 2 \pi q/N} \vert P=1,q \rangle$ and $\hat{T}_y(2\pi l_B^2/L_x) \vert P=1,q  \rangle = e^{i 2\pi P/N_{\phi}} \vert P=1, q \rangle$ such that
\begin{eqnarray}
    pl_B =  \sqrt{\left(\frac{q L_y}{3N l_B}\right)^2 + \left(\frac{2 \pi P l_B}{L_y}\right)^2}
    \label{eq:rotonmin}
\end{eqnarray}
}

\rev{As long as our result from perturbation theory (Eq.~\eqref{eq:neutraldisp1}) holds, the magneto-roton minimum would always be located at $q/N = 1/2,~P=1$. Of course, in the highly anisotropic TT limit, we would not expect this result to line up with the composite fermion calculations (indeed, $\eqref{eq:rotonmin}$ diverges in the $L_y \rightarrow 0$ limit), but we can nevertheless estimate the location of the minimum in 2D as follows. As we increase $L_y$, we would expect the minimum to move closer to that of the isotropic limit, up until the point where the hopping coefficients are too large for perturbation theory. By calculating the value of $L_y$ when the magneto-roton minimum moves away from the bulk value, we can also estimate when our approach starts to fail. As a function of $L_y$, Eq.~\eqref{eq:rotonmin} has its global minimum when $L_y = 6.14$, after which it starts to diverge linearly with $L_y$. At this point the scaled wavevector of the magneto-roton minimum is $p l_B = 1.44$, which is remarkably close to the value of 1.40 obtained from the microscopic CF calculation~\cite{Balram2017}. 
}

\subsection{\label{ssec:STM}Perturbation theory for the charged sector}

So far we have limited our discussion to states with exactly $\nu = 1/3$ filling fraction. Now we consider states with one electron added or removed - this would physically correspond to the states produced in STM experiments. Adding or removing an extra electron will increase the total y-momentum depending on the position of the inserted or removed electron, thus breaking the center of mass conservation.

Adding or removing an electron into the system would require an energy of
\begin{eqnarray}
    E_e = \sum_{k=1}^{\infty} (V_{3k-1,0} + V_{3k-2,0})
\end{eqnarray}
or
\begin{eqnarray}
    E_h = -\sum_{k = 1}^{\infty} 2V_{3k,0}
\end{eqnarray}
Note that with our original $V_{k0}$ this summation is divergent, so we need to add the regularizing positive background as described in Section~\ref{ssec:CoulombPot}. Using our regularized Coulomb potential with $L_y = 4l_B$, we find $E_e = -0.213 (e^2/l_B)$ and $E_h = 0.884(e^2/l_B)$. 

We can still use the dipole description to calculate the unperturbed energies in the charged sector, but this time we have to add an interaction energy between the charge and the dipoles. The interaction between an added (or removed) electron at $d_i$ and another dipole at $d_j$ can be written as
\begin{eqnarray}
    U_{e_i;p_j}(d_j - d_i) = -\mathrm{sgn}(e_i p_j) \Delta_{\vert p\vert} V_{j - i,0}
\end{eqnarray}
where $e_i = -1 (+1)$ for an added (removed) electron.

The state with the bare electron added is obviously not the ground state of the relevant sector. If we consider states in the same $P$ sector as the one with only the extra electron added in, we can find that by creating dipole pairs around the electron that point away from the electron we can reduce the Coulomb energy of the states. For example, the energy difference between $\vert \rev{(-)}_{-1} (+)_{3}; e_{1} \rangle \equiv \vert ...0010~\overset{\curvearrowleft}{\textcolor{blue}{\underline{1}0}}~\textcolor{blue}{\underset{\uparrow}{\mathbf{1}}}0~\overset{\curvearrowright}{\textcolor{red}{\underline{0}1}}~01001 \rangle$ and $\vert 0;e_1 \rangle = \vert ...001001\textcolor{blue}{\underset{\uparrow}{\mathbf{1}}}01001001... \rangle$. is given by $2E_{ph} + \Delta^2 V_{40} - \Delta_{-1} V_{20} + \Delta_{1} V_{20}$, with $-\Delta_{-1} V_{20} + \Delta_{1} V_{20} = V_{30} - V_{10} < 0$, which, for small enough $L_y$, would be larger than $2E_{ph} + \Delta^2 V_{40}$ so that the energy change is negative. The true ground state, with the electrons evenly distributed, is likely to be inaccessible from this electron injection state since it will take multiple hopping terms.

We can then apply perturbation theory to calculate the correction to the energies of these excited states in the charged sector. The state with an electron or a hole, however, will have different total dipole moment depending on where the particle is inserted, so unlike for the neutral excitations we will not have broadened energy levels. The broadening for the neutral excitations (Eqs.~\eqref{eq:neutraldisp1}-\eqref{eq:neutraldispP}) is caused by the breaking of the degeneracy between states that are translations of each other, forming $x$-momentum eigenstates with different energies. However, for the charged excitations the dipole moment conservation prevents a translation of the excitation without creating or annihilating a dipole, which can be understood more formally as follows. The total dipole moment is given by $P = \sum_k kn_k \mod{N_{\phi}}$. A translation by $l$ sites changes this dipole moment by $\Delta P = (\sum_k (k+l)n_k - \sum_k kn_k) \mod{N_{\phi}} = lN \mod{N_{\phi}}$ where $N = \sum_k n_k$ is the total number of particles. Now, for the zero charge sector, we have $N = N_{\phi}/3$, which means a translation by $3$ will not change the total dipole moment. As a result the $\hat{V}_{km}$ hoppings are able to break the degeneracy between the different translationally invariant states. Meanwhile, for the charged sector, we have $N = N_{\phi}/3 \pm 1$ so a translation by $l$ will always change the dipole moment, unless $lN$ happens to be multiple of $N_{\phi}$. For a very large $N_{\phi}$ we would then also need a very large $l$, so in the thermodynamic limit there is no physically reasonable way of connecting the different degenerate states with just a finite number of the $\hat{V}_{km}$ hoppings. As a result the $x$-momentum eigenstates will not form a broad dispersive band.

This means that the spectrum for the charged states would remain dispersionless, which suggests that they would be difficult to access through conventional transport measurements. STM experiments, however, would be able to access these states and also measure their actual sharpness. Indeed, the results from Hu et.al. \cite{Hu2023} shows that their measured $\nu=2/3$ state has a sharp peak in the spectrum. To see how our approach would predict these STM states, we consider states that are created by physically adding an electron at a coordinate $x$ to the ground state, $\hat{\Psi}^{\dagger}(x) \vert GS \rangle$. The measured STM conductivity is proportional to the local density of states (LDOS) \cite{Tersoff1985, Papic2018}
\begin{eqnarray}
    \frac{dI}{dV}(\mathbf{r},E) &&\propto \mathrm{LDOS}(\mathbf{r},E) \nonumber\\ &&= \sum_n \vert \langle n \vert \hat{\Psi}^{\dagger}(\mathbf{r}) \vert \psi \rangle \vert^2 \delta(E-E_n) 
\end{eqnarray}
The state that is created by $\hat{\Psi}^{\dagger}(\mathbf{r})$ has a physical electron in the lowest Landau level. This is not the Landau gauge wave-function, which wraps around the cylinder, but something that is maximally localized in both the $x$ and $y$ direction. This cannot be a simple delta function either in the lowest Landau level, as the commutation relation of the LLL-projected position operators sets a limit to the possible localization of the wave function \cite{Girvin2019}. We have $[x_{LLL},y_{LLL}] = il_B^2$ such that
\begin{equation}
    \sigma_{x} \sigma_y \geq l_B^2
\end{equation}
In the symmetric gauge, the most localized Gaussian wave function which obeys this limit is the $m=0$ state, $\frac{1}{\sqrt{2\pi} l_B} e^{-(x^2 + y^2)/4l_B^2}$. Since we work in the Landau gauge, we expect the wave function of the electron created by $\hat{\Psi}^{\dagger}$ to be proportional up to a phase to this wave function.

Using the Poisson summation formula, we can show that the wave function of an electron localized at $\mathbf{r'}=(x',y')$ on the cylinder (ignoring normalization for now),
\begin{eqnarray}
    \psi(\mathbf{r},\mathbf{r'}) \sim \sum_n e^{-\frac{(x-x')^2 + (y-nL_y)^2}{4l_B^2}} e^{-\frac{i(x+x')(y-y'-nL_y)}{2l_B^2}} \nonumber\\ \label{eq:LocWF}
\end{eqnarray}
(note the summation to ensure periodicity in the $y$ direction, and the phase factor to ensure compatibility with the Landau gauge) can be expressed in terms of the lowest Landau level wave function
\begin{eqnarray}
    \psi(\mathbf{r},\mathbf{r'}) &&\propto  \sum_{k} e^{-\frac{x^2+x'^2}{2l_B^2}} e^{-\frac{2\pi k l_B^2}{L_y}(x+x')} e^{-\frac{4\pi^2 l_B^2 k^2}{L_y^2}} e^{-i \frac{2\pi k}{L_y}(y'-y)} \nonumber\\ \\
    &&=\sum_{k}  \phi^*_{k}(x',y') \phi_k(x,y)
\end{eqnarray}
We can then write the creation operator $\hat{\Psi}^{\dagger}(\mathbf{r}')$ as
\begin{eqnarray}
    \hat{\Psi}^{\dagger}(\mathbf{r'}) = Z \sum_k \phi^*_k(x',y') \hat{c}_k^{\dagger}
\end{eqnarray}
such that $\langle \mathbf{r} \vert \hat{\Psi}^{\dagger}(0) \vert 0 \rangle = \psi(\mathbf{r},0)$. Here, $Z$ is a normalization constant that ensures $\{ \Psi^{\dagger},\Psi\} = 1$. In terms of the Jacobi theta function $\vartheta(z,q) \equiv \sum_{n} q^{n^2} e^{2\pi i nz}$, $Z = \vartheta(0,e^{-4\pi^2 l_B^2/L_y^2})^{-1/2}$.

Some remarks on these electron injection operators: note that due to the extra phase term $e^{-i(x+x')(y-y'-nL_y)/2l_B^2}$ the wave function of the injected electron is not translation invariant. This is to be expected since we already broke translation invariance when we fixed the gauge potential to be $\mathbf{A} = Bx~\hat{\mathbf{y}}$. Furthermore, this fixing of the gauge also sets the periodic boundary condition to be $\psi(x,y+L_y) = \psi(x,y)$ with no extra phase factor from the flux, as can be seen from Eq.~\eqref{eq:LocWF}.    

Using this expression for $\hat{\Psi}^{\dagger}(\mathbf{r})$ we can expand the LDOS as (omitting normalization for brevity)
\begin{eqnarray}
    &&\mathrm{LDOS}(\mathbf{r},E) \nonumber\\
    && = \sum_n \left\vert \sum_k e^{-\frac{1}{2l_B^2}(x - \frac{2\pi kl_B^2}{L_y})^2} e^{i\frac{2\pi k}{L_y}y'} \langle n \vert \hat{c}^{\dagger}_k \vert GS\rangle \right\vert^2 \delta(E-E_n) \nonumber \\
    && = \sum_n \sum_k e^{-\frac{1}{l_B^2}(x - \frac{2\pi kl_B^2}{L_y})^2} \vert \langle n \vert \hat{c}_k^{\dagger} \vert GS \rangle \vert^2 \delta(E-E_n) \nonumber \\
    && = \sum_{k \in 3\mathbb{Z}} \sum_{i=0,1,2} e^{-\frac{4\pi^2 l_B^2 }{L_y^2}(k+i-\frac{xL_y}{2\pi l_B^2})^2}  \sum_n \vert \langle n\vert \hat{c}^{\dagger}_i \vert GS \rangle \vert^2  \delta(E-E_n) \nonumber\\ \label{eq:LDOSgeneral}
\end{eqnarray}
where in the second line we used the orthogonality of $\hat{c}^{\dagger}_k \vert GS \rangle$ and in the third line we used the translation symmetry modulo 3 lattice sites of the ground state. If, instead of collapsing to one of the triply degenerate ground states, the ground state stays in an equal superposition, as we would expect in a homogeneous bulk system, then the LDOS can be rewritten as 
\begin{eqnarray}
    \mathrm{LDOS}(\mathbf{r},E) = &&\frac{1}{3}\left( \sum_{k\in \mathbb{Z}} e^{-\frac{4\pi^2 l_B^2 }{L_y^2}(k-\frac{xL_y}{2\pi l_B^2})^2} \right) \nonumber\\ &&\times\sum_n \sum_{i=0,1,2} \vert \langle n \vert \hat{c}^{\dagger}_i \vert GS \rangle \vert^2 \delta(E-E_n) 
\end{eqnarray}
where $\vert GS \rangle$ is any of the triply degenerate ground state.

We can then expand the wave functions $\vert GS \rangle,$ $\vert n \rangle \in \{\vert e_1;(+)_i(-)_j\}\rangle$ in first order perturbation theory to calculate the local density of states. Naively we might expect an injected electron to split into three independent $-e/3$ quasiparticles with a broad excitation peak. However, from our previous discussions it is clear that the quasiparticles are confined by the center of mass conservation, which prevents mixing of the states generated by different injection position $\hat{c}^{\dagger}_i$. So, instead of forming a broad peak, the injected state would at best split into several discrete states that are formed by the application of the $\hat{V}_{km}$ hopping to the initial injection state $\hat{c}^{\dagger}_1 \vert GS \rangle$. 

Since the matrix elements for hopping to these states are proportional to $V_{km}$, the dominant states would be those that are accessible from the initial state by just one application of $\hat{V}_{k1}$, so here we will consider only the first order perturbation of $\vert n \rangle$ and $\vert GS \rangle$ by $\hat{V}_{k1}$. Without loss of generality, we set $\vert GS \rangle = \vert ...\underset{k=0}{1}00100100... \rangle$ Expanding the ground state, we have
\begin{eqnarray}
    \vert GS \rangle = &&\left(1-\frac{N}{2} \sum_{k \in (3\mathbb{Z}-1)}\frac{V_{k1}^2}{(2E_{ph} + \Delta^2 V_{k0})^2}\right)\vert GS \rangle^{(0)} \nonumber \\ &&- \sum_{n}\sum_{k \in (3\mathbb{Z}-1)} \frac{V_{k1}}{2E_{ph} + \Delta^2 V_{k0}} \vert (+)_n (-)_{n + k} \rangle 
\end{eqnarray}
Injecting an electron into the $k=0$ lattice site,
\begin{eqnarray}
    \hat{c}^{\dagger}_0 \vert GS \rangle = - \sum_{k \in (3\mathbb{Z}-1)} \frac{V_{k1}}{2E_{ph} + \Delta^2 V_{k0}} (\vert e_1; (-)_{k} \rangle + \vert e_{-1}; (+)_{k-1} \rangle) \nonumber \\
\end{eqnarray}
while for the $k=1$ lattice site,
\begin{eqnarray}
    \hat{c}^{\dagger}_1 \vert GS \rangle = &&\left(1-\frac{N}{2} \sum_{k \in (3\mathbb{Z}-1)}\frac{V_{k1}^2}{(2E_{ph} + \Delta^2 V_{k0})^2}\right)\vert e_1 \rangle \nonumber \\&&- \sum_{n}\sum_{k \in (3\mathbb{Z}-1)} \frac{V_{k1}}{2E_{ph} + \Delta^2 V_{k0}} \vert e_1; (+)_n (-)_{n + k} \rangle \nonumber \\
\end{eqnarray}    

\begin{widetext}
For $\hat{c}^{\dagger}_0$, the states with non-zero matrix elements are then $\vert e_1;(-)_k \rangle$ and $\vert e_{-1};(+)_k \rangle$, which gives the same matrix elements by symmetry. For $\hat{c}^{\dagger}_1$, the relevant states $\vert n \rangle$ with a non-zero matrix elements are $\vert e_1 \rangle, \vert e_1; (+)_n (-)_{n+k}\rangle, \vert e_2 \rangle$ and $\vert e_2; (-)_p\rangle$. Their first order corrections are
\begin{eqnarray}
    \vert e_1 \rangle ^{(1)}&&= -\left(\sum_{n \neq 0,-k} \sum_{k \in (3\mathbb{Z}-1)} \frac{V_{k1}^2}{2(2E_{ph} + \Delta^2V_{k0} - \Delta_k {\Delta_1}V_{n-1,0})^2} + \frac{V_{k1}^2}{2(E_{ph} - \Delta_{1}V_{k-1,0})^2} \right)\vert e_1\rangle \nonumber \\ &&- \sum_{n \neq 0,-k} \sum_{k \in (3\mathbb{Z}-1)} \frac{V_{k1}}{2E_{ph} + \Delta^2V_{k0} - \Delta_k {\Delta_1}V_{n-1,0}} \vert e_1; (+)_n (-)_{n+k} \rangle \nonumber \\
    &&- \sum_{k \in (3 \mathbb{Z}-1)} \frac{V_{k1}}{E_{ph} - \Delta_{1}V_{k-1,0}} \vert e_2; (-)_k \rangle \\
    \vert e_2; (-)_p \rangle^{(1)} &&=  - \frac{V_{p1}^2}{2(E_{ph } + \Delta_1 V_{k,0})^2}\vert e_2; (-)_p \rangle + \frac{V_{p1}}{E_{ph} + \Delta_1V_{k,0}} \vert e_1 \rangle + (...) \vert e_2; (-)_p (+)_m (-)_{m\pm k} \rangle \\
    \vert e_1; (+)_m (-)_{m+p}\rangle^{(1)} &&= - \frac{V^2_{p1}}{2(2E_{ph} + \Delta^2 V_{p0}- \Delta_k {\Delta_1}V_{n-1,0})^2} \vert e_1; (+)_m (-)_{m+p} \rangle \nonumber\\ &&+ \frac{V_{p1}}{2E_{ph} + \Delta^2 V_{p0}- \Delta_k {\Delta_1}V_{n-1,0}}\vert e_1^{(0)} \rangle + (...) \vert e_1; (+)_m (-)_{m+p} (+)_n (-)_{n+k} \rangle
\end{eqnarray}
where these are normalized such that $\langle \psi \vert \psi \rangle = 1 +  \mathcal{O}(N^2 V^4)$. The terms with $(...)$ as coefficients are orthogonal to all the other states so these coefficients are omitted for brevity. 
We can then write down the matrix elements as follows
\begin{eqnarray}
    \vert \langle e_1;(-)_p  \vert \hat{c}^{\dagger}_0 \vert GS\rangle \vert^2 &&= \vert \langle e_{-1};(+)_p \vert \hat{c}^{\dagger}_0 \vert GS \rangle \vert^2 = \frac{V_{k1}^2}{2E_{ph} + \Delta^2 V_{k0}}\\
    \vert \langle e_1 \vert \hat{c}_1^{\dagger} \vert GS \rangle \vert ^2 &&= 1 - \sum_{k \in (3\mathbb{Z}-1)}\frac{V_{k1}^2}{(2E_{ph} + \Delta^2 V_{k0})} - \nonumber \\ &&\sum_{n \neq 0,-k} \sum_{k \in (3\mathbb{Z}-1)} \left( \frac{V_{k1} (\Delta_k \Delta_{1} V_{n-1,0})}{(2E_{ph} + \Delta^2 V_{k0})(2E_{ph} + \Delta^2 V_{k0} - \Delta_k \Delta_1 V_{n-1,0})}\right)^2 \nonumber \\ &&- \sum_{k \in (3\mathbb{Z}-1)} \frac{V_{k1}^2}{(E_{ph} - \Delta_1 V_{k0})^2} \\
    \vert \langle e_1; (+)_m (-)_{m+p} \vert c_1^{\dagger} \vert GS \rangle \vert ^2 &&= \left( \frac{V_{p1} (\Delta_p \Delta_{1} V_{n-1,0})}{(2E_{ph} + \Delta^2 V_{p0})(2E_{ph} + \Delta^2 V_{p0} - \Delta_p \Delta_1 V_{n-1,0})}\right)^2 \\
    \vert \langle e_2; (-)_p \vert c_1^{\dagger}\vert GS \rangle \vert^2 && = \frac{V_{p1}^2}{(E_{ph} - \Delta_1 V_{k0})^2}
\end{eqnarray}
\end{widetext}

For a negative bias voltage, the STM will instead remove an electron/inject a hole into the system. Following the same arguments as before, but replacing $\hat{\Psi}^{\dagger}(\mathbf{r})$ with its conjugate, we have
\begin{eqnarray}
    &&\mathrm{LDOS}(\mathbf{r},E) \nonumber \\
    &&= \sum_n \vert \langle n \vert \hat{\Psi}(\mathbf{r}) \vert GS \rangle \vert ^2 \delta(E-E_n) \nonumber \\
    &&= \sum_n \sum_k \left(\sum_i e^{-\frac{4\pi^2 l_B^2 }{L_y^2}(k+i-\frac{xL_y}{2\pi l_B^2})^2} \vert \langle n \vert \hat{c}_i \vert GS \rangle \vert^2 \right) \delta(E-E_n)  \nonumber\\
    &&=  \frac{1}{3}\left( \sum_k e^{-\frac{4\pi^2 l_B^2 }{L_y^2}(k-\frac{xL_y}{2\pi l_B^2})^2}  \right) \sum_n \sum_{i=0,1,2}  \vert\langle n \vert \hat{c}_i \vert GS \rangle \vert^2  \delta(E-E_n) \nonumber \\
 \end{eqnarray}
The calculation for the matrix elements follows that of the electron. 

\begin{figure}
\includegraphics[width=0.49\textwidth]{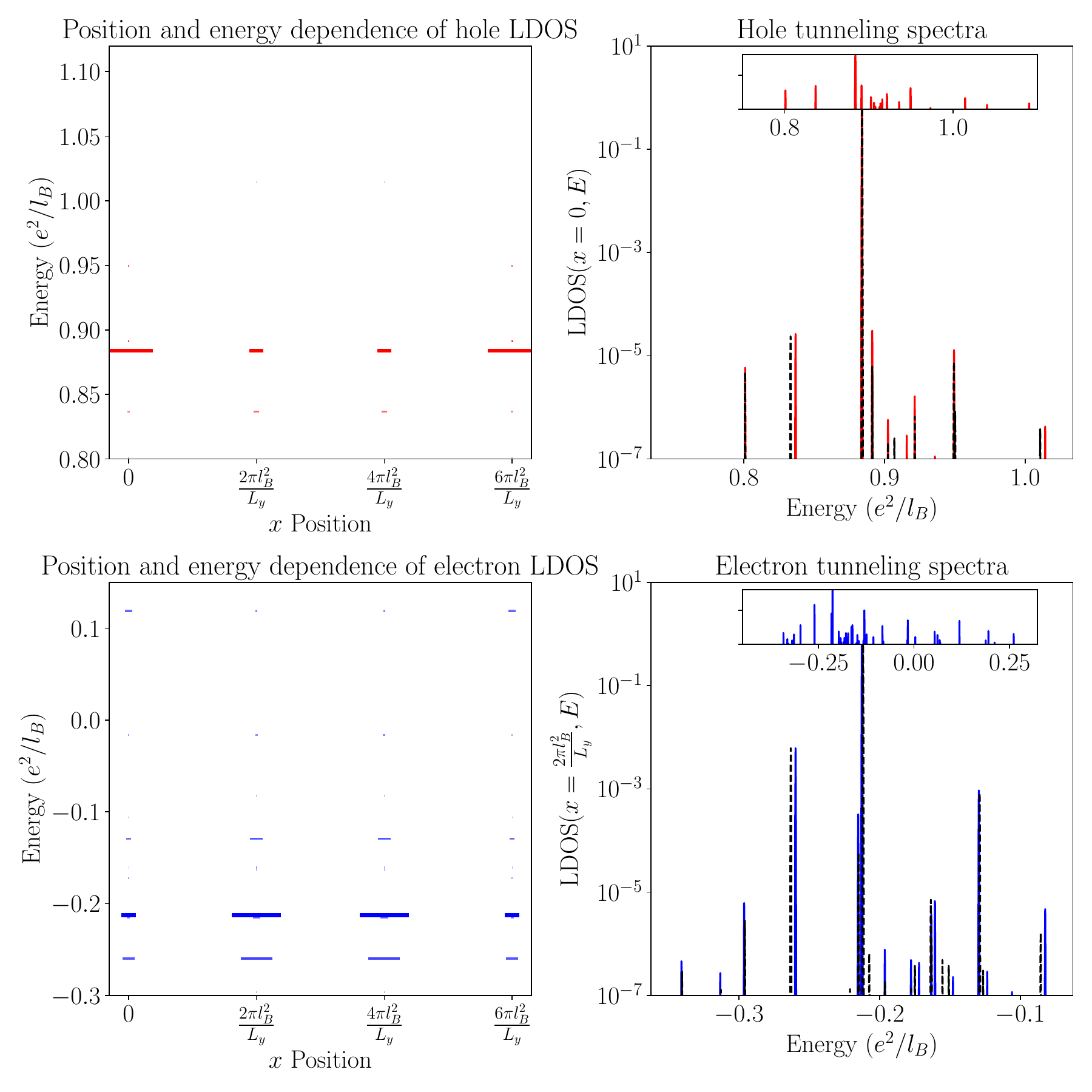}%
\caption{Calculation of the LDOS spectra for $L_y = 4l_B$. The energy scales are measured from the ground state. Note the negative charge gap for the electron due to the positive background. While we could include a chemical potential term such that the electron and hole spectra can be plotted on a single axis with $E_h<E_e$, we chose not to since they belong to different charge and dipole moment sectors. (Left) The position and energy of the peaks in the LDOS spectra (perturbation theory results), assuming the ground state is in the $\vert ...100100100...\rangle$ state. The size of the markers are proportional to the strength of the peaks. The boldest stripe mark the strongest peak. (Right) The LDOS peaks as a function of energy at position $x = 2\pi l_B^2/L_y$ ($x=0$) for electron (hole) injection. We also plot our ED results with $N_{\phi} = 24$ lattice sites (dashed black lines) for comparison. \textbf{Note the log scale}. The energies for the ED results are shifted by $3(V_{10,0} - V_{11,0})/2$ to account for size effects from the periodic boundary condition. The inset shows the spectra from perturbation theory calculation for the full range of energy, including those that are too high to be reached using ED.   
\label{fig:SpecLDOS}}
\end{figure}

We plot the expected spectrum at $x=0$ ($x=2\pi l_B^2/L_y$) $\sum_n \vert \langle n \vert c^{(\dagger)} \vert GS\rangle \vert^2  \delta(E-E_n)$ for holes (electron) on the right side of Figure \ref{fig:SpecLDOS}, using $L_y = 4l_B$. Again, we compare results from perturbation theory (solid lines) with exact diagonalization results with $N_{\phi} = 24$ lattice sites (dashed lines), A few remarks on Figure \ref{fig:SpecLDOS}: first, to account for size effects, we shift the energy of the ED lines by $3(V_{10,0} - V_{11,0})/2$, which compensates the difference in potential experienced by two dipoles separated by $N_{\phi}/2 = 12$. Secondly, we see there are a small number of lines that are present from our calculation but missing in the ED results and vice versa. Lines that are missing in the ED results are those coming from $\vert (+)_i (-)_{i+k} \rangle$ with $\vert k  \vert>12$ which are longer than the half-length of our ED system, while lines that are missing from the perturbation calculations are states that can only be reached by more than one hopping, which are not accounted for by our first-order perturbation - note that these peaks are always smaller than $10^{-6}$. We see that with these taken into account, the perturbation calculations and the ED results are in good agreement.

On the left side of Figure~\ref{fig:SpecLDOS} we show the position and energy dependence of the LDOS. The size of the lines (except the largest ones, which would be much larger than the other) is proportional to the strength of the peaks. Note that because the ground state is collapsed to $\vert...100100.. \rangle$ we obtained an inhomogeneous pattern with period $6\pi l_B^2/L_y$. On the right we showed the magnitude and energy of the peaks (in log scale) for certain values of $x$: $2\pi l_B^2/L_y$ for the electron excitation and $0$ for the hole excitation. These peaks correspond to the different excitations in the charged sector.  

To better understand the relevant excitations, we can rewrite the excited states in terms of quasihole-quasiparticle excitations. As discussed in \cite{Bergholtz2008}, the quasiparticles (quasiholes) of our system show up as an occupation pattern of  $...101...$ (or $...0001...$ for quasiholes), and are domain walls between regions with unperturbed state $...010010010...$ and those that are moved to the left (right) $...100100100...$ ($...001001001...$). However, we also have states which contains $...0110...$, forming a domain wall between $...100100100...$ and $...001001001...$. These we interpret as two quasiparticles stacked on the same momentum with total charge $-2e/3$. The bare electron injection state $\vert e_1 \rangle$ is then described by two stacked quasiparticles at $k=0$ and a single quasiparticle at $k=2$ -- for a total charge of $-e$, as expected:
\begin{eqnarray}
    \vert e_1 \rangle = \vert ... 1001\underset{-\frac{2e}{3}}\vert \textcolor{blue}{\underset{\uparrow}{\mathbf{1}}}0 \underset{-\frac{e}{3}}\vert1 001 ...\rangle = \vert (-\frac{2e}{3})_{0} (-\frac{e}{3})_2 \rangle
\end{eqnarray}
Similarly, the bare hole injection state is composed of three adjacent quasiholes
\begin{eqnarray}
    \vert h_0 \rangle &&= \vert ... 100100 \underset{+\frac{e}{3}} \vert \textcolor{red}{\underset{\downarrow}{\mathbf{0}}} \underset{+\frac{e}{3}}\vert0 \underset{+\frac{e}{3}}\vert01001 ... \rangle \nonumber \\&&= \vert (+\frac{e}{3})_{-1}(+\frac{e}{3})_0 (+\frac{e}{3})_1 \rangle
\end{eqnarray}
The hoppings would then either move these quasiparticles around or create nearby dipole pair states. There is an energy cost to creating dipoles which penalizes the weight of these states, so among these states the most significant ones are the ones that split or move the quasiparticles around such that the reduced interaction energy compensates this energy cost - provided, of course, that these state are reachable from the initial state by a short ranged hopping. For example, the four most prominent states for $\hat{c}^{\dagger}_1$ are
\begin{eqnarray}
    \vert e_2;(-)_{-1} \rangle &&= \vert ..10 \underset{-\frac{e}{3}}\vert\overset{\curvearrowleft}{\textcolor{blue}{\underline{1}0}}~\overset{\curvearrowright}{\textcolor{blue}{\underset{\uparrow}0\mathbf{1}}} \underset{-\frac{2e}{3}}\vert1001..\rangle \label{eq:dipQuasi1} \\
    \vert e_1;(+)_{-3} (-)_{-1}\rangle &&= \vert ... 100 \underset{+\frac{e}{3}}\vert \overset{\curvearrowright}{\textcolor{red}{\underline{0}1}} \underset{-\frac{2e}{3}} \vert \overset{\curvearrowleft}{\textcolor{blue}{\underline{1}0}} \underset{-\frac{e}{3}} \vert\textcolor{blue}{\underset{\uparrow}{\mathbf{1}}}0 \underset{-\frac{e}{3}} \vert 1001 ...\rangle \nonumber \\ \\
    \vert e_1; (+)_3 (-)_{-4} \rangle &&= \vert... \underset{-\frac{e}{3}}\vert \overset{\curvearrowleft}{\textcolor{blue}{\underline{1}0}}~0 \underset{+\frac{e}{3}}\vert01 \underset{-\frac{2e}{3}} \vert\textcolor{blue}{\underset{\uparrow}{\mathbf{1}}}0~\overset{\curvearrowright}{\textcolor{red}{\underline{0}1}}~0 \underset{-\frac{e}{3}}\vert1 ... \rangle \\
    \vert e_1; (+)_3 (-)_{-1} \rangle &&= \vert ... 10\underset{-\frac{e}{3}} \vert \overset{\curvearrowleft}{\textcolor{blue}{\underline{1}0 }}\underset{-\frac{e}{3}} \vert\textcolor{blue}{\underset{\uparrow}{\mathbf{1}}}0~\overset{\curvearrowright}{\textcolor{red}{\underline{0}1}}~0 \underset{-\frac{e}{3}}\vert100 ...\rangle \label{eq:dipQuasi4} 
\end{eqnarray}
(The arrow indicates where the electron is initially injected). 

Our calculation of the tunneling spectral function shows multiple sharp peaks, which agrees with both experimental results and previous theoretical calculations based on the composite fermion approach \cite{Jain2005}. The sharpness of the peaks implies that the injected electrons form a long-lived bound state because, as we see from Eqs. \eqref{eq:dipQuasi1}-\eqref{eq:dipQuasi4}, the quasiparticles formed from the electron injection are unable to separate quickly through the electron-electron hopping even though configurations with widely separated quasiparticles have a much lower energy. Our approach clearly shows that the lack of broadening \rev{\sout{is not due to energetic constraints but}} due to the dipole moment/$y$-momentum conservation, which prevents the formation of 'hoppable' states that are $x$ momentum eigenstates. \rev{Quasiparticles located at different orbital/$x$-position would have different dipole moments, which means that to move quasiparticles around we would have to create (or remove) a dipole, incurring an energy cost. This prevents the existence of a large number of easily hoppable states with similar energy as the injection state, in contrast with the neutral excitations, where without the hopping terms we have a large (extensively many) number of states with degenerate energy which then forms a broadened band when we turn on the hopping.} This phenomena is analogous to angular momentum conservation in the $L_y \rightarrow \infty$ system, which would also prevent broadening by forming quasiparticle bound states \cite{He1993, Jain2005}.  

Most of the spectral weight is concentrated on the main peak - the 'bare' state created by injecting an electron or a hole. The small spectral weight of the other peaks is somewhat expected since the hopping element required to split the electron into separate quasiparticle is suppressed in our thin-cylinder system - we would expect the secondary peak to grow in size as $L_y$ is increased. However, the general pattern of the peak should provide hints to the structure of the spectrum in the $L_y \rightarrow \infty$ system. After the main peak, the weight of the remaining 1st order peaks are ordered based on the magnitude of the $V_{k1}$ hopping required to reach them. Since we expect local hoppings to stay larger than long-ranged hoppings even for large $L_y$ this suggests that in the $L_y \rightarrow \infty$ system we would see the quasiparticles remain clustered around the initial injection site, with only a small number of excited quasiparticle configurations accessible. 

To get a clearer picture of the spectrum without the clutter of states created by the longer range hoppings, it is useful to redo the calculation using the Haldane pseudopotential $V_{km} \sim (k^2 - m^2) e^{-2\pi^2 l_B^2 (k^2+m^2)/L_y^2}$. For $L_y \sim 4 l_B$, we would have $V_{10} \gg V_{20} > V_{21} \gg$ everything else, so instead of treating $\hat{V}_{k1}$ as the perturbation, we take $\hat{V}_{10}$ (nearest neighbor interaction) as the unperturbed Hamiltonian and $\hat{V}_{20} + \hat{V}_{21}$ as perturbation of the same order $\lambda = e^{-6\pi^2 l_B^2/L_y^2 }$ (for $L_y = 4l_B, V_{20} \sim 5 V_{21}$). Everything else is less than $\lambda^2$ so they can be discarded. With just $V_{10}$ as the unperturbed interaction, the excited states are highly degenerate. We have three states that have the same energy as the bare electron state $\vert e_1 \rangle$:
\begin{eqnarray}
    \vert e_2;(-)_{-1} \rangle &&= \vert ..10 \underset{-\frac{e}{3}}\vert\overset{\curvearrowleft}{\textcolor{blue}{\underline{1}0}}~\overset{\curvearrowright}{\textcolor{blue}{\underset{\uparrow}0\mathbf{1}}} \underset{-\frac{2e}{3}}\vert1001..\rangle \label{eq:dipQuasi1} \\
    \vert e_1;(+)_{-3} (-)_{-1}\rangle &&= \vert ... 100 \underset{+\frac{e}{3}}\vert \overset{\curvearrowright}{\textcolor{red}{\underline{0}1}} \underset{-\frac{2e}{3}} \vert \overset{\curvearrowleft}{\textcolor{blue}{\underline{1}0}} \underset{-\frac{e}{3}} \vert\textcolor{blue}{\underset{\uparrow}{\mathbf{1}}}0 \underset{-\frac{e}{3}} \vert 1001 ...\rangle \nonumber \\ \\
    \vert e_2; (-)_{-1} (+)_3 (-)_5\rangle &&= \vert ..10 \underset{-\frac{e}{3}}\vert\overset{\curvearrowleft}{\textcolor{blue}{\underline{1}0}}~ \overset{\curvearrowright}{\textcolor{blue}{\underset{\uparrow}0\mathbf{1}}} \underset{-\frac{e}{3}}\vert \overset{\curvearrowright}{\textcolor{red}{\underline{0}1}}\underset{-\frac{2e}{3}}\vert\overset{\curvearrowleft}{\textcolor{blue}{\underline{1}0}}~0\underset{+\frac{e}{3}}\vert01..\rangle 
    \nonumber \\\label{eq:HaldQuasi3}
\end{eqnarray}
Applying the perturbation up to first order, we see that a superposition of $\vert e_2; (-)_{-1} \rangle$ and $\vert e_1\rangle$ forms two similar peaks separated by $2V_{21}$ around $E = V_{10} + V_{20}$ while$\vert e_1;(+)_{-3}(-)_{-1} \rangle $ and $\vert e_2; (-)_{-1} (+)_3 (-)_5\rangle$ forms a smaller pair of peaks at $V_{10} + 2V_{20}$. Other than the double-peak, the spectrum has generally similar features with our calculation in Figure 5.

\subsection{\label{ssec:2Dlimit}Connection to the two-dimensional limit}


We have obtained the energetics and tunneling spectra of the $\nu=1/3$ system on a torus or cylinder, but the question remains as to which of our predictions would remain relevant for the $L_y \rightarrow \infty$ planar system. The small direction of the thin-torus geometry gives one a small parameter with which to control perturbation theory and organize a hierarchy of effects, when there is no such small parameter for the interacting system in 2D. We note that, since non-variational numerical methods such as exact diagonalization and density-matrix renormalization group with large bond dimension are carried out in finite geometries, one could worry similarly that results from those might not extend to the planar case. Our main goal in this section is to build on the intuition that the locality of physics in the FQHE state means that there is a continuity between a large cylinder 
and the plane, since local physics will be insensitive to the boundary conditions--while the boundary conditions do affect topological properties such as ground state degeneracy, the different ground states are not connected by local operators. In particular, the charged excitations of the system, the quasielectron and quasiholes, are essentially localized since the dipole moment conservation only allows movement of the quasiparticles by emitting or absorbing a dipole, which costs energy. This implies that for a large enough cylinder ($L_y \gtrsim l_B$) the injected electron cannot 'see' the periodic boundary so the cylinder would be indistinguishable from an infinite plane. 

While we chose a small enough $L_y$ that suppresses the larger hopping terms, note that this $L_y$ is already larger than $l_B$, (Most of our calculation are done with $L_y = 4l_B$). Furthermore, evidence shows that the thin torus limit gives plenty of information about larger systems, since we can show that the thin torus ground state is adiabatically connected to the bulk states \cite{Bergholtz2008, Haldane1985a}. It is well known that for Haldane pseudopotentials the $\nu = 1/m$, odd $m$ Tao-Thouless ground state is the limiting form of the Laughlin wave function when $L_y \rightarrow 0$ \cite{Bergholtz2008, Chen2015, Mazaheri2015}. While we are using a different potential, it is known that the FQHE state is insensitive to the specific form of the potential, and furthermore, since Haldane pseudopotentials are much more short ranged than our Coulomb potential, it is reasonable to assume that our potential would capture more of the interaction effects. 

To understand how the thin torus $\nu = 1/m$ state is adiabatically connected to the bulk states for odd $m$ but not for even $m$, let us take a closer look at what would happen in our system as we increase $L_y$. As $L_y$ increases past $2\pi l_B$, the concavity condition would no longer hold for $V_{k0}$ such that $\Delta^2 V_{10} < 0 $. Then, even ignoring hopping terms, the energetics would favor a collapsed "striped" state such as $\vert ..001100001100001100.. \rangle$ over the $m$-periodic Tao-Thouless ground state (the hopping terms would prefer these kind of states since they're more hoppable, $001100001100 \leftrightarrow 100001100001$ ) The ground state of both $\nu=1/2n$ and $1/{(2n+1)}$ states would then be dominated by these kind of collapsed state \cite{Bergholtz2005}. However, unlike the $\nu =1/2n$ states, the total dipole moment of the striped $\nu=1/(2n+1)$ states is the same as the total dipole moment of the original ground state, while for the $\nu=1/2n$ states they have different dipole moments (we can't go from $1010101$ to $01100110$ without shifting some electrons around), which means a closing of the gap is necessary as $L_y$ is increased. To see this, consider that in the TT state each electron is separated by an odd (even) number of zeroes for even (odd) $m$ respectively. This means the dipole conserving hopping terms are unable to collapse the TT state into the stripe state directly for even $m$. Note that the dipole description still works for these kind of states. We can write the ground state in terms of dipoles as (for the $\nu=1/3$ state) $ \vert ...(+)_i (-)_{i+2} (+)_{i+6}(-)_{i+8}... \rangle \pm \vert ...(+)_{i+3} (-)_{i+5} (+)_{i+9}(-)_{i+11}... \rangle$ so we can see that qualitatively the excitations would work similarly to our previous one, only instead of adding dipole pairs we add missing dipole pairs. We then expect the spectra to have a similar structure to ours.

One remaining question is how our sharp LDOS spectra (Fig. \ref{fig:SpecLDOS}) would fare as $L_y$ is increased. As we explained in the previous subsection, the single sharp peak - which is consistent with the observed peak for the $\nu = 2/3$ state in the experimental result of \cite{Hu2023} - is well explained by the localization of charged excitations. Since this localization is a consequence of the dipole moment conservation, this will remain true even when $L_y \rightarrow \infty$. With larger and larger $L_y$, the hopping coefficients would increase such that the injected electron is more likely to fractionalize into other states, but due to the dipole moment conservation the number of these states is going to stay constrained even in the thermodynamic limit (as argued in the previous subsection, there are no band of $x$ momentum eigenstates $\sum e^{iqn} \hat{T}^n\vert e_i \rangle$ to hop into). The only states that are reachable would all be similar in energy, as can be seen through the following heuristic argument: in the $n$-th order perturbation expansion, the magnitude of a secondary peak that might cause a significant broadening of the main peak would be proportional to $(V_{km}^n/(E^{n-1} \Delta E))^2$, where $\Delta E$ is the distance from the main peak while $E$ is the average distance from other hoppable states. $E$ would be proportional to $V_{k0}$, so we have 
\begin{eqnarray}
    \left( \frac{V_{km}^n}{E^{n-1} \Delta E}\right) &&\sim \left( \frac{V_{km}}{V_{k0}}\right)^{(n-1)} \left( \frac{V_{km}}{\Delta E}\right) \nonumber \\&&\sim e^{-\frac{2\pi^2 l_B^2 m^2 n}{L_y^2}} \frac{e^2}{L_y \Delta E}
\end{eqnarray}
Now, since $e^{-\frac{2\pi^2 l_B^2 mn}{L_y^2}}\rightarrow1$ for large $L_y$, we see that this quantity should not increase monotonically with $L_y$. We see then for this magnitude to be $\mathcal{O}(1) \times e^2/l_B$ we need $\Delta E \sim L_y^{-1}$. Thus in the bulk limit, if the higher-order peaks are to contribute to the main peak, they would have to be very close to it, creating a sharp peak instead of a broad one. To test our argument, we calculated the LDOS spectra using exact diagonalization on the $N_{\phi} = 24$ torus for larger aspect ratio $\alpha = L_x/L_y$ (Fig. \ref{fig:SpecAlpha}). We found that when $\alpha$ is increased the spectra is now distributed into several smaller peaks, but as we argued above all the significant peaks are concentrated in a narrow energy window. Note the small scale of the energy scale: for the magnetic field used in \cite{Hu2023} $e^2/l_B \sim 0.015~\mathrm{eV}$ so the width of these peaks are just around $1.5~\mathrm{meV}$, which is consistent with the results in Ref. \onlinecite{Hu2023}. With the limited energy resolution of the STM experimental set-up, these peaks will likely show up as one sharp peak.

\section{\label{sec:conc}Discussion and Outlook}

\begin{figure}
\includegraphics[width=0.49\textwidth]{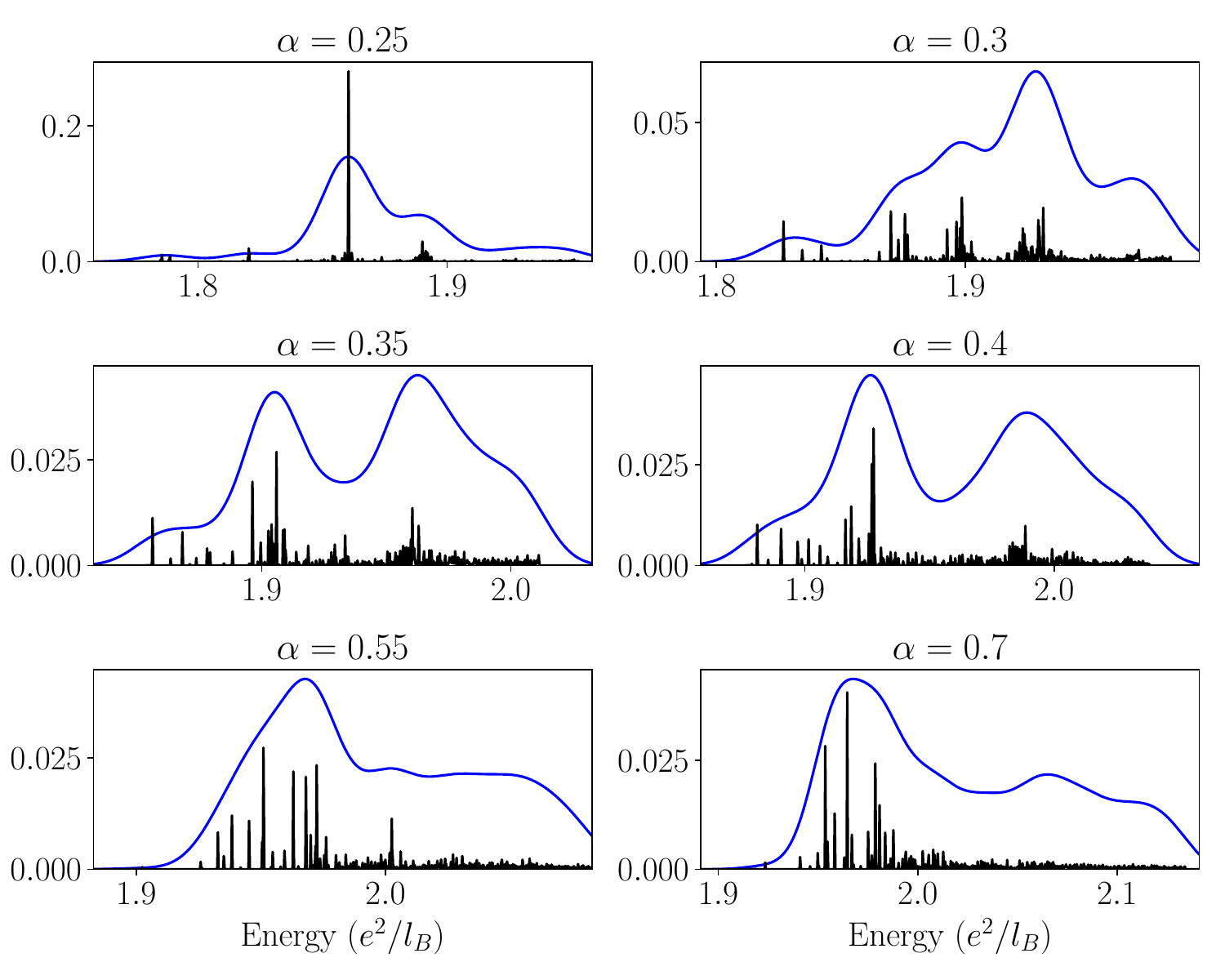}%
\caption{Exact diagonalization results for the electron injection LDOS spectra for varying values of torus aspect ratio $\alpha = L_y/L_x$ on a torus with $N_{\phi} = 24$ lattice sites. The blue line is a Gaussian envelope of width $0.01e^2/l_B$ added to smooth out the peaks as in \cite{Gattu2024, Pu2024}. The thin torus case corresponds to $\alpha \sim 0.1$. 
\label{fig:SpecAlpha}}
\end{figure}

Recent theoretical works have shown the possibility of using STM measurements as a 'fingerprint' to determine the nature of the FQHE state under study \cite{Papic2018, Gattu2024, Pu2024}, and we discuss the relationship between those results, obtained using different techniques, our thin-cylinder results, and numerical results on tori of various aspect ratios. Gattu et.al. \cite{Gattu2024} and Pu et.al. \cite{Pu2024} use a composite fermion (CF) \cite{Jain2005, Pu2017} wavefunction ansatz on a sphere to calculate the STM spectra of the $\nu=1/3$ and $\nu=2/5$ states and obtain the characteristic STM spectra for these states. Their result shows two prominent peaks for electron injection into the $\nu = 1/3$ state, while for the $\nu=2/5$ state they found three different peaks. Their approach is similar to our calculation of the LDOS spectra - we can in fact identify some of the CF configurations from their work with a corresponding thin cylinder quasiparticle configuration. In their work, the first peak is identified with a configuration with two CF in the first $\Lambda$L (CF Landau level) and one CF in the second $\Lambda$L. The CF in the second $\Lambda$L is stacked with another CF in the lower $\Lambda$L, i.e. they have the same momentum. We can identify the single CF with a single quasiparticle in our system, while the two stacked CF, which in total has double the momentum of the single CF, is identified with a stacked quasiparticle of total charge $-2e/3$, so that they have the same momentum. This first peak is then identifiable as a superposition of the bare electron state $\vert e_1 \rangle$ or the nearby $\vert e_2; (-)_{-1} \rangle$ state. The other peak, however, contains CF configurations with no clear analogue in our system. Our exact diagonalization calculations for larger aspect ratio (Fig. \ref{fig:SpecAlpha}) did not find a clear second peak that is identifiable as the second peak in CF theory. At around $\alpha =0.4-0.55$ there is a group of smaller peaks a short distance apart from the main grouping, but this does not look like a uniquely identifiable second peak as one would expect from the results of Refs. \onlinecite{Gattu2024, Pu2024}, especially since this peak appears to merge with the central group of peaks at larger $\alpha$. Of course, our numerical calculations are limited by the small system size of $N_{\phi}=24$ (8 electrons) which makes the toroidal nature of our system more significant. Conceivably, in a larger system with more similarity to the sphere or the plane, one could obtain a clear second peak. The numerical results on small systems suggest, however, that experiments with finite resolution may well see a single peak whose origin can essentially be understood from the thin-cylinder limit. 

We have shown how the thin-cylinder limit allows us to analytically derive the dispersion of the neutral excitations and the LDOS of the charged sector. We calculated the energies and dispersions of the excitation in the neutral sector, including that of a magnetoroton mode, while in the charged sector we demonstrated that the STM spectra would form sharp peaks due to dipole moment conservation. Our analytical results were checked against exact diagonalization for a small system of electrons on the torus, and similarly could be used to interpret other kinds of numerical studies in this geometry, including the density-matrix renormalization group approach.

There are several directions in which one could extend this work. We can generalize our result to filling fractions other than $\nu = 1/m$ - by expressing the excited states in terms of dipoles as we have done here it is straightforward, if tedious, to write down the second order perturbation correction to the energies. We can also apply the thin cylinder limit to more exotic systems, such as the fractional Chern insulator, building on several earlier works \cite{Bernevig2012}. The dipole conservation of our system also leads to non-trivial hydrodynamics, as explored in \cite{Han2023}. Indeed, numerics has shown that the FQH system exhibit fracton hydrodynamics with a $\sim t^{-0.25}$ decay of the density-density correlation \cite{Zerba2024} for strong enough interaction, while in the thin torus limit it exhibits Hilbert space fragmentation \cite{Zerba2024, Sala2020}. It would be interesting to apply our perturbation theory approach to obtain a better understanding of this crossover and how this might relate to the neutral excitations that we have found. 

\rev{A potentially interesting question remains about the choice of the unperturbed Hamiltonian to which we apply our perturbation theory, which might allow the calculation to approach the bulk limit more closely. For example, in Ref.~\onlinecite{Nakamura2012} an exact $\nu=1/3$ state solution is found for a specific choice of potential. We might then introduce deformation to this Hamiltonian (such that it would approximate the Coulomb Hamiltonian, for example) as a perturbation and see its effects on the unperturbed states.}

\rev{We can also relax the concavity condition on the $V_{k0}$ coefficients and take an unperturbed $H_0$ with non-concave $V_{k0}$s, similar to the one obtained when we take $L_y$ to be large enough. Once the non-concavity is large enough, the unperturbed ground state would have the form of $\vert ...001100001100...\rangle - \vert...100001100001...\rangle$ instead of a charge density wave. (In our dipole notation, this would be written as $\vert \prod_i(+)_i(-)_{i+2} \rangle - \vert \prod_i(+)_{i+3} (-)_{i+5} \rangle $ ). Applying the hopping terms as a perturbation, we would then expect something that would look similar in structure to our current results, with low-lying neutral excitations created by removing $P$ of the dipoles -- with one important difference: each level is now doubly degenerate in the thermodynamic limit, because we have two possible configurations of the ground state for each different center of mass value.}

\rev{We note, however, that with our choice of potential (and with other physical potentials), when the $L_y$ is large enough to break the concavity condition, the hopping coefficients are already of the same order as the $V_{k0}$ coefficients -- which casts doubt on the validity of using this kind of perturbation theory approach to describe a physical system. Still, a more detailed calculation with these choice of potentials could possibly reveal more features of the thin torus state with $L_y$ closer to the bulk value. Furthermore, there is an alternative way of breaking the concavity condition without increasing $L_y$: moving to higher Landau levels could also break the concavity condition. Studying the thin-torus behavior beyond the concavity condition would then also help describe states that involve higher Landau levels.}

\rev{It is also interesting to consider the connections of the thin cylinder limit to matrix product state (MPS) representations of FQHE on a cylinder. Density matrix renormalization group (DMRG) methods have been used to calculate the MPS for several FQHE states from a microscopic Hamiltonian and thereby extract the entanglement entropy and other topological quantities \cite{Zaletel2013}. For certain model wave functions (such as the Laughlin wave function or the Moore-Read wave function), there is in fact an \textit{exact} MPS representation on a cylinder obtainable from the corresponding 1 + 1D CFT of the 2 + 1D FQHE state \cite{Zaletel2012}. As argued in Ref.~\onlinecite{Zaletel2012}, this MPS representation is closely related to the thin torus limit expansion: truncating the corresponding CFT to the lowest energy state recovers the ultrathin ground state, while including the states with momentum $\vert P \vert \leq 1$ recovers the exact thin torus MPS found in Ref. \onlinecite{Nakamura2012}.} 

\rev{The relation between the MPS obtained from DMRG and the exact MPS from the model wavefunction in Ref. \onlinecite{Zaletel2012}, is, in a sense, similar to the relation between our perturbation theory and the exact solution for the truncated Haldane pseudpotential Hamiltonian found in Ref.~\onlinecite{Nakamura2012}. Both the DMRG MPS and our perturbation theory generalizes an exact expression for a particular solvable point at the cost of being approximate -- for our perturbation theory, the approximation comes when we discard higher order terms, while for DMRG, the approximation is done by limiting the bond dimension of the MPS to a finite number. MPS methods have a major advantage over our naive perturbation theory approach, namely that the number of bond dimension required only scales linearly in $L_y$~\cite{Zaletel2012}. To extend our result to larger $L_y$ then, one direction that could be taken in a future work is to find an analytic way to write down our perturbation expansion in MPS form, analogous to the exact MPS found in \cite{Nakamura2012}. }

We hope that the systematic derivation of the STM spectra and other observables in the thin-cylinder limit will aid in the interpretation of new experiments as they continue to emerge, enabled by the remarkable progress in topological materials in the past few years.

\begin{acknowledgments}
    The authors acknowledge useful conversations with M. Zaletel, A. Seidel, and Y.C. Wang. This work was supported by the U.S. Department of Energy, Office of Science, National Quantum Information Sciences Research Centers, Quantum Science Center.
\end{acknowledgments}

\section*{Data Availability}
The data that support the findings of this article are openly available \cite{data}
\appendix
\rev{\section{\label{ssec:CoulombPotAppendix}Interaction coefficients and the thin cylinder limit}}
 
\rev{In this work, we would like to specialize to the thin \textit{cylinder} limit, where we take the thermodynamic limit $L_x \rightarrow \infty$ and $N_{\phi} \rightarrow \infty$. Some difficulties are present in this limit; of note is that a straightforward summation of Eq.~\ref{eq:VkmFourier} will diverge logarithmically for Coulomb-like potentials, since the Gaussian factors $e^{-2\pi^2 l_B^2 n^2/L_x^2}$ fall off very slowly for very large $L_x$. However, now that we have large $L_x$ we can take the continuum limit \rev{in the $x$-direction. \sout{Introducing} With $q_x = 2\pi n/L_x$ \sout{and $q_y = 2\pi k/L_y$ we have} we can rewrite the summation over $x$ as an integral. The coefficients can then be calculated as}
\begin{widetext}
\begin{eqnarray}
    V_{km} = \frac{1}{2\pi L_y} \int_0^{\infty} d^2 \mathbf{q} e^{-q^2 l_B^2/2}  \tilde{V}(\mathbf{q}) \left[\delta\left(q_y - \frac{2\pi}{L_y}m\right)\cos{\frac{2\pi k l_B^2}{L_y} q_x} - \delta\left(q_y - \frac{2\pi}{L_y}k\right) \cos{\frac{2\pi m l_B^2}{L_y}q_x}\right]  
\end{eqnarray}
\end{widetext}
The Gaussian $e^{-q^2 l_B^2/2}$ will suppress the integrand for large $q_x$, but now for $m=0$ or $k=0$ we have an IR divergence for a Coulomb potential with $\tilde{V}(\mathbf{q}) \sim 1/\vert \mathbf{q} \vert$. To deal with this, instead of using a "periodized" Coulomb potential on the torus $V(\mathbf{r}) \sim \sum_{\mathbf{d} = aL_x + bL_y,~a,b \in \mathbb{Z}} 1/\vert\mathbf{r} + \mathbf{d} \vert $ we just used the 3D Coulomb interaction between points on the surface of a cylinder, 
\begin{eqnarray}
    V(\mathbf{r}) = \frac{e^2}{\epsilon\sqrt{x^2 + \frac{L_y^2}{2\pi^2}(1 - \cos{\frac{2\pi y}{L_y}})}}
\end{eqnarray}
With this choice of potential, $\tilde{V}(q_x,0)$ diverges slower than $q_x^{-1/2}$ as $q_x \rightarrow 0$ so that the integral is convergent. 
}
\rev{The $\sim 1/k$ decay of $V_{k0}$ created another complication for the thermodynamic limit: static interaction terms $\sum_k V_{k0}$ diverges logarithmically with system size. To deal with this divergence we need to add a positive background to the system. We add to each site a positive $+e\nu$ charge and we include the interaction between the electron and these positive charges into the coefficients. Like the electrons, these positive charges are Gaussians of width $l_B$ centered on $x = \frac{2\pi n l_B^2}{L_y}$, so to each $V_{k0}$ we add the interaction with the $1/\nu = 3$ nearest positive charges to obtain the regularized coefficients $\tilde{V}_{k0}$.
\begin{eqnarray}
    \bar{V}_{k0} = &&V_{k0} -  \frac{e^2\nu}{\epsilon \sqrt{2\pi} l_B L_y^2} \int_{-\infty}^{\infty} dx  \int_{-L_y}^{L_y} dy~ \nonumber\\ &&\times \sum_{n=-\frac{1-\nu}{2\nu}}^{\frac{1-\nu}{2\nu}} \frac{e^{-\frac{x^2}{2l_B^2}}(L_y - \vert y \vert)}{\sqrt{(x_1 - \frac{2 \pi (k-n)l_B^2}{L_y})^2 + \frac{L_y^2}{2\pi^2}(1 - \cos{\frac{2\pi y}{L_y}})}}  \nonumber\\
\end{eqnarray}
These coefficients now fall off much quicker than $1/k$, so that a sum over $\bar{V}_{k0}$ is convergent. For our choice of potential and geometry, the combined interaction between the electrons and the background is actually attractive such that we would have a negative (positive) electron (hole) charge gap. While Ref. \cite{Lemm2024} shows that without a neutralizing background the charge gap is always greater than the neutral gap, we note that their class of Hamiltonians doesn't include the positive background charge that might be expected in a realistic system. Estimating the observed (positive) charge gap above the ground state could be done, if needed, by using a finite system and selecting a background charge and confining potential to ensure that a particular charge number is the many-particle ground state. The relative energies within a particular charge sector are more general and will be the focus here.}

\section{Expression for the second order correction to energy}

\begin{widetext}
Here we provide the detailed expression for the second order corrections. The diagonal element of the perturbation matrix would depend on the matrix elements with states that are reacheable by a single hopping $\hat{V}_{km}$.
The possible hoppings that can be taken from the dipole-pair state are:
\begin{align}
    1000101010001 \leftrightarrow 1001001001001:~ & \vert\langle (+)_0 (-)_k \vert V_{k1} \vert GS \rangle \vert^2\\
    1001000110001 \leftrightarrow 1000101010001:~ &\vert \langle (+)_0 (-)_k \vert V^{\dagger}_{n1} \vert (+)_{n} (-)_k \rangle \vert^2\\
    0100001001001 \leftrightarrow 0100001000110: ~ & \vert\langle (+)_0 (-)_k \vert V_{p1} \vert (+)_0 (-)_k (+)_n (-)_{n+p} \rangle\vert^2\\
    100010101000100 \leftrightarrow 100001110000100:~ &\vert \langle (+)_0 (-)_k \vert V_{k-2,1} \vert (+2)_0 (-2)_{k-1} \rangle \vert^2  \\
    100101000010100 \leftrightarrow 100110000010010:~ & \vert \langle (+)_0 (-)_k \vert V_{p1} \vert (+)_0  (-2)_{k-1} (+)_{p-k} \rangle\vert^2   
\end{align}
Then the contribution to the diagonal element of the matrix from each of these hoppings are
\begin{align}
    \vert \langle (-)_{3i} (+)_{3i\pm k} \vert V_{k1} \vert GS \rangle \vert^2  &\rightarrow \frac{V_{k1}^2}{2 E_{ph} + \Delta^2 V_{k0}} \\
    \vert \langle (-)_{3i} (+)_{3i \pm k } \vert V_{3n,1} \vert ((-)_{3i \pm 3n} (+)_{3i \pm k} \rangle \vert^2 &\rightarrow 2\sum_{n \neq 0,-k} \frac{V_{3n,1}^2}{\Delta^2 V_{k0} - \Delta^2 V_{k+3n,0}}  \\
    \vert \langle (-)_{3i} (+)_{3i \pm k} \vert V_{p1} \vert l_{3i} (+)_{3i \pm k} (-)_{n} (+)_{n \pm p} \rangle \vert^2\\ \rightarrow -\sum_{(p+k)\mathrm{mod}3=0} \sum_{(n+k)\mathrm{mod}3=2} &\frac{V_{p1}^2 (1-\delta_{n+p,k}-\delta_{n+p,k+1})(1-\delta_{n,k}-\delta_{n,k+1})}{2 E_{ph} + \Delta^2 V_{p0} + \Delta_{\mathrm{sgn}(n)p} \Delta_{-\mathrm{sgn}(n)k}(\Delta^2 V_{n,0}) } \nonumber\\
    -\sum_{(p+k)\mathrm{mod}3=1} \sum_{(n+k)\mathrm{mod}3=2} &\frac{V_{p1}^2 (1-\delta_{n+p,k}-\delta_{n+p,k+1})(1-\delta_{n,k}-\delta_{n,k+1})}{2 E_{ph} + \Delta^2 V_{p0} - \Delta_{\mathrm{sgn}(n)p} \Delta_{-\mathrm{sgn}(n)k}(\Delta^2 V_{n,0}) }\\
    \vert\langle (+)_{3i} (-)_{3i+ k} \vert V_{31} \vert (+2)_{3i} (-2)_{3i-1+ k} \rangle \vert^2 &\rightarrow \frac{V_{k\pm 2,1}^2}{2E_{dip}(2) - \Delta_2 \Delta_{-2} V_{k0} - 2E_{ph} + \Delta^2 V_{k0}}  \\
    \vert \langle (+)_{3i} (-)_{3i+k} \vert V_{p1} \vert (+2)_{3i} (-)_{3i+k} (-)_{3i+p} \rangle \vert^2 &\rightarrow \sum_{p\in (3\mathbb{Z}-1)} \frac{2V_{p-1,1}^2(1-\delta_{p,-1} - \delta_{p,k})}{E_{dip}(2) -\Delta^2 V_{k0} -\Delta^2 V_{p-k,0} - \Delta_{-2} \Delta_{1} V_{k0} - \Delta_{-2} \Delta_{1} V_{p0} } \\
\end{align}
The expression of the diagonal elements, relative to the ground state, is then
\begin{align}
    U^k_{ii} - E_{GS}^{(2)} &=\frac{V_{k1}^2}{2 E_{ph} + \Delta^2 V_{k0}} + 2\sum_{n=1} \frac{V_{3n,1}^2}{\Delta^2 V_{k0} - \Delta^2 V_{k+3n,0}} + 2 \sum_{1\leq n < k/3} \frac{V_{3n,1}^2}{\Delta^2 V_k - \Delta^2 V_{k-3n}}  \nonumber \\ 
    &+ \sum_{p=2,4,5,7,...} \frac{2V_{p1}^2}{2E_{ph} + \Delta^2 V_{p0}}+\sum_{(p+k)\mathrm{mod}3=0} \sum_{(n+k)\mathrm{mod}3=2} \frac{V_{p1}^2 \Delta_{\mathrm{sgn}(n)p} \Delta_{-\mathrm{sgn}(n)k}(\Delta^2 V_{n,0})(1-\delta_{p1})}{2 E_{ph} + \Delta^2 V_{p0} + \Delta_{\mathrm{sgn}(n)p} \Delta_{-\mathrm{sgn}(n)k}(\Delta^2 V_{n,0}) }\nonumber\\
    &+\sum_{(p+k)\mathrm{mod}3=1} \sum_{(n+k)\mathrm{mod}3=2} \frac{V_{p1}^2 (- \Delta_{\mathrm{sgn}(n)p} \Delta_{-\mathrm{sgn}(n)k}(\Delta^2 V_{n,0})) (1-\delta_{p1})}{2 E_{ph} + \Delta^2 V_{p0} - \Delta_{\mathrm{sgn}(n)p} \Delta_{-\mathrm{sgn}(n)k}(\Delta^2 V_{n,0}) } \nonumber \\
    &+\frac{V_{k\pm 2,1}^2}{2E_{dip}(2) - \Delta_2 \Delta_{-2} V_{k0}} \nonumber \\
    &+\sum_{p\in (3\mathbb{Z}+1)} \frac{2V_{p1}^2(1-\delta_{p,1} - \delta_{p,k-1})}{E_{dip}(2) - \Delta^2 V_{p-k+1,0} - \Delta_{-2} \Delta_{1} V_{k0} - \Delta_{-\mathrm{sgn}(p)2} \Delta_{\mathrm{sgn}(p)1} V_{p+1,0} }  \nonumber\\ 
    &+\sum_{p\in (3\mathbb{Z}-1)} \frac{2V_{p1}^2(1-\delta_{p,-1}-\delta_{p,2}-\delta_{p,k+1})}{E_{dip}(2) - \Delta^2 V_{p-k-1,0} - \Delta_{-2} \Delta_{1} V_{k+1,0} - \Delta_{-\mathrm{sgn}(p)2} \Delta_{\mathrm{sgn}(p)1} V_{p,0} }    
\end{align}
which converges since $V_{p1} \sim p^{-3}$ and $\Delta_{\mathrm{sgn}(n)p} \Delta_{-\mathrm{sgn}(n)k}(\Delta^2 V_{n,0})$ falls off quickly with $n$.

For the $P=1$ sector, we have a similar calculation. As in the ground state perturbation, the diagonal element comes from jumps to the state with an extra dipole pair, but now we also need to include jumps to the states where the original dipole is shifted again:
\begin{eqnarray}
    \vert \langle (+)_0 \vert V_{p1}^2 \vert (+)_0 (+)_n (-)_{n+p} \rangle \vert^2 &&\rightarrow \frac{V_{p1}^2}{2E_{ph} + \Delta^2 V_{p0} - \Delta^2 V_{n0} + \Delta^2 V_{n+p,0}}  \\
    \vert \langle (+)_0 \vert V_{p1}^2 \vert (+)_0 (-)_{n-p} (+)_{n} \rangle \vert^2
    && \rightarrow \frac{V_{p1}^2}{2E_{ph} + \Delta^2 V_{p0} - \Delta^2 V_{n0} + \Delta^2 V_{n-p,0}} \\
    \vert \langle (+)_0 \vert V_{p1}^2 \vert (+2)_0 (-)_p \rangle \vert^2 &&
    \rightarrow \frac{V_{p-1,1}^2}{E_{dip}(2) - \Delta_{-2} \Delta_{1} V_{p0}}
\end{eqnarray}
so then, 
\begin{eqnarray}
    U_{ii} - E_{GS}^{(2)} = &&-\sum_{n \in 3 \mathbb{Z}} \sum_{p = 2,5,8,.} \frac{V_{p1}^2(1-\delta_{n,-p})}{2E_{ph} + \Delta^2 V_{p0} - \Delta^2 V_{n0} + \Delta^2 V_{n+p,0}} \nonumber  \\ &&- \sum_{n \in 3 \mathbb{Z}} \sum_{p=4,7,...} \frac{V_{p1}^2 (1-\delta_{n,p})}{2E_{ph} + \Delta^2 V_{p0} - \Delta^2 V_{n0} + \Delta^2 V_{n-p,0}} \nonumber\\
    &&-\sum_{p=2,4,5,7,...} \frac{V_{p1}^2}{E_{dip}(2) - \Delta_{-\mathrm{sgn}(p)2} \Delta_{\mathrm{sgn}(p)1} V_{p0}} + N \sum_{k=2,4,5,7,...} \frac{V_{k1}}{2E_{ph} + \Delta^2 V_{k0}} \\
    =&& \sum_{p=2,4,5,7,...}  \frac{2V_{p1}^2}{2E_{ph} + \Delta^2 V_{k0}} + \sum_{n\in 3\mathbb{Z}} \sum_{p\in (3\mathbb{Z}-1)} \frac{V_{p1}^2(\Delta^2 V_{n+p,0} - \Delta^2 V_{n0})(1-\delta_{n,-p})}{(2E_{ph} + \Delta^2 V_{p0} - \Delta^2 V_{n0} + \Delta^2 V_{n+p,0})(2E_{ph} + \Delta^2 V_{k0})} \nonumber \\
    &&-\sum_{p=4,7,...} \frac{V_{p1}^2}{E_{dip}(2) - \Delta_{2} \Delta_{-1} V_{\vert p+1\vert,0}} -\sum_{p=5,8,...} \frac{V_{p1}^2}{E_{dip}(2) - \Delta_{-2} \Delta_{1} V_{\vert p-1\vert,0}}
\end{eqnarray}
while the off-diagonal elements are given by
\begin{eqnarray}
    \langle (+)_i \vert V_{p1} \vert (+)_i (+)_{j} (-)_{j+p} \rangle \langle (+)_i (+)_j (-)_{j+p} \vert V_{i-j,1} \vert (+)_j \rangle && \rightarrow \frac{V_{p1} V_{i-j+p,1}}{2E_{ph} + \Delta^2 V_{p0} - \Delta^2 V_{j0} + \Delta^2 V_{j+p,0}} \\
\end{eqnarray}
such that
\begin{eqnarray}
    U_{ij} = \sum_{p\in (3\mathbb{Z}-1)} \frac{V_{p1} V_{3(j-i)+p,1} (1-\delta_{3(j-i)+p+1,0})}{2E_{ph} + \Delta^2 V_{p0} - \Delta^2 V_{3(j-i),0} + \Delta^2 V_{3(j-i)+p,0}}
\end{eqnarray}
Again, diagonalizing this matrix would give the second order correction to the propagating $P=1$ states
\begin{eqnarray}
    E_{P=1}^{(2)}(q) = U_{ii} - E_{GS}^{(2)} +\sum_{n} \sum_{p\in (3\mathbb{Z}-1)} \frac{V_{p1} V_{3n+p,1} (1-\delta_{3(j-i)+p+1,0})}{2E_{ph} + \Delta^2 V_{p0} - \Delta^2 V_{3n,0} + \Delta^2 V_{3n+p,0}} \cos{\rev{\frac{2\pi qn}{N}}}
\end{eqnarray}    
\end{widetext}



\begin{thebibliography}{59}%
\makeatletter
\providecommand \@ifxundefined [1]{%
 \@ifx{#1\undefined}
}%
\providecommand \@ifnum [1]{%
 \ifnum #1\expandafter \@firstoftwo
 \else \expandafter \@secondoftwo
 \fi
}%
\providecommand \@ifx [1]{%
 \ifx #1\expandafter \@firstoftwo
 \else \expandafter \@secondoftwo
 \fi
}%
\providecommand \natexlab [1]{#1}%
\providecommand \enquote  [1]{``#1''}%
\providecommand \bibnamefont  [1]{#1}%
\providecommand \bibfnamefont [1]{#1}%
\providecommand \citenamefont [1]{#1}%
\providecommand \href@noop [0]{\@secondoftwo}%
\providecommand \href [0]{\begingroup \@sanitize@url \@href}%
\providecommand \@href[1]{\@@startlink{#1}\@@href}%
\providecommand \@@href[1]{\endgroup#1\@@endlink}%
\providecommand \@sanitize@url [0]{\catcode `\\12\catcode `\$12\catcode `\&12\catcode `\#12\catcode `\^12\catcode `\_12\catcode `\%12\relax}%
\providecommand \@@startlink[1]{}%
\providecommand \@@endlink[0]{}%
\providecommand \url  [0]{\begingroup\@sanitize@url \@url }%
\providecommand \@url [1]{\endgroup\@href {#1}{\urlprefix }}%
\providecommand \urlprefix  [0]{URL }%
\providecommand \Eprint [0]{\href }%
\providecommand \doibase [0]{https://doi.org/}%
\providecommand \selectlanguage [0]{\@gobble}%
\providecommand \bibinfo  [0]{\@secondoftwo}%
\providecommand \bibfield  [0]{\@secondoftwo}%
\providecommand \translation [1]{[#1]}%
\providecommand \BibitemOpen [0]{}%
\providecommand \bibitemStop [0]{}%
\providecommand \bibitemNoStop [0]{.\EOS\space}%
\providecommand \EOS [0]{\spacefactor3000\relax}%
\providecommand \BibitemShut  [1]{\csname bibitem#1\endcsname}%
\let\auto@bib@innerbib\@empty
\bibitem [{\citenamefont {Tsui}\ \emph {et~al.}(1982)\citenamefont {Tsui}, \citenamefont {Stormer},\ and\ \citenamefont {Gossard}}]{Tsui1982}%
  \BibitemOpen
  \bibfield  {author} {\bibinfo {author} {\bibfnamefont {D.~C.}\ \bibnamefont {Tsui}}, \bibinfo {author} {\bibfnamefont {H.~L.}\ \bibnamefont {Stormer}},\ and\ \bibinfo {author} {\bibfnamefont {A.~C.}\ \bibnamefont {Gossard}},\ }\bibfield  {title} {\bibinfo {title} {Two-dimensional magnetotransport in the extreme quantum limit},\ }\href {https://doi.org/10.1103/physrevlett.48.1559} {\bibfield  {journal} {\bibinfo  {journal} {Physical Review Letters}\ }\textbf {\bibinfo {volume} {48}},\ \bibinfo {pages} {1559} (\bibinfo {year} {1982})}\BibitemShut {NoStop}%
\bibitem [{\citenamefont {Girvin}(1999)}]{Girvin1999}%
  \BibitemOpen
  \bibfield  {author} {\bibinfo {author} {\bibfnamefont {S.~M.}\ \bibnamefont {Girvin}},\ }\href {https://arxiv.org/abs/cond-mat/9907002} {\bibinfo {title} {The quantum {H}all effect: Novel excitations and broken symmetries}} (\bibinfo {year} {1999}),\ \Eprint {https://arxiv.org/abs/cond-mat/9907002} {arXiv:cond-mat/9907002 [cond-mat.mes-hall]} \BibitemShut {NoStop}%
\bibitem [{\citenamefont {Klitzing}\ \emph {et~al.}(1980)\citenamefont {Klitzing}, \citenamefont {Dorda},\ and\ \citenamefont {Pepper}}]{Klitzing1980}%
  \BibitemOpen
  \bibfield  {author} {\bibinfo {author} {\bibfnamefont {K.~v.}\ \bibnamefont {Klitzing}}, \bibinfo {author} {\bibfnamefont {G.}~\bibnamefont {Dorda}},\ and\ \bibinfo {author} {\bibfnamefont {M.}~\bibnamefont {Pepper}},\ }\bibfield  {title} {\bibinfo {title} {New method for high-accuracy determination of the fine-structure constant based on quantized {H}all resistance},\ }\href {https://doi.org/10.1103/physrevlett.45.494} {\bibfield  {journal} {\bibinfo  {journal} {Physical Review Letters}\ }\textbf {\bibinfo {volume} {45}},\ \bibinfo {pages} {494} (\bibinfo {year} {1980})}\BibitemShut {NoStop}%
\bibitem [{\citenamefont {Thouless}\ \emph {et~al.}(1982)\citenamefont {Thouless}, \citenamefont {Kohmoto}, \citenamefont {Nightingale},\ and\ \citenamefont {den Nijs}}]{Thouless1982}%
  \BibitemOpen
  \bibfield  {author} {\bibinfo {author} {\bibfnamefont {D.~J.}\ \bibnamefont {Thouless}}, \bibinfo {author} {\bibfnamefont {M.}~\bibnamefont {Kohmoto}}, \bibinfo {author} {\bibfnamefont {M.~P.}\ \bibnamefont {Nightingale}},\ and\ \bibinfo {author} {\bibfnamefont {M.}~\bibnamefont {den Nijs}},\ }\bibfield  {title} {\bibinfo {title} {Quantized {H}all conductance in a two-dimensional periodic potential},\ }\href {https://doi.org/10.1103/physrevlett.49.405} {\bibfield  {journal} {\bibinfo  {journal} {Physical Review Letters}\ }\textbf {\bibinfo {volume} {49}},\ \bibinfo {pages} {405} (\bibinfo {year} {1982})}\BibitemShut {NoStop}%
\bibitem [{\citenamefont {Arovas}\ \emph {et~al.}(1984)\citenamefont {Arovas}, \citenamefont {Schrieffer},\ and\ \citenamefont {Wilczek}}]{Arovas1984}%
  \BibitemOpen
  \bibfield  {author} {\bibinfo {author} {\bibfnamefont {D.}~\bibnamefont {Arovas}}, \bibinfo {author} {\bibfnamefont {J.~R.}\ \bibnamefont {Schrieffer}},\ and\ \bibinfo {author} {\bibfnamefont {F.}~\bibnamefont {Wilczek}},\ }\bibfield  {title} {\bibinfo {title} {Fractional statistics and the quantum {H}all effect},\ }\href {https://doi.org/10.1103/physrevlett.53.722} {\bibfield  {journal} {\bibinfo  {journal} {Physical Review Letters}\ }\textbf {\bibinfo {volume} {53}},\ \bibinfo {pages} {722} (\bibinfo {year} {1984})}\BibitemShut {NoStop}%
\bibitem [{\citenamefont {Laughlin}(1983)}]{Laughlin1983}%
  \BibitemOpen
  \bibfield  {author} {\bibinfo {author} {\bibfnamefont {R.~B.}\ \bibnamefont {Laughlin}},\ }\bibfield  {title} {\bibinfo {title} {Anomalous quantum {H}all effect: An incompressible quantum fluid with fractionally charged excitations},\ }\href {https://doi.org/10.1103/physrevlett.50.1395} {\bibfield  {journal} {\bibinfo  {journal} {Physical Review Letters}\ }\textbf {\bibinfo {volume} {50}},\ \bibinfo {pages} {1395} (\bibinfo {year} {1983})}\BibitemShut {NoStop}%
\bibitem [{\citenamefont {Haldane}(1983)}]{Haldane1983}%
  \BibitemOpen
  \bibfield  {author} {\bibinfo {author} {\bibfnamefont {F.~D.~M.}\ \bibnamefont {Haldane}},\ }\bibfield  {title} {\bibinfo {title} {Fractional quantization of the {H}all effect: A hierarchy of incompressible quantum fluid states},\ }\href {https://doi.org/10.1103/physrevlett.51.605} {\bibfield  {journal} {\bibinfo  {journal} {Physical Review Letters}\ }\textbf {\bibinfo {volume} {51}},\ \bibinfo {pages} {605} (\bibinfo {year} {1983})}\BibitemShut {NoStop}%
\bibitem [{\citenamefont {Halperin}(1984)}]{Halperin1984}%
  \BibitemOpen
  \bibfield  {author} {\bibinfo {author} {\bibfnamefont {B.~I.}\ \bibnamefont {Halperin}},\ }\bibfield  {title} {\bibinfo {title} {Statistics of quasiparticles and the hierarchy of fractional quantized {H}all states},\ }\href {https://doi.org/10.1103/physrevlett.52.1583} {\bibfield  {journal} {\bibinfo  {journal} {Physical Review Letters}\ }\textbf {\bibinfo {volume} {52}},\ \bibinfo {pages} {1583} (\bibinfo {year} {1984})}\BibitemShut {NoStop}%
\bibitem [{\citenamefont {Jain}(1989)}]{Jain1989}%
  \BibitemOpen
  \bibfield  {author} {\bibinfo {author} {\bibfnamefont {J.~K.}\ \bibnamefont {Jain}},\ }\bibfield  {title} {\bibinfo {title} {Composite-fermion approach for the fractional quantum {H}all effect},\ }\href {https://doi.org/10.1103/physrevlett.63.199} {\bibfield  {journal} {\bibinfo  {journal} {Physical Review Letters}\ }\textbf {\bibinfo {volume} {63}},\ \bibinfo {pages} {199} (\bibinfo {year} {1989})}\BibitemShut {NoStop}%
\bibitem [{\citenamefont {Hu}\ \emph {et~al.}(2023)\citenamefont {Hu}, \citenamefont {Tsui}, \citenamefont {He}, \citenamefont {Kamber}, \citenamefont {Wang}, \citenamefont {Mohammadi}, \citenamefont {Watanabe}, \citenamefont {Taniguchi}, \citenamefont {Papic}, \citenamefont {Zaletel},\ and\ \citenamefont {Yazdani}}]{Hu2023}%
  \BibitemOpen
  \bibfield  {author} {\bibinfo {author} {\bibfnamefont {Y.}~\bibnamefont {Hu}}, \bibinfo {author} {\bibfnamefont {Y.-C.}\ \bibnamefont {Tsui}}, \bibinfo {author} {\bibfnamefont {M.}~\bibnamefont {He}}, \bibinfo {author} {\bibfnamefont {U.}~\bibnamefont {Kamber}}, \bibinfo {author} {\bibfnamefont {T.}~\bibnamefont {Wang}}, \bibinfo {author} {\bibfnamefont {A.~S.}\ \bibnamefont {Mohammadi}}, \bibinfo {author} {\bibfnamefont {K.}~\bibnamefont {Watanabe}}, \bibinfo {author} {\bibfnamefont {T.}~\bibnamefont {Taniguchi}}, \bibinfo {author} {\bibfnamefont {Z.}~\bibnamefont {Papic}}, \bibinfo {author} {\bibfnamefont {M.~P.}\ \bibnamefont {Zaletel}},\ and\ \bibinfo {author} {\bibfnamefont {A.}~\bibnamefont {Yazdani}},\ }\bibfield  {title} {\bibinfo {title} {High-resolution tunneling spectroscopy of fractional quantum {H}all states},\ }\href@noop {} {\bibfield  {journal} {\bibinfo  {journal} {arXiv:2308.05789v2}\ } (\bibinfo {year} {2023})},\ \Eprint {https://arxiv.org/abs/2308.05789} {arXiv:2308.05789
  [cond-mat.mes-hall]} \BibitemShut {NoStop}%
\bibitem [{\citenamefont {Bergholtz}\ and\ \citenamefont {Karlhede}(2005)}]{Bergholtz2005}%
  \BibitemOpen
  \bibfield  {author} {\bibinfo {author} {\bibfnamefont {E.~J.}\ \bibnamefont {Bergholtz}}\ and\ \bibinfo {author} {\bibfnamefont {A.}~\bibnamefont {Karlhede}},\ }\bibfield  {title} {\bibinfo {title} {Half-filled lowest {L}andau level on a thin torus},\ }\href {https://doi.org/10.1103/physrevlett.94.026802} {\bibfield  {journal} {\bibinfo  {journal} {Physical Review Letters}\ }\textbf {\bibinfo {volume} {94}},\ \bibinfo {pages} {026802} (\bibinfo {year} {2005})}\BibitemShut {NoStop}%
\bibitem [{\citenamefont {Bergholtz}\ and\ \citenamefont {Karlhede}(2008)}]{Bergholtz2008}%
  \BibitemOpen
  \bibfield  {author} {\bibinfo {author} {\bibfnamefont {E.~J.}\ \bibnamefont {Bergholtz}}\ and\ \bibinfo {author} {\bibfnamefont {A.}~\bibnamefont {Karlhede}},\ }\bibfield  {title} {\bibinfo {title} {Quantum {H}all system in {T}ao-{T}houless limit},\ }\href {https://doi.org/10.1103/physrevb.77.155308} {\bibfield  {journal} {\bibinfo  {journal} {Physical Review B}\ }\textbf {\bibinfo {volume} {77}},\ \bibinfo {pages} {155308} (\bibinfo {year} {2008})}\BibitemShut {NoStop}%
\bibitem [{\citenamefont {Tao}\ and\ \citenamefont {Thouless}(1983)}]{Tao1983}%
  \BibitemOpen
  \bibfield  {author} {\bibinfo {author} {\bibfnamefont {R.}~\bibnamefont {Tao}}\ and\ \bibinfo {author} {\bibfnamefont {D.~J.}\ \bibnamefont {Thouless}},\ }\bibfield  {title} {\bibinfo {title} {Fractional quantization of {H}all conductance},\ }\href {https://doi.org/10.1103/physrevb.28.1142} {\bibfield  {journal} {\bibinfo  {journal} {Physical Review B}\ }\textbf {\bibinfo {volume} {28}},\ \bibinfo {pages} {1142} (\bibinfo {year} {1983})}\BibitemShut {NoStop}%
\bibitem [{\citenamefont {Seidel}\ \emph {et~al.}(2005)\citenamefont {Seidel}, \citenamefont {Fu}, \citenamefont {Lee}, \citenamefont {Leinaas},\ and\ \citenamefont {Moore}}]{Seidel2005}%
  \BibitemOpen
  \bibfield  {author} {\bibinfo {author} {\bibfnamefont {A.}~\bibnamefont {Seidel}}, \bibinfo {author} {\bibfnamefont {H.}~\bibnamefont {Fu}}, \bibinfo {author} {\bibfnamefont {D.-H.}\ \bibnamefont {Lee}}, \bibinfo {author} {\bibfnamefont {J.~M.}\ \bibnamefont {Leinaas}},\ and\ \bibinfo {author} {\bibfnamefont {J.}~\bibnamefont {Moore}},\ }\bibfield  {title} {\bibinfo {title} {Incompressible quantum liquids and new conservation laws},\ }\href {https://doi.org/10.1103/physrevlett.95.266405} {\bibfield  {journal} {\bibinfo  {journal} {Physical Review Letters}\ }\textbf {\bibinfo {volume} {95}},\ \bibinfo {pages} {266405} (\bibinfo {year} {2005})}\BibitemShut {NoStop}%
\bibitem [{\citenamefont {Nakamura}\ \emph {et~al.}(2010)\citenamefont {Nakamura}, \citenamefont {Bergholtz},\ and\ \citenamefont {Suorsa}}]{Nakamura2010}%
  \BibitemOpen
  \bibfield  {author} {\bibinfo {author} {\bibfnamefont {M.}~\bibnamefont {Nakamura}}, \bibinfo {author} {\bibfnamefont {E.~J.}\ \bibnamefont {Bergholtz}},\ and\ \bibinfo {author} {\bibfnamefont {J.}~\bibnamefont {Suorsa}},\ }\bibfield  {title} {\bibinfo {title} {Link between the hierarchy of fractional quantum {H}all states and {H}aldane’s conjecture for quantum spin chains},\ }\href {https://doi.org/10.1103/physrevb.81.165102} {\bibfield  {journal} {\bibinfo  {journal} {Physical Review B}\ }\textbf {\bibinfo {volume} {81}},\ \bibinfo {pages} {165102} (\bibinfo {year} {2010})}\BibitemShut {NoStop}%
\bibitem [{\citenamefont {Haldane}\ and\ \citenamefont {Rezayi}(1985)}]{Haldane1985a}%
  \BibitemOpen
  \bibfield  {author} {\bibinfo {author} {\bibfnamefont {F.~D.~M.}\ \bibnamefont {Haldane}}\ and\ \bibinfo {author} {\bibfnamefont {E.~H.}\ \bibnamefont {Rezayi}},\ }\bibfield  {title} {\bibinfo {title} {Periodic {L}aughlin-{J}astrow wave functions for the fractional quantized hall effect},\ }\href {https://doi.org/10.1103/physrevb.31.2529} {\bibfield  {journal} {\bibinfo  {journal} {Physical Review B}\ }\textbf {\bibinfo {volume} {31}},\ \bibinfo {pages} {2529} (\bibinfo {year} {1985})}\BibitemShut {NoStop}%
\bibitem [{\citenamefont {Chen}\ and\ \citenamefont {Seidel}(2015)}]{Chen2015}%
  \BibitemOpen
  \bibfield  {author} {\bibinfo {author} {\bibfnamefont {L.}~\bibnamefont {Chen}}\ and\ \bibinfo {author} {\bibfnamefont {A.}~\bibnamefont {Seidel}},\ }\bibfield  {title} {\bibinfo {title} {Algebraic approach to the study of zero modes of {H}aldane pseudopotentials},\ }\href {https://doi.org/10.1103/physrevb.91.085103} {\bibfield  {journal} {\bibinfo  {journal} {Physical Review B}\ }\textbf {\bibinfo {volume} {91}},\ \bibinfo {pages} {085103} (\bibinfo {year} {2015})}\BibitemShut {NoStop}%
\bibitem [{\citenamefont {Trugman}\ and\ \citenamefont {Kivelson}(1985)}]{Trugman1985}%
  \BibitemOpen
  \bibfield  {author} {\bibinfo {author} {\bibfnamefont {S.~A.}\ \bibnamefont {Trugman}}\ and\ \bibinfo {author} {\bibfnamefont {S.}~\bibnamefont {Kivelson}},\ }\bibfield  {title} {\bibinfo {title} {Exact results for the fractional quantum {H}all effect with general interactions},\ }\href {https://doi.org/10.1103/physrevb.31.5280} {\bibfield  {journal} {\bibinfo  {journal} {Physical Review B}\ }\textbf {\bibinfo {volume} {31}},\ \bibinfo {pages} {5280} (\bibinfo {year} {1985})}\BibitemShut {NoStop}%
\bibitem [{\citenamefont {Haldane}(2011)}]{Haldane2011}%
  \BibitemOpen
  \bibfield  {author} {\bibinfo {author} {\bibfnamefont {F.~D.~M.}\ \bibnamefont {Haldane}},\ }\bibfield  {title} {\bibinfo {title} {Geometrical description of the fractional quantum {H}all effect},\ }\href {https://doi.org/10.1103/physrevlett.107.116801} {\bibfield  {journal} {\bibinfo  {journal} {Physical Review Letters}\ }\textbf {\bibinfo {volume} {107}},\ \bibinfo {pages} {116801} (\bibinfo {year} {2011})}\BibitemShut {NoStop}%
\bibitem [{\citenamefont {Chen}\ \emph {et~al.}(2019)\citenamefont {Chen}, \citenamefont {Bandyopadhyay}, \citenamefont {Yang},\ and\ \citenamefont {Seidel}}]{Chen2019}%
  \BibitemOpen
  \bibfield  {author} {\bibinfo {author} {\bibfnamefont {L.}~\bibnamefont {Chen}}, \bibinfo {author} {\bibfnamefont {S.}~\bibnamefont {Bandyopadhyay}}, \bibinfo {author} {\bibfnamefont {K.}~\bibnamefont {Yang}},\ and\ \bibinfo {author} {\bibfnamefont {A.}~\bibnamefont {Seidel}},\ }\bibfield  {title} {\bibinfo {title} {Composite fermions in {F}ock space: Operator algebra, recursion relations, and order parameters},\ }\href {https://doi.org/10.1103/physrevb.100.045136} {\bibfield  {journal} {\bibinfo  {journal} {Physical Review B}\ }\textbf {\bibinfo {volume} {100}},\ \bibinfo {pages} {045136} (\bibinfo {year} {2019})}\BibitemShut {NoStop}%
\bibitem [{\citenamefont {Rezayi}\ and\ \citenamefont {Haldane}(1994)}]{Rezayi1994}%
  \BibitemOpen
  \bibfield  {author} {\bibinfo {author} {\bibfnamefont {E.~H.}\ \bibnamefont {Rezayi}}\ and\ \bibinfo {author} {\bibfnamefont {F.~D.~M.}\ \bibnamefont {Haldane}},\ }\bibfield  {title} {\bibinfo {title} {Laughlin state on stretched and squeezed cylinders and edge excitations in the quantum {H}all effect},\ }\href {https://doi.org/10.1103/physrevb.50.17199} {\bibfield  {journal} {\bibinfo  {journal} {Physical Review B}\ }\textbf {\bibinfo {volume} {50}},\ \bibinfo {pages} {17199} (\bibinfo {year} {1994})}\BibitemShut {NoStop}%
\bibitem [{\citenamefont {Nakamura}\ \emph {et~al.}(2012)\citenamefont {Nakamura}, \citenamefont {Wang},\ and\ \citenamefont {Bergholtz}}]{Nakamura2012}%
  \BibitemOpen
  \bibfield  {author} {\bibinfo {author} {\bibfnamefont {M.}~\bibnamefont {Nakamura}}, \bibinfo {author} {\bibfnamefont {Z.-Y.}\ \bibnamefont {Wang}},\ and\ \bibinfo {author} {\bibfnamefont {E.~J.}\ \bibnamefont {Bergholtz}},\ }\bibfield  {title} {\bibinfo {title} {Exactly solvable fermion chain describing $\nu =1/3$ fractional quantum {H}all state},\ }\href {https://doi.org/10.1103/physrevlett.109.016401} {\bibfield  {journal} {\bibinfo  {journal} {Physical Review Letters}\ }\textbf {\bibinfo {volume} {109}},\ \bibinfo {pages} {016401} (\bibinfo {year} {2012})}\BibitemShut {NoStop}%
\bibitem [{\citenamefont {Papić}(2014)}]{Papic2014}%
  \BibitemOpen
  \bibfield  {author} {\bibinfo {author} {\bibfnamefont {Z.}~\bibnamefont {Papić}},\ }\bibfield  {title} {\bibinfo {title} {Solvable models for unitary and nonunitary topological phases},\ }\href {https://doi.org/10.1103/physrevb.90.075304} {\bibfield  {journal} {\bibinfo  {journal} {Physical Review B}\ }\textbf {\bibinfo {volume} {90}},\ \bibinfo {pages} {075304} (\bibinfo {year} {2014})}\BibitemShut {NoStop}%
\bibitem [{\citenamefont {Sala}\ \emph {et~al.}(2020)\citenamefont {Sala}, \citenamefont {Rakovszky}, \citenamefont {Verresen}, \citenamefont {Knap},\ and\ \citenamefont {Pollmann}}]{Sala2020}%
  \BibitemOpen
  \bibfield  {author} {\bibinfo {author} {\bibfnamefont {P.}~\bibnamefont {Sala}}, \bibinfo {author} {\bibfnamefont {T.}~\bibnamefont {Rakovszky}}, \bibinfo {author} {\bibfnamefont {R.}~\bibnamefont {Verresen}}, \bibinfo {author} {\bibfnamefont {M.}~\bibnamefont {Knap}},\ and\ \bibinfo {author} {\bibfnamefont {F.}~\bibnamefont {Pollmann}},\ }\bibfield  {title} {\bibinfo {title} {Ergodicity breaking arising from {H}ilbert space fragmentation in dipole-conserving hamiltonians},\ }\href {https://doi.org/10.1103/physrevx.10.011047} {\bibfield  {journal} {\bibinfo  {journal} {Physical Review X}\ }\textbf {\bibinfo {volume} {10}},\ \bibinfo {pages} {011047} (\bibinfo {year} {2020})}\BibitemShut {NoStop}%
\bibitem [{\citenamefont {Zerba}\ \emph {et~al.}(2025)\citenamefont {Zerba}, \citenamefont {Seidel}, \citenamefont {Pollmann},\ and\ \citenamefont {Knap}}]{Zerba2024}%
  \BibitemOpen
  \bibfield  {author} {\bibinfo {author} {\bibfnamefont {C.}~\bibnamefont {Zerba}}, \bibinfo {author} {\bibfnamefont {A.}~\bibnamefont {Seidel}}, \bibinfo {author} {\bibfnamefont {F.}~\bibnamefont {Pollmann}},\ and\ \bibinfo {author} {\bibfnamefont {M.}~\bibnamefont {Knap}},\ }\bibfield  {title} {\bibinfo {title} {Emergent fracton hydrodynamics of ultracold atoms in partially filled landau levels},\ }\href {https://doi.org/10.1103/PRXQuantum.6.020321} {\bibfield  {journal} {\bibinfo  {journal} {PRX Quantum}\ }\textbf {\bibinfo {volume} {6}},\ \bibinfo {pages} {020321} (\bibinfo {year} {2025})}\BibitemShut {NoStop}%
\bibitem [{\citenamefont {Jansen}\ \emph {et~al.}(2008)\citenamefont {Jansen}, \citenamefont {Lieb},\ and\ \citenamefont {Seiler}}]{Jansen2008}%
  \BibitemOpen
  \bibfield  {author} {\bibinfo {author} {\bibfnamefont {S.}~\bibnamefont {Jansen}}, \bibinfo {author} {\bibfnamefont {E.~H.}\ \bibnamefont {Lieb}},\ and\ \bibinfo {author} {\bibfnamefont {R.}~\bibnamefont {Seiler}},\ }\bibfield  {title} {\bibinfo {title} {Symmetry breaking in {L}aughlin’s state on a cylinder},\ }\href {https://doi.org/10.1007/s00220-008-0576-4} {\bibfield  {journal} {\bibinfo  {journal} {Communications in Mathematical Physics}\ }\textbf {\bibinfo {volume} {285}},\ \bibinfo {pages} {503} (\bibinfo {year} {2008})}\BibitemShut {NoStop}%
\bibitem [{\citenamefont {Jansen}(2012)}]{Jansen2012}%
  \BibitemOpen
  \bibfield  {author} {\bibinfo {author} {\bibfnamefont {S.}~\bibnamefont {Jansen}},\ }\bibfield  {title} {\bibinfo {title} {Fermionic and bosonic {L}aughlin state on thick cylinders},\ }\bibfield  {journal} {\bibinfo  {journal} {Journal of Mathematical Physics}\ }\textbf {\bibinfo {volume} {53}},\ \href {https://doi.org/10.1063/1.4768250} {10.1063/1.4768250} (\bibinfo {year} {2012})\BibitemShut {NoStop}%
\bibitem [{\citenamefont {Rotondo}\ \emph {et~al.}(2016)\citenamefont {Rotondo}, \citenamefont {Molinari}, \citenamefont {Ratti},\ and\ \citenamefont {Gherardi}}]{Rotondo2016}%
  \BibitemOpen
  \bibfield  {author} {\bibinfo {author} {\bibfnamefont {P.}~\bibnamefont {Rotondo}}, \bibinfo {author} {\bibfnamefont {L.~G.}\ \bibnamefont {Molinari}}, \bibinfo {author} {\bibfnamefont {P.}~\bibnamefont {Ratti}},\ and\ \bibinfo {author} {\bibfnamefont {M.}~\bibnamefont {Gherardi}},\ }\bibfield  {title} {\bibinfo {title} {Devil’s staircase phase diagram of the fractional quantum {H}all effect in the thin-torus limit},\ }\href {https://doi.org/10.1103/physrevlett.116.256803} {\bibfield  {journal} {\bibinfo  {journal} {Physical Review Letters}\ }\textbf {\bibinfo {volume} {116}},\ \bibinfo {pages} {256803} (\bibinfo {year} {2016})}\BibitemShut {NoStop}%
\bibitem [{\citenamefont {Weerasinghe}\ and\ \citenamefont {Seidel}(2014)}]{Weerasinghe2014}%
  \BibitemOpen
  \bibfield  {author} {\bibinfo {author} {\bibfnamefont {A.}~\bibnamefont {Weerasinghe}}\ and\ \bibinfo {author} {\bibfnamefont {A.}~\bibnamefont {Seidel}},\ }\bibfield  {title} {\bibinfo {title} {Thin torus perturbative analysis of elementary excitations in the {G}affnian and {H}aldane-{R}ezayi quantum hall states},\ }\href {https://doi.org/10.1103/physrevb.90.125146} {\bibfield  {journal} {\bibinfo  {journal} {Physical Review B}\ }\textbf {\bibinfo {volume} {90}},\ \bibinfo {pages} {125146} (\bibinfo {year} {2014})}\BibitemShut {NoStop}%
\bibitem [{\citenamefont {Mazaheri}\ \emph {et~al.}(2015)\citenamefont {Mazaheri}, \citenamefont {Ortiz}, \citenamefont {Nussinov},\ and\ \citenamefont {Seidel}}]{Mazaheri2015}%
  \BibitemOpen
  \bibfield  {author} {\bibinfo {author} {\bibfnamefont {T.}~\bibnamefont {Mazaheri}}, \bibinfo {author} {\bibfnamefont {G.}~\bibnamefont {Ortiz}}, \bibinfo {author} {\bibfnamefont {Z.}~\bibnamefont {Nussinov}},\ and\ \bibinfo {author} {\bibfnamefont {A.}~\bibnamefont {Seidel}},\ }\bibfield  {title} {\bibinfo {title} {Zero modes, bosonization, and topological quantum order: The laughlin state in second quantization},\ }\href {https://doi.org/10.1103/physrevb.91.085115} {\bibfield  {journal} {\bibinfo  {journal} {Physical Review B}\ }\textbf {\bibinfo {volume} {91}},\ \bibinfo {pages} {085115} (\bibinfo {year} {2015})}\BibitemShut {NoStop}%
\bibitem [{\citenamefont {Weerasinghe}\ \emph {et~al.}(2016)\citenamefont {Weerasinghe}, \citenamefont {Mazaheri},\ and\ \citenamefont {Seidel}}]{Weerasinghe2016}%
  \BibitemOpen
  \bibfield  {author} {\bibinfo {author} {\bibfnamefont {A.}~\bibnamefont {Weerasinghe}}, \bibinfo {author} {\bibfnamefont {T.}~\bibnamefont {Mazaheri}},\ and\ \bibinfo {author} {\bibfnamefont {A.}~\bibnamefont {Seidel}},\ }\bibfield  {title} {\bibinfo {title} {Bounds for low-energy spectral properties of center-of-mass conserving positive two-body interactions},\ }\href {https://doi.org/10.1103/physrevb.93.155135} {\bibfield  {journal} {\bibinfo  {journal} {Physical Review B}\ }\textbf {\bibinfo {volume} {93}},\ \bibinfo {pages} {155135} (\bibinfo {year} {2016})}\BibitemShut {NoStop}%
\bibitem [{\citenamefont {Nachtergaele}\ \emph {et~al.}(2021)\citenamefont {Nachtergaele}, \citenamefont {Warzel},\ and\ \citenamefont {Young}}]{Nachtergaele2021}%
  \BibitemOpen
  \bibfield  {author} {\bibinfo {author} {\bibfnamefont {B.}~\bibnamefont {Nachtergaele}}, \bibinfo {author} {\bibfnamefont {S.}~\bibnamefont {Warzel}},\ and\ \bibinfo {author} {\bibfnamefont {A.}~\bibnamefont {Young}},\ }\bibfield  {title} {\bibinfo {title} {Spectral gaps and incompressibility in a $\nu = 1/3$ fractional quantum {H}all system},\ }\href {https://doi.org/10.1007/s00220-021-03997-0} {\bibfield  {journal} {\bibinfo  {journal} {Communications in Mathematical Physics}\ }\textbf {\bibinfo {volume} {383}},\ \bibinfo {pages} {1093} (\bibinfo {year} {2021})}\BibitemShut {NoStop}%
\bibitem [{\citenamefont {Girvin}\ \emph {et~al.}(1986)\citenamefont {Girvin}, \citenamefont {MacDonald},\ and\ \citenamefont {Platzman}}]{Girvin1986}%
  \BibitemOpen
  \bibfield  {author} {\bibinfo {author} {\bibfnamefont {S.~M.}\ \bibnamefont {Girvin}}, \bibinfo {author} {\bibfnamefont {A.~H.}\ \bibnamefont {MacDonald}},\ and\ \bibinfo {author} {\bibfnamefont {P.~M.}\ \bibnamefont {Platzman}},\ }\bibfield  {title} {\bibinfo {title} {Magneto-roton theory of collective excitations in the fractional quantum {H}all effect},\ }\href {https://doi.org/10.1103/physrevb.33.2481} {\bibfield  {journal} {\bibinfo  {journal} {Physical Review B}\ }\textbf {\bibinfo {volume} {33}},\ \bibinfo {pages} {2481} (\bibinfo {year} {1986})}\BibitemShut {NoStop}%
\bibitem [{\citenamefont {Repellin}\ \emph {et~al.}(2014)\citenamefont {Repellin}, \citenamefont {Neupert}, \citenamefont {Papić},\ and\ \citenamefont {Regnault}}]{Repellin2014}%
  \BibitemOpen
  \bibfield  {author} {\bibinfo {author} {\bibfnamefont {C.}~\bibnamefont {Repellin}}, \bibinfo {author} {\bibfnamefont {T.}~\bibnamefont {Neupert}}, \bibinfo {author} {\bibfnamefont {Z.}~\bibnamefont {Papić}},\ and\ \bibinfo {author} {\bibfnamefont {N.}~\bibnamefont {Regnault}},\ }\bibfield  {title} {\bibinfo {title} {Single-mode approximation for fractional {C}hern insulators and the fractional quantum {H}all effect on the torus},\ }\href {https://doi.org/10.1103/physrevb.90.045114} {\bibfield  {journal} {\bibinfo  {journal} {Physical Review B}\ }\textbf {\bibinfo {volume} {90}},\ \bibinfo {pages} {045114} (\bibinfo {year} {2014})}\BibitemShut {NoStop}%
\bibitem [{\citenamefont {Repellin}\ \emph {et~al.}(2015)\citenamefont {Repellin}, \citenamefont {Neupert}, \citenamefont {Bernevig},\ and\ \citenamefont {Regnault}}]{Repellin2015}%
  \BibitemOpen
  \bibfield  {author} {\bibinfo {author} {\bibfnamefont {C.}~\bibnamefont {Repellin}}, \bibinfo {author} {\bibfnamefont {T.}~\bibnamefont {Neupert}}, \bibinfo {author} {\bibfnamefont {B.~A.}\ \bibnamefont {Bernevig}},\ and\ \bibinfo {author} {\bibfnamefont {N.}~\bibnamefont {Regnault}},\ }\bibfield  {title} {\bibinfo {title} {Projective construction of the $\mathbb{Z}_k$ {R}ead-{R}ezayi fractional quantum {H}all states and their excitations on the torus geometry},\ }\href {https://doi.org/10.1103/physrevb.92.115128} {\bibfield  {journal} {\bibinfo  {journal} {Physical Review B}\ }\textbf {\bibinfo {volume} {92}},\ \bibinfo {pages} {115128} (\bibinfo {year} {2015})}\BibitemShut {NoStop}%
\bibitem [{\citenamefont {Shibata}\ and\ \citenamefont {Yoshioka}(2001)}]{Shibata2001}%
  \BibitemOpen
  \bibfield  {author} {\bibinfo {author} {\bibfnamefont {N.}~\bibnamefont {Shibata}}\ and\ \bibinfo {author} {\bibfnamefont {D.}~\bibnamefont {Yoshioka}},\ }\bibfield  {title} {\bibinfo {title} {Ground-state phase diagram of 2d electrons in a high landau level: A density-matrix renormalization group study},\ }\href {https://doi.org/10.1103/PhysRevLett.86.5755} {\bibfield  {journal} {\bibinfo  {journal} {Phys. Rev. Lett.}\ }\textbf {\bibinfo {volume} {86}},\ \bibinfo {pages} {5755} (\bibinfo {year} {2001})}\BibitemShut {NoStop}%
\bibitem [{\citenamefont {Zaletel}\ \emph {et~al.}(2013)\citenamefont {Zaletel}, \citenamefont {Mong},\ and\ \citenamefont {Pollmann}}]{Zaletel2013}%
  \BibitemOpen
  \bibfield  {author} {\bibinfo {author} {\bibfnamefont {M.~P.}\ \bibnamefont {Zaletel}}, \bibinfo {author} {\bibfnamefont {R.~S.~K.}\ \bibnamefont {Mong}},\ and\ \bibinfo {author} {\bibfnamefont {F.}~\bibnamefont {Pollmann}},\ }\bibfield  {title} {\bibinfo {title} {Topological characterization of fractional quantum hall ground states from microscopic hamiltonians},\ }\href {https://doi.org/10.1103/PhysRevLett.110.236801} {\bibfield  {journal} {\bibinfo  {journal} {Phys. Rev. Lett.}\ }\textbf {\bibinfo {volume} {110}},\ \bibinfo {pages} {236801} (\bibinfo {year} {2013})}\BibitemShut {NoStop}%
\bibitem [{\citenamefont {Shen}\ \emph {et~al.}(2025)\citenamefont {Shen}, \citenamefont {Lin}, \citenamefont {Lin}, \citenamefont {Xiao},\ and\ \citenamefont {Cao}}]{Shen2025}%
  \BibitemOpen
  \bibfield  {author} {\bibinfo {author} {\bibfnamefont {L.}~\bibnamefont {Shen}}, \bibinfo {author} {\bibfnamefont {M.}~\bibnamefont {Lin}}, \bibinfo {author} {\bibfnamefont {C.~Y.-Y.}\ \bibnamefont {Lin}}, \bibinfo {author} {\bibfnamefont {D.}~\bibnamefont {Xiao}},\ and\ \bibinfo {author} {\bibfnamefont {T.}~\bibnamefont {Cao}},\ }\href {https://arxiv.org/abs/2503.13294} {\bibinfo {title} {Realization of fermionic laughlin state on a quantum processor}} (\bibinfo {year} {2025}),\ \Eprint {https://arxiv.org/abs/2503.13294} {arXiv:2503.13294 [quant-ph]} \BibitemShut {NoStop}%
\bibitem [{\citenamefont {Girvin}\ and\ \citenamefont {Yang}(2019)}]{Girvin2019}%
  \BibitemOpen
  \bibfield  {author} {\bibinfo {author} {\bibfnamefont {S.~M.}\ \bibnamefont {Girvin}}\ and\ \bibinfo {author} {\bibfnamefont {K.}~\bibnamefont {Yang}},\ }\href {https://doi.org/10.1017/9781316480649} {\emph {\bibinfo {title} {Modern Condensed Matter Physics}}}\ (\bibinfo  {publisher} {Cambridge University Press},\ \bibinfo {year} {2019})\BibitemShut {NoStop}%
\bibitem [{\citenamefont {Haldane}(1985)}]{Haldane1985}%
  \BibitemOpen
  \bibfield  {author} {\bibinfo {author} {\bibfnamefont {F.~D.~M.}\ \bibnamefont {Haldane}},\ }\bibfield  {title} {\bibinfo {title} {Many-particle translational symmetries of two-dimensional electrons at rational {L}andau-level filling},\ }\href {https://doi.org/10.1103/physrevlett.55.2095} {\bibfield  {journal} {\bibinfo  {journal} {Physical Review Letters}\ }\textbf {\bibinfo {volume} {55}},\ \bibinfo {pages} {2095} (\bibinfo {year} {1985})}\BibitemShut {NoStop}%
\bibitem [{\citenamefont {Shankar}\ and\ \citenamefont {Murthy}(1997)}]{Shankar1997}%
  \BibitemOpen
  \bibfield  {author} {\bibinfo {author} {\bibfnamefont {R.}~\bibnamefont {Shankar}}\ and\ \bibinfo {author} {\bibfnamefont {G.}~\bibnamefont {Murthy}},\ }\bibfield  {title} {\bibinfo {title} {Towards a field theory of fractional quantum {H}all states},\ }\href {https://doi.org/10.1103/physrevlett.79.4437} {\bibfield  {journal} {\bibinfo  {journal} {Physical Review Letters}\ }\textbf {\bibinfo {volume} {79}},\ \bibinfo {pages} {4437} (\bibinfo {year} {1997})}\BibitemShut {NoStop}%
\bibitem [{\citenamefont {Read}(1998)}]{Read1998}%
  \BibitemOpen
  \bibfield  {author} {\bibinfo {author} {\bibfnamefont {N.}~\bibnamefont {Read}},\ }\bibfield  {title} {\bibinfo {title} {Lowest-landau-level theory of the quantum {H}all effect: The {F}ermi-liquid-like state of bosons at filling factor one},\ }\href {https://doi.org/10.1103/physrevb.58.16262} {\bibfield  {journal} {\bibinfo  {journal} {Physical Review B}\ }\textbf {\bibinfo {volume} {58}},\ \bibinfo {pages} {16262} (\bibinfo {year} {1998})}\BibitemShut {NoStop}%
\bibitem [{\citenamefont {Pasquier}\ and\ \citenamefont {Haldane}(1998)}]{Pasquier1998}%
  \BibitemOpen
  \bibfield  {author} {\bibinfo {author} {\bibfnamefont {V.}~\bibnamefont {Pasquier}}\ and\ \bibinfo {author} {\bibfnamefont {F.}~\bibnamefont {Haldane}},\ }\bibfield  {title} {\bibinfo {title} {A dipole interpretation of the $\nu=1/2$ state},\ }\href {https://doi.org/10.1016/s0550-3213(98)00069-8} {\bibfield  {journal} {\bibinfo  {journal} {Nuclear Physics B}\ }\textbf {\bibinfo {volume} {516}},\ \bibinfo {pages} {719} (\bibinfo {year} {1998})}\BibitemShut {NoStop}%
\bibitem [{\citenamefont {Griffiths}\ and\ \citenamefont {Schroeter}(2018)}]{Griffiths2018}%
  \BibitemOpen
  \bibfield  {author} {\bibinfo {author} {\bibfnamefont {D.~J.}\ \bibnamefont {Griffiths}}\ and\ \bibinfo {author} {\bibfnamefont {D.~F.}\ \bibnamefont {Schroeter}},\ }\href {https://doi.org/10.1017/9781316995433} {\emph {\bibinfo {title} {Introduction to Quantum Mechanics}}}\ (\bibinfo  {publisher} {Cambridge University Press},\ \bibinfo {year} {2018})\BibitemShut {NoStop}%
\bibitem [{\citenamefont {Golkar}\ \emph {et~al.}(2016)\citenamefont {Golkar}, \citenamefont {Nguyen}, \citenamefont {Roberts},\ and\ \citenamefont {Son}}]{Golkar2016}%
  \BibitemOpen
  \bibfield  {author} {\bibinfo {author} {\bibfnamefont {S.}~\bibnamefont {Golkar}}, \bibinfo {author} {\bibfnamefont {D.~X.}\ \bibnamefont {Nguyen}}, \bibinfo {author} {\bibfnamefont {M.~M.}\ \bibnamefont {Roberts}},\ and\ \bibinfo {author} {\bibfnamefont {D.~T.}\ \bibnamefont {Son}},\ }\bibfield  {title} {\bibinfo {title} {Higher-spin theory of the magnetorotons},\ }\href {https://doi.org/10.1103/PhysRevLett.117.216403} {\bibfield  {journal} {\bibinfo  {journal} {Phys. Rev. Lett.}\ }\textbf {\bibinfo {volume} {117}},\ \bibinfo {pages} {216403} (\bibinfo {year} {2016})}\BibitemShut {NoStop}%
\bibitem [{\citenamefont {Balram}\ and\ \citenamefont {Pu}(2017)}]{Balram2017}%
  \BibitemOpen
  \bibfield  {author} {\bibinfo {author} {\bibfnamefont {A.~C.}\ \bibnamefont {Balram}}\ and\ \bibinfo {author} {\bibfnamefont {S.}~\bibnamefont {Pu}},\ }\bibfield  {title} {\bibinfo {title} {Positions of the magnetoroton minima in the fractional quantum hall effect},\ }\href {https://doi.org/10.1140/epjb/e2017-80177-5} {\bibfield  {journal} {\bibinfo  {journal} {The European Physical Journal B}\ }\textbf {\bibinfo {volume} {90}},\ \bibinfo {pages} {124} (\bibinfo {year} {2017})}\BibitemShut {NoStop}%
\bibitem [{\citenamefont {Simon}\ and\ \citenamefont {Halperin}(1993)}]{Simon1993}%
  \BibitemOpen
  \bibfield  {author} {\bibinfo {author} {\bibfnamefont {S.~H.}\ \bibnamefont {Simon}}\ and\ \bibinfo {author} {\bibfnamefont {B.~I.}\ \bibnamefont {Halperin}},\ }\bibfield  {title} {\bibinfo {title} {Finite-wave-vector electromagnetic response of fractional quantized hall states},\ }\href {https://doi.org/10.1103/PhysRevB.48.17368} {\bibfield  {journal} {\bibinfo  {journal} {Physical Review B}\ }\textbf {\bibinfo {volume} {48}},\ \bibinfo {pages} {17368} (\bibinfo {year} {1993})}\BibitemShut {NoStop}%
\bibitem [{\citenamefont {Tersoff}\ and\ \citenamefont {Hamann}(1985)}]{Tersoff1985}%
  \BibitemOpen
  \bibfield  {author} {\bibinfo {author} {\bibfnamefont {J.}~\bibnamefont {Tersoff}}\ and\ \bibinfo {author} {\bibfnamefont {D.~R.}\ \bibnamefont {Hamann}},\ }\bibfield  {title} {\bibinfo {title} {Theory of the scanning tunneling microscope},\ }\href {https://doi.org/10.1103/physrevb.31.805} {\bibfield  {journal} {\bibinfo  {journal} {Physical Review B}\ }\textbf {\bibinfo {volume} {31}},\ \bibinfo {pages} {805} (\bibinfo {year} {1985})}\BibitemShut {NoStop}%
\bibitem [{\citenamefont {Papić}\ \emph {et~al.}(2018)\citenamefont {Papić}, \citenamefont {Mong}, \citenamefont {Yazdani},\ and\ \citenamefont {Zaletel}}]{Papic2018}%
  \BibitemOpen
  \bibfield  {author} {\bibinfo {author} {\bibfnamefont {Z.}~\bibnamefont {Papić}}, \bibinfo {author} {\bibfnamefont {R.~S.~K.}\ \bibnamefont {Mong}}, \bibinfo {author} {\bibfnamefont {A.}~\bibnamefont {Yazdani}},\ and\ \bibinfo {author} {\bibfnamefont {M.~P.}\ \bibnamefont {Zaletel}},\ }\bibfield  {title} {\bibinfo {title} {Imaging anyons with scanning tunneling microscopy},\ }\href {https://doi.org/10.1103/physrevx.8.011037} {\bibfield  {journal} {\bibinfo  {journal} {Physical Review X}\ }\textbf {\bibinfo {volume} {8}},\ \bibinfo {pages} {011037} (\bibinfo {year} {2018})}\BibitemShut {NoStop}%
\bibitem [{\citenamefont {Jain}\ and\ \citenamefont {Peterson}(2005)}]{Jain2005}%
  \BibitemOpen
  \bibfield  {author} {\bibinfo {author} {\bibfnamefont {J.~K.}\ \bibnamefont {Jain}}\ and\ \bibinfo {author} {\bibfnamefont {M.~R.}\ \bibnamefont {Peterson}},\ }\bibfield  {title} {\bibinfo {title} {Reconstructing the electron in a fractionalized quantum fluid},\ }\href {https://doi.org/10.1103/physrevlett.94.186808} {\bibfield  {journal} {\bibinfo  {journal} {Physical Review Letters}\ }\textbf {\bibinfo {volume} {94}},\ \bibinfo {pages} {186808} (\bibinfo {year} {2005})}\BibitemShut {NoStop}%
\bibitem [{\citenamefont {He}\ \emph {et~al.}(1993)\citenamefont {He}, \citenamefont {Platzman},\ and\ \citenamefont {Halperin}}]{He1993}%
  \BibitemOpen
  \bibfield  {author} {\bibinfo {author} {\bibfnamefont {S.}~\bibnamefont {He}}, \bibinfo {author} {\bibfnamefont {P.~M.}\ \bibnamefont {Platzman}},\ and\ \bibinfo {author} {\bibfnamefont {B.~I.}\ \bibnamefont {Halperin}},\ }\bibfield  {title} {\bibinfo {title} {Tunneling into a two-dimensional electron system in a strong magnetic field},\ }\href {https://doi.org/10.1103/physrevlett.71.777} {\bibfield  {journal} {\bibinfo  {journal} {Physical Review Letters}\ }\textbf {\bibinfo {volume} {71}},\ \bibinfo {pages} {777} (\bibinfo {year} {1993})}\BibitemShut {NoStop}%
\bibitem [{\citenamefont {Gattu}\ \emph {et~al.}(2024)\citenamefont {Gattu}, \citenamefont {Sreejith},\ and\ \citenamefont {Jain}}]{Gattu2024}%
  \BibitemOpen
  \bibfield  {author} {\bibinfo {author} {\bibfnamefont {M.}~\bibnamefont {Gattu}}, \bibinfo {author} {\bibfnamefont {G.~J.}\ \bibnamefont {Sreejith}},\ and\ \bibinfo {author} {\bibfnamefont {J.~K.}\ \bibnamefont {Jain}},\ }\bibfield  {title} {\bibinfo {title} {Scanning tunneling microscopy of fractional quantum {H}all states: Spectroscopy of composite-fermion bound states},\ }\href {https://doi.org/10.1103/physrevb.109.l201123} {\bibfield  {journal} {\bibinfo  {journal} {Physical Review B}\ }\textbf {\bibinfo {volume} {109}},\ \bibinfo {pages} {l201123} (\bibinfo {year} {2024})}\BibitemShut {NoStop}%
\bibitem [{\citenamefont {Pu}\ \emph {et~al.}(2024)\citenamefont {Pu}, \citenamefont {Balram}, \citenamefont {Hu}, \citenamefont {Tsui}, \citenamefont {He}, \citenamefont {Regnault}, \citenamefont {Zaletel}, \citenamefont {Yazdani},\ and\ \citenamefont {Papić}}]{Pu2024}%
  \BibitemOpen
  \bibfield  {author} {\bibinfo {author} {\bibfnamefont {S.}~\bibnamefont {Pu}}, \bibinfo {author} {\bibfnamefont {A.~C.}\ \bibnamefont {Balram}}, \bibinfo {author} {\bibfnamefont {Y.}~\bibnamefont {Hu}}, \bibinfo {author} {\bibfnamefont {Y.-C.}\ \bibnamefont {Tsui}}, \bibinfo {author} {\bibfnamefont {M.}~\bibnamefont {He}}, \bibinfo {author} {\bibfnamefont {N.}~\bibnamefont {Regnault}}, \bibinfo {author} {\bibfnamefont {M.~P.}\ \bibnamefont {Zaletel}}, \bibinfo {author} {\bibfnamefont {A.}~\bibnamefont {Yazdani}},\ and\ \bibinfo {author} {\bibfnamefont {Z.}~\bibnamefont {Papić}},\ }\bibfield  {title} {\bibinfo {title} {Fingerprints of composite fermion lambda levels in scanning tunneling microscopy},\ }\href {https://doi.org/10.1103/physrevb.110.l081107} {\bibfield  {journal} {\bibinfo  {journal} {Physical Review B}\ }\textbf {\bibinfo {volume} {110}},\ \bibinfo {pages} {l081107} (\bibinfo {year} {2024})}\BibitemShut {NoStop}%
\bibitem [{\citenamefont {Pu}\ \emph {et~al.}(2017)\citenamefont {Pu}, \citenamefont {Wu},\ and\ \citenamefont {Jain}}]{Pu2017}%
  \BibitemOpen
  \bibfield  {author} {\bibinfo {author} {\bibfnamefont {S.}~\bibnamefont {Pu}}, \bibinfo {author} {\bibfnamefont {Y.-H.}\ \bibnamefont {Wu}},\ and\ \bibinfo {author} {\bibfnamefont {J.~K.}\ \bibnamefont {Jain}},\ }\bibfield  {title} {\bibinfo {title} {Composite fermions on a torus},\ }\href {https://doi.org/10.1103/physrevb.96.195302} {\bibfield  {journal} {\bibinfo  {journal} {Physical Review B}\ }\textbf {\bibinfo {volume} {96}},\ \bibinfo {pages} {195302} (\bibinfo {year} {2017})}\BibitemShut {NoStop}%
\bibitem [{\citenamefont {Bernevig}\ and\ \citenamefont {Regnault}(2012)}]{Bernevig2012}%
  \BibitemOpen
  \bibfield  {author} {\bibinfo {author} {\bibfnamefont {B.~A.}\ \bibnamefont {Bernevig}}\ and\ \bibinfo {author} {\bibfnamefont {N.}~\bibnamefont {Regnault}},\ }\href {https://arxiv.org/abs/1204.5682} {\bibinfo {title} {Thin-torus limit of fractional topological insulators}} (\bibinfo {year} {2012}),\ \Eprint {https://arxiv.org/abs/1204.5682} {arXiv:1204.5682 [cond-mat.str-el]} \BibitemShut {NoStop}%
\bibitem [{\citenamefont {Han}\ \emph {et~al.}(2023)\citenamefont {Han}, \citenamefont {Lake},\ and\ \citenamefont {Ro}}]{Han2023}%
  \BibitemOpen
  \bibfield  {author} {\bibinfo {author} {\bibfnamefont {J.~H.}\ \bibnamefont {Han}}, \bibinfo {author} {\bibfnamefont {E.}~\bibnamefont {Lake}},\ and\ \bibinfo {author} {\bibfnamefont {S.}~\bibnamefont {Ro}},\ }\bibfield  {title} {\bibinfo {title} {Scaling and localization in multipole-conserving diffusion},\ }\href@noop {} {\bibfield  {journal} {\bibinfo  {journal} {arXiv:2304.03276}\ } (\bibinfo {year} {2023})},\ \Eprint {https://arxiv.org/abs/2304.03276} {arXiv:2304.03276 [cond-mat.stat-mech]} \BibitemShut {NoStop}%
\bibitem [{\citenamefont {Zaletel}\ and\ \citenamefont {Mong}(2012)}]{Zaletel2012}%
  \BibitemOpen
  \bibfield  {author} {\bibinfo {author} {\bibfnamefont {M.~P.}\ \bibnamefont {Zaletel}}\ and\ \bibinfo {author} {\bibfnamefont {R.~S.~K.}\ \bibnamefont {Mong}},\ }\bibfield  {title} {\bibinfo {title} {Exact matrix product states for quantum hall wave functions},\ }\href {https://doi.org/10.1103/PhysRevB.86.245305} {\bibfield  {journal} {\bibinfo  {journal} {Phys. Rev. B}\ }\textbf {\bibinfo {volume} {86}},\ \bibinfo {pages} {245305} (\bibinfo {year} {2012})}\BibitemShut {NoStop}%
\bibitem [{dat()}]{data}%
  \BibitemOpen
  \href@noop {} {}\bibinfo {note} {The data that support the findings of this article are openly available at \url{https://github.com/JMAdhidewata/24_fqhe_torus}}\BibitemShut {NoStop}%
\bibitem [{\citenamefont {Lemm}\ \emph {et~al.}(2024)\citenamefont {Lemm}, \citenamefont {Nachtergaele}, \citenamefont {Warzel},\ and\ \citenamefont {Young}}]{Lemm2024}%
  \BibitemOpen
  \bibfield  {author} {\bibinfo {author} {\bibfnamefont {M.}~\bibnamefont {Lemm}}, \bibinfo {author} {\bibfnamefont {B.}~\bibnamefont {Nachtergaele}}, \bibinfo {author} {\bibfnamefont {S.}~\bibnamefont {Warzel}},\ and\ \bibinfo {author} {\bibfnamefont {A.}~\bibnamefont {Young}},\ }\href {https://arxiv.org/abs/2410.11645} {\bibinfo {title} {The charge gap is greater than the neutral gap in fractional quantum hall systems}} (\bibinfo {year} {2024}),\ \Eprint {https://arxiv.org/abs/2410.11645} {arXiv:2410.11645 [cond-mat.str-el]} \BibitemShut {NoStop}%
\end{thebibliography}
%

\end{document}